\documentclass[usenatbib]{mn2e}

\usepackage{subfigure}
\usepackage{graphicx}

\title[A Uniformly Derived Catalogue of Exoplanets from Radial Velocities]{A Uniformly Derived Catalogue of Exoplanets from Radial Velocities}
\author[Morgan D.~J.~Hollis, Sreekumar T.~Balan, Greg Lever, and Ofer Lahav]{Morgan D.~J.~Hollis $^{1}$\thanks{E-mail: mdjh@star.ucl.ac.uk}, Sreekumar T.~Balan$^{2}$\thanks{E-mail: st452@mrao.cam.ac.uk}, Greg Lever$^{1}$\thanks{E-mail: gl319@cam.ac.uk}, and Ofer Lahav$^{1}$\thanks{E-mail: lahav@star.ucl.ac.uk}\\
$^{1}$Department of Physics and Astronomy, University College London, Gower Street, London  WC1E 6BT, UK\\
$^{2}$Astrophysics Group, Cavendish Laboratory, J J Thomson Avenue, Cambridge CB3 0HE, UK}
\begin{document}
\pagerange{\pageref{firstpage}--\pageref{lastpage}} \pubyear{2011}
\maketitle
\label{firstpage}

\begin{abstract}
A new catalogue of extrasolar planets is presented by re-analysing a selection of published radial velocity data sets using \textsc{exofit}~\citep{Balan2009}. All objects are treated on an equal footing within a Bayesian framework, to give orbital parameters for 94 exoplanetary systems. Model selection (between one- and two-planet solutions) is then performed, using both a visual flagging method and a standard chi-square analysis, with agreement between the two methods for $99\%$ of the systems considered. The catalogue is to be made available online, and this `proof of concept' study may be maintained and extended in the future to incorporate all systems with publicly available radial velocity data, as well as transit and microlensing data.
\end{abstract}

\section{Introduction}
\label{sec:intro}
Since the discovery of the first extrasolar planet in 1995~\citep{Mayor1995}, the research on extrasolar planets has undergone exponential expansion. A wide range of search methods have been developed during this period, resulting in the discovery of more than 700 planets to date, the majority of which have been from the radial velocity method. Traditional data reduction methods use a periodogram \citep{lomb1973,scargle1982} to fix the orbital period and then the Levenberg-Marquardt minimisation~\citep{levenberg1944,marquardt1964} to fit the other orbital parameters. A catalogue of exoplanets has already been published by \citet{butler2006} using this method to extract the orbital parameters of exoplanets. Recently, Bayesian MCMC methods have been introduced by \citet{gregory2005,ford2005,ford2007} as a replacement for the traditional data reduction pipeline. \textsc{exofit} \citep{Balan2009} is a freely available tool for estimating orbital parameters of extrasolar planets from radial velocity data using a Bayesian framework. Here are analysed 94 previously published data sets using \textsc{exofit}, forming a new, uniformly derived catalogue of exoplanets from a Bayesian perspective. 

Statistical properties of the distribution of orbital parameters are critical for explaining the planetary formation process. It has been argued that there is now a statistically significant number of known companions to make inferences about the correlation between orbital elements. Early discussions on this subject can be found in a series of articles on the statistical properties of exoplanets by \citet{urdy2003,Santos2003,eggenberger2004} and \citet{Halbwachs2005}. The statistical discussion in this article is informed by the comparison to the published catalogues at \textit{http://www.exoplanet.eu}~\citep{Schneider2011} and \textit{http://exoplanets.org}~\citep{Wright2011}.

The rest of this article is structured as follows: in Sections~\ref{sec:bayes_framework} and \ref{sec:exofit} the Bayesian framework and \textsc{exofit} software package are introduced. The data analysis pipeline is described in Section~\ref{sec:data_analysis}, model selection is discussed in Section~\ref{sec:model}, and the catalogue is presented in Section~\ref{sec:catalogue}, and in the tables at the end of the article. The statistical properties of the distributions of various orbital parameters are discussed in Section~\ref{sec:param_comparison} and the results are summarised in Section~\ref{sec:discussion}.

\section{Bayesian Framework}
\label{sec:bayes_framework}
The Bayesian framework provides a transparent way of making probabilistic inferences from data. It is based on Bayes' theorem, which states that for a given model $H$ with a set of parameters $\mathbf{\Theta}$ and data $\mathbf{D}$, the posterior probability distribution of parameters $\Pr(\mathbf{\Theta}|\mathbf{D},H)$ is proportional to the prior probability distribution $\Pr(\mathbf{\Theta}|H)$ times the likelihood of data, $\Pr(\mathbf{D}|\mathbf{\Theta},H)$. Using standard mathematical notation one can write,

\begin{equation}
\Pr(\mathbf{\Theta}|\mathbf{D},H)=\frac{\Pr(\mathbf{D}|\mathbf{\Theta},H)\Pr(\mathbf{\Theta}|H)}{\Pr(\mathbf{D}|H)}.
\label{eqn:bayes}
\end{equation}

The denominator of the right hand side of the above equation is called the Bayesian Evidence. Since it is the estimation of parameters that is of interest here, this term can be considered as a normalising constant and Equation~\ref{eqn:bayes} can be written as,

\begin{equation}
\Pr(\mathbf{\Theta}|\mathbf{D},H) \propto \Pr(\mathbf{D}|\mathbf{\Theta},H)\Pr(\mathbf{\Theta}|H).
\label{eqn:posterior_prob}
\end{equation}

The key step in the Bayesian approach is to obtain the posterior distribution of parameters accurately. The inferences are then derived from the posterior distribution. The Markov Chain Monte Carlo (MCMC) method is a widely employed technique for simulating the posterior distribution (the left hand side of Equation~\ref{eqn:posterior_prob}). The basic steps in Bayesian parameter estimation can be summarised as follows: 
\begin{enumerate}
\item model the observed data, i.e. construct the likelihood function,
\item choose the prior probability distributions of parameters,
\item obtain the posterior probability distribution,
\item make inferences based on the posterior probability distribution.
\end{enumerate}

\textsc{exofit} is a software package that estimates the orbital parameters of extrasolar planets, following the steps outlined above. It should be noted that \textsc{exofit} does not perform any Bayesian model selection - for a discussion of the relation of this aspect of the Bayesian framework to this study, the reader is directed to Section~\ref{sec:model}.

\section{Exofit}
\label{sec:exofit}
\textsc{exofit}~\citep{Balan2009} is a publicly available tool for extracting orbital parameters of exoplanets from radial velocity measurements. It uses the MCMC method to simulate the posterior probability distribution of the orbital parameters. The likelihood of data $\Pr(\mathbf{D}|\mathbf{\Theta},H)$ in Equation~\ref{eqn:posterior_prob} connects the mathematical model to the observed data. The radial velocity model and the corresponding Gaussian likelihood function are given in \citet{Balan2009}, where the choice of likelihood is based on \citet{gregory2005}. The prior probabilities are as used by \citet{ford2007}. \textsc{exofit} then generates samples from the posterior distributions of the orbital parameters in the mathematical model, which can be analysed with the aid of any statistical software. Details of the algorithmic structure of the code, including methods of controlling chain mixing and assessing convergence, can be found in~\citet{Balan2009}. For further information, the reader is directed to the \textsc{exofit} user's guide.

\section{Data Analysis}
\label{sec:data_analysis} 
As of 2009 August 21, when the data sets were extracted from the literature (\citet{butler2006}, and references therein), there were 295 planetary systems detected using the radial velocity method according to \textit{http://www.exoplanet.eu}, with 346 individual planets in total. Some radial velocity data sets from the literature have fewer than ten entries and as such are not appropriate for use with Bayesian inference methods, as a radial velocity data set will need to include at least half an orbital period of a potential planetary companion. Hence data sets with not enough measurements to give accurate orbital solutions were not included. Also, the radial velocity data of any systems with more than 2 confirmed planets were ignored since at present, \textsc{exofit} can only search for either one or two planets. 

Many more and different radial velocity data sets and stellar mass estimates are now available (though some not publicly), but for the sake of uniformity the original radial velocity data were used (i.e. those publicly available, frozen as of 2009 August 21 when the original data were collected). At a later stage the results can be improved by updating the original data sets to those which are now available, as well as incorporating the data from the many hundreds of additional planets that have been detected since the start of this study. 

To enable accurate calculation of the derivable orbital parameters, the masses as well as the radial velocities of the associated stars were needed. These values were all taken from the published literature at \textit{http://www.exoplanet.eu}, frozen as of 2011 March 01. The input for \textsc{exofit} is in the form of a simple text file with radial velocity, uncertainty, and the time of observation (in Julian Date format), where the radial velocity values must be in $ms^{-1}$. The Julian Date of the observation is offset to zero within \textsc{exofit}. 

The publicly available statistical data analysis package, R, from the R Project for Statistical Computing (\textit{http://www.r-project.org}), was used to analyse the output of \textsc{exofit}. The output from R includes the mean, median and standard deviation of the orbital parameters extracted from the posterior distribution samples produced by \textsc{exofit}. The modal values are also produced, but will only have significance in the event of the posterior having more than one peak. Posterior distribution plots can also be produced with R, and the marginal distributions of each parameter can be found by plotting a histogram of the samples from the posterior. The full posterior distribution is helpful in analysing correlations between various parameters. Even though parameter degeneracy is present in the orbital solutions, highly degenerate solutions are less common. 

The calculation time required for \textsc{exofit} depends on computational resources available to the user. It scales linearly with the number of radial velocity entries input to the code, and also depends significantly on the ease with which \textsc{exofit} can converge the data. If \textsc{exofit} is presented with data from a non-converging posterior distribution, it will take much longer than a larger data set with convergent orbital solutions. In technical terms, the mean average calculation time using \textsc{exofit} on 26 radial velocity data sets ranging from 10 to 50 data entries for a 1-planet search was 44 seconds per data entry. For a 2-planet search using 30 data sets with between 11 and 256 entries, the calculation time increased to 3 minutes and 40 seconds per data entry. These times were achieved on a 2.80 GHz dual core linux system. Multiple runs were performed in order to confirm the orbital solutions for each system - these analyses were carried out using the UCL Legion High Performance Computing Facility, details of which can be found at \textit{http://www.ucl.ac.uk/isd/common/research-computing/services/legion-upgrade}.

\section{Model Selection}
\label{sec:model}
One of the most challenging aspects of the statistical inference procedure is the model selection problem. For the analysis of the radial velocity data the question of model selection refers to the selection of the correct number of planets to fit the observed data. \cite{ford2007, gregory2007} employed thermodynamic integration for calculating the Evidence and selecting the optimal number of planets that fit the data. On the other hand,~\cite{feroz2011a, feroz2011b} approached the situation as an object detection problem.

One of the most commonly employed model selection procedures makes use of the chi-square statistic. This is one of the most prominent methods for estimating the goodness of fit and it has been applied to many astronomical problems including the analysis of radial velocity data (see e.g. \citet{butler2006}). Bayesian inference also offers a straightforward way of performing statistical model selection, based on Equation~\ref{eqn:bayes} and the Evidence. Even though this approach is conceptually simple, its implementation is in general computationally expensive, and \textsc{exofit} does not currently have the functionality to perform such Bayesian model selection. 

Hence, in the present analysis, we make use of the traditional chi-square statistic as well as a visual flagging approach, discussing the relationship between the two in Section \ref{sec:discussion}. We limit ourselves to 1 or 2 planets, as per the current capabilities of the code, but this may of course be extended in later studies. The rationale behind the visual flagging is that one can identify the poor fits to the data by comparing the predicted radial velocity curves for the 1-planet and the 2-planet solutions. The method involves assigning a `visual quality flag' by eye to each system, where `1' signifies that the 1-planet fit is best, `2' means that the 2-planet fit is best, and `3' means that both 1- and 2-planet solutions provide equally good (or equally bad) fits. The results of this classification are shown in Table~\ref{tab:flags}, next to the number of planets currently confirmed to exist in that system, taken by comparing the values on both \textit{http://www.exoplanet.eu} and \textit{http://exoplanets.org} (as these catalogues do not agree with each other in some cases). 

Table~\ref{tab:flags} also shows the log likelihood ratio of the reduced chi-square value of the 1-planet fit to that of the 2-planet fit, where we define the log likelihood ratio,

\begin{equation}
R\,\equiv\,-\frac{1}{2}(\chi_{1p}^{2}\,-\,\chi_{2p}^{2})
\label{eq:llr}
\end{equation}

Hence a value of $R\,>\,0$ indicates that the 1-planet fit was best (had a smaller reduced chi-square value), and $R\,<\,0$ indicates that the 2-planet model provided the best fit to the data. For all bar one system (HD8574), every `1' and `2' quality flag assigned to the fits by eye was in agreement with the calculated value of $R$, endorsing our method of assignation by visual inspection (see Figure~\ref{multifig:llr}). For only a few systems there were not sufficient degrees of freedom to calculate a value for $R$ (due to e.g. only having 11 datapoints for the 12 parameters), denoted by `-' in the table.

\section{Catalogue of extrasolar planets}
\label{sec:catalogue}
In this paper the catalogue of extrasolar planets generated using \textsc{exofit} is presented in Tables~\ref{tab:catalogue1p} and \ref{tab:catalogue2p}. These contain the best estimates of the orbital parameters for both 1- and 2-planet fits for all systems analysed. The orbital parameters used to fit to the model were the systematic velocity offset of the data $V$, the orbital period of the planet $T$, the radial velocity semi-amplitude $K$, the orbital eccentricity $e$, the argument of periastron $\omega$, and $\chi$, parameterising the fraction of the orbit at the start time of the data along which the planet has travelled from the point of periastron passage. The final parameter, $s$, is a measure of all extra signal in the data after the planetary fits have been accounted for, and hence a high value could indicate the presence of an additional planet, or noisy data due to stellar activity, or the combined noise from all sources. The reader is referred to~\citet{Balan2009} for a more complete description of this parameter, which is not considered in any more detail in this study. It should also be noted that the orbital parameter $\chi$ is not related to the statistical measure $\chi^2$ used for determining the log likelihood ratio in Equation~\ref{eq:llr}. 

The direct output values from \textsc{exofit} shown in the tables are the medians of the parameter posterior distributions, and the associated $68.3\%$ confidence regions. The other displayed derivable parameters of the systems (mass and semi-major axis) were calculated by transforming the orbital parameter posteriors using the standard relations, 

\begin{equation}
a_{p}\,=\,\frac{m_{*}a_{*}sin i}{m_{p}sin i},
\label{eqn:smaxis}
\end{equation}

and,
\begin{equation}
m_{p}sin i\,\approx\,\frac{K_{*}m_{*}^{2/3}T^{1/3}\sqrt{1-e^{2}}}{(2 \pi G)^{1/3}},
\label{eqn:mass}
\end{equation}

assuming $m_p\,\ll\,m_*$. The final values for these derivable quantities were again taken to be the medians with $68.3\%$ confidence regions, and are also displayed in the tables.

\subsection{Choice of priors}
\label{subsec:priors}
\begin{figure}
\caption{The resulting marginal posterior distributions of the orbital parameters for a 1-planet fit to the BD-17 63 data, and the corresponding radial velocity curve.}
\label{multifig:BD_17_0063_1p_sols}
\smallskip
\scriptsize

\begin{centering}
\subfigure[Posterior distributions of orbital parameters for BD-17 63.]{
\includegraphics[width=0.9\columnwidth]{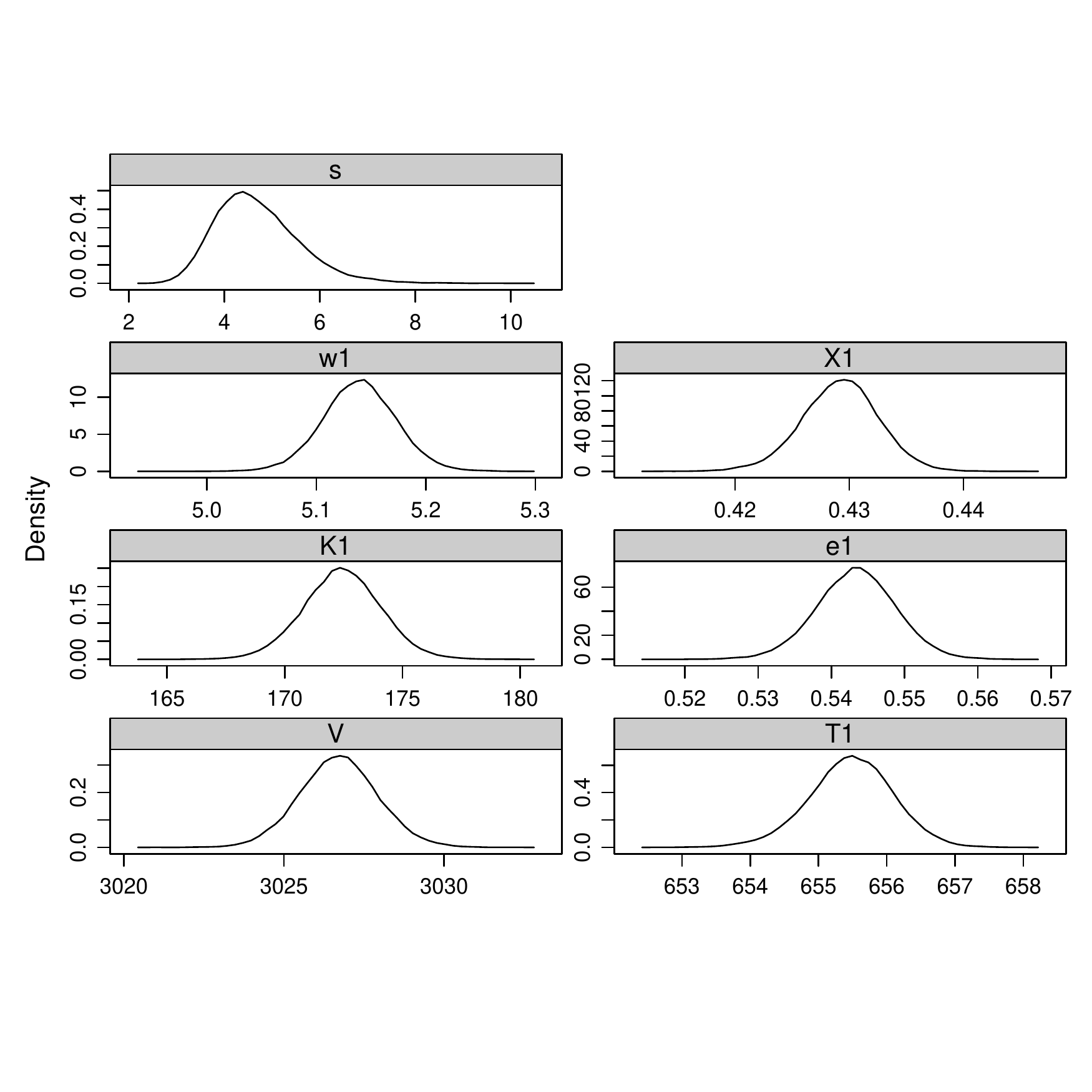}
\label{fig:BD_17_0063_1p_dens}
}

\subfigure[Radial velocity plot for BD-17 63.]{
\includegraphics[width=0.9\columnwidth]{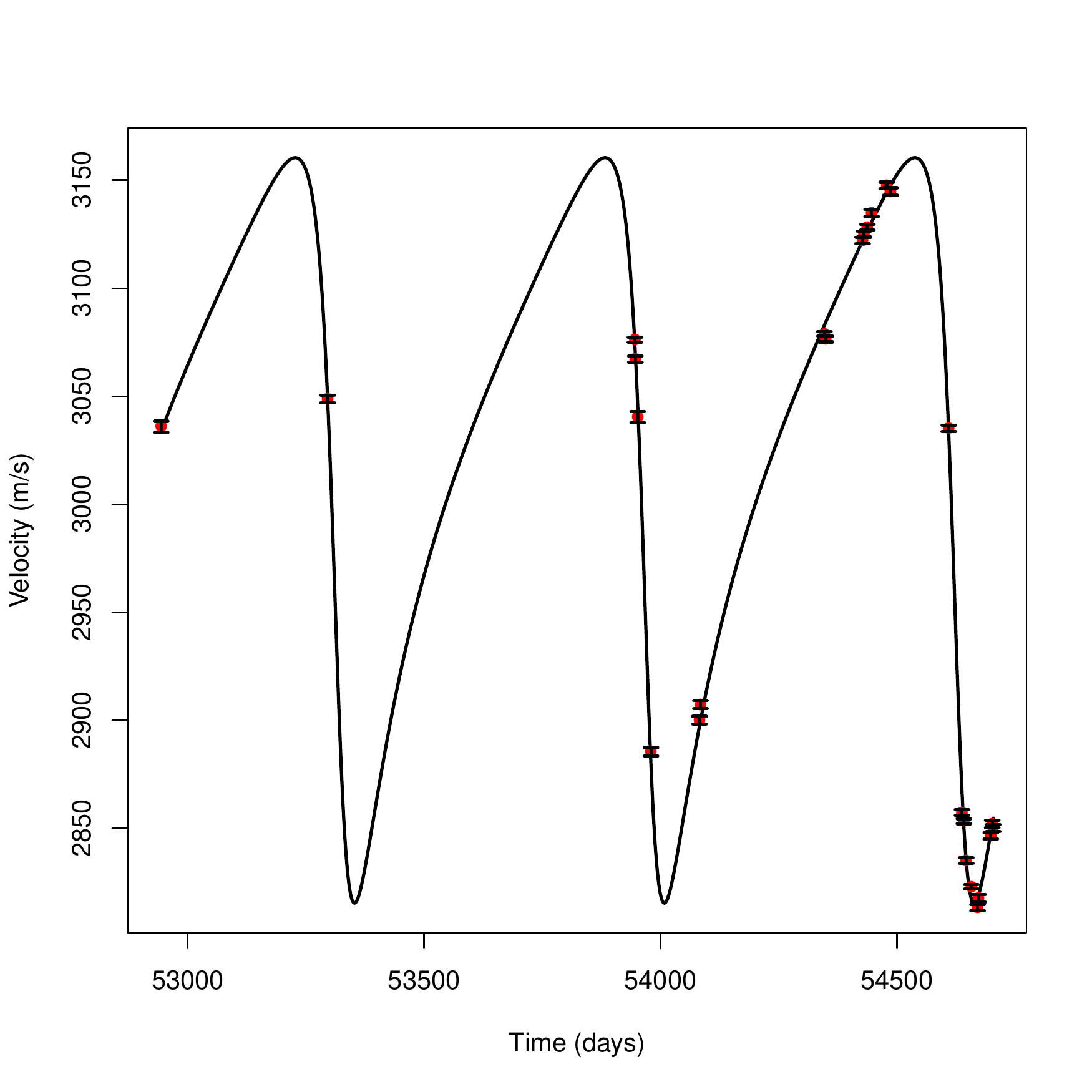}
\label{fig:BD_17_0063_1p_orb}
}

\end{centering}
\end{figure}

\begin{figure}
\caption{The resulting marginal posterior distributions of the orbital parameters for a 2-planet fit to the HD108874 data, and the corresponding radial velocity curve.}
\label{multifig:HD108874_VOGT_2p_sols}
\smallskip
\scriptsize

\begin{centering}
\subfigure[Posterior distributions of orbital parameters for HD108874.]{
\includegraphics[width=0.9\columnwidth]{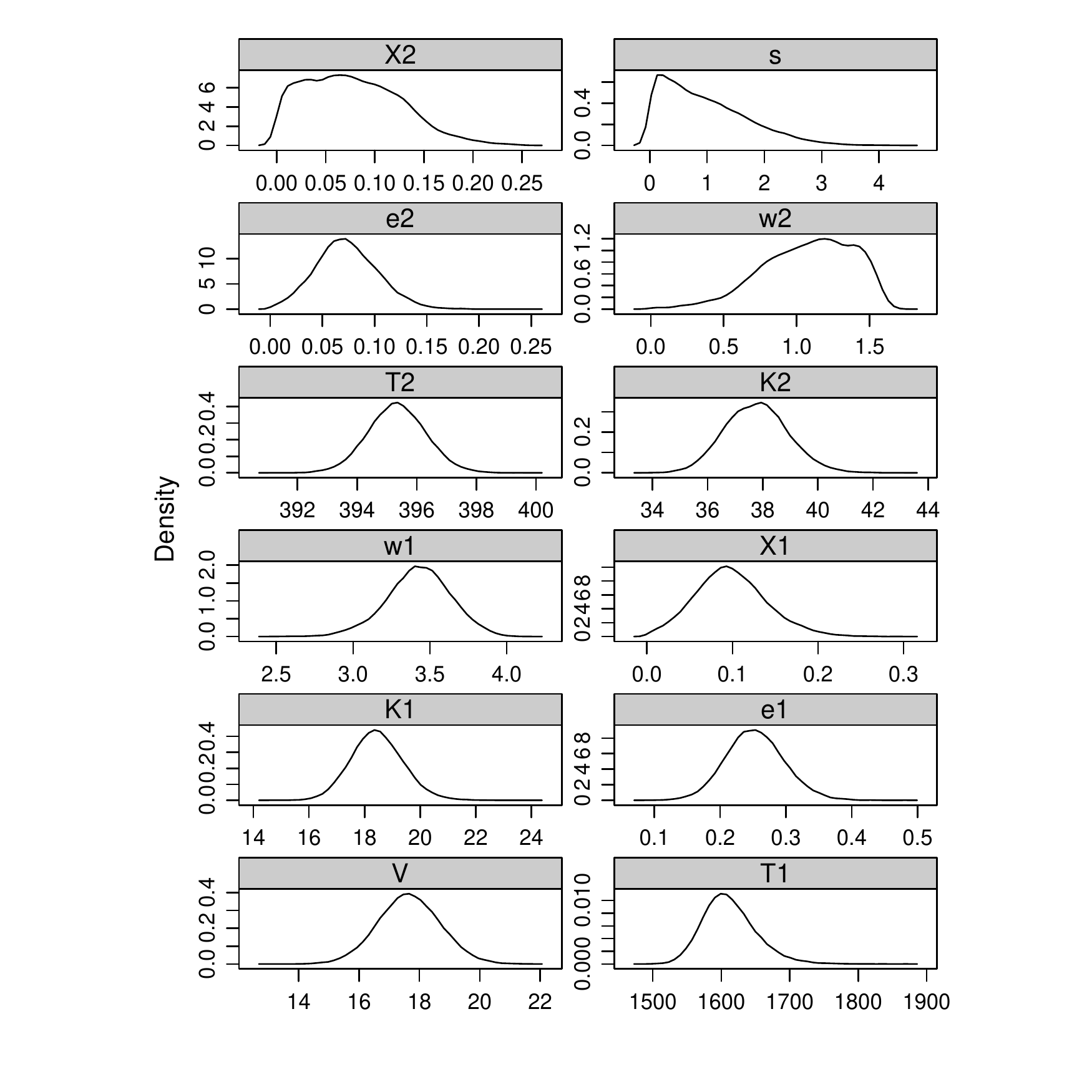}
\label{fig:HD108874_VOGT_2p_dens}
}

\subfigure[Radial velocity plot for HD108874.]{
\includegraphics[width=0.9\columnwidth]{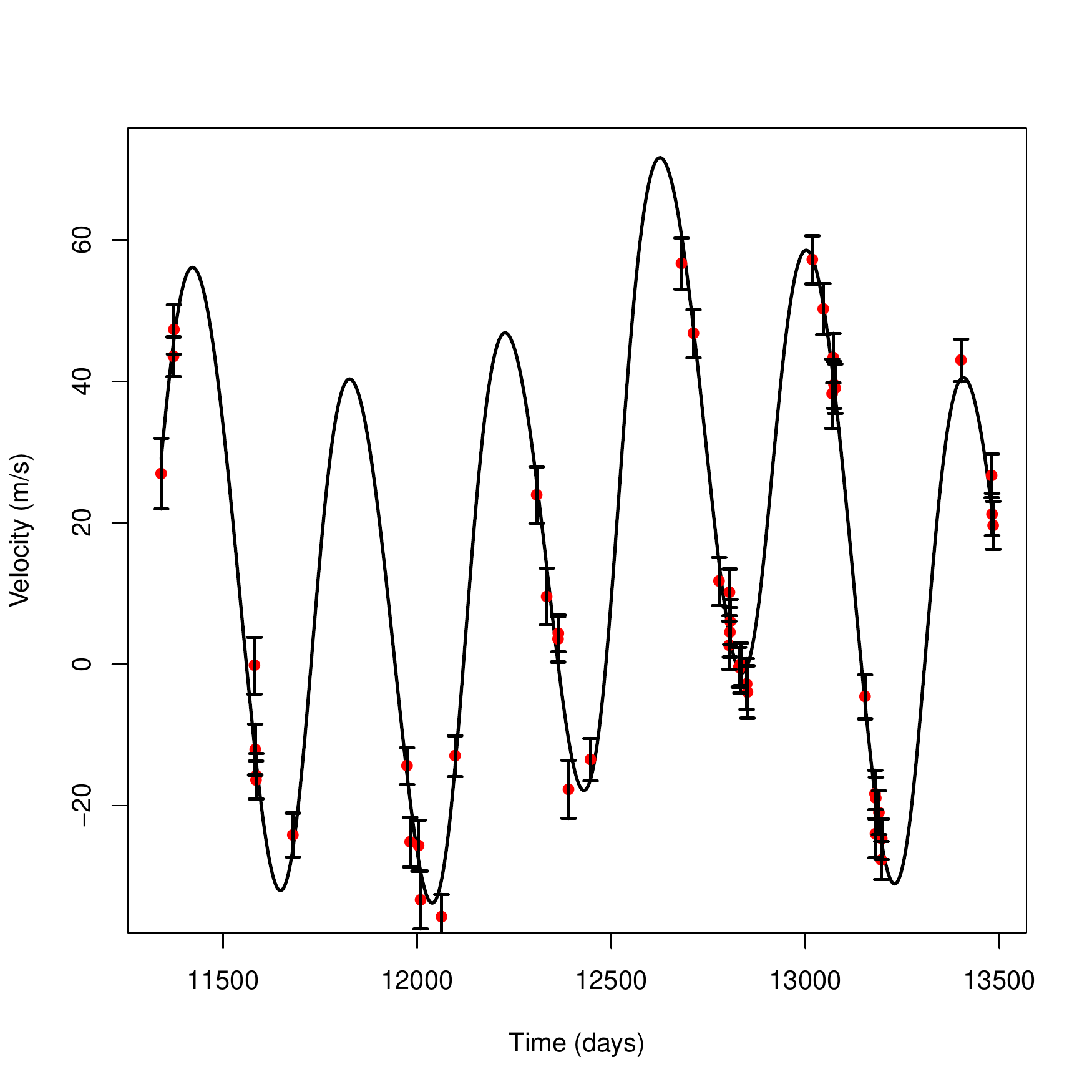}
\label{fig:HD108874_VOGT_2p_orb}
}

\end{centering}
\end{figure}

The prior distributions and ranges used for the initial analysis were as shown in Tables~\ref{tab:prior1p} and \ref{tab:prior2p}. The prior for the systematic velocity was dependent on the system - the mean of the input RV data was calculated and used as the initial value, and the allowed range was $10\,kms^{-1}$ symmetrically about this. For some systems different sets of prior boundaries were used in a second round of analysis - these stars and the prior boundaries applied are listed in Table~\ref{tab:diff_prior}. Systems which did not return good fits using the normal prior boundaries were re-run with these `tight' priors, where the period of the planet was constrained to be within a range given by,

\begin{equation}
T\,\in\,[ T_{pub} - 2\sigma_{pub} , T_{pub} + 2\sigma_{pub} ],
\label{eqn:constraints}
\end{equation}

where $T_{pub}$ is the published value of the period and $\sigma_{pub}$ is the published error on the period, both taken from \textit{http://www.exoplanet.eu} on 2011 August 01. Further constraints may also be applied, for example, systems with near zero eccentricities require tight priors on the orbital parameter $\chi$ in order to avoid multimodal distributions (see Section~\ref{sec:discussion}), whilst systems with eccentricities close to unity need tight priors on the orbital period $T$ in order to achieve convergence of the MCMC chains. Examples of the output of \textsc{exofit} are shown in Figures~\ref{multifig:BD_17_0063_1p_sols} and \ref{multifig:HD108874_VOGT_2p_sols}. 


\begin{table}
\caption[Choice of priors for the 1-planet model]
{The assumed orbital parameter prior distributions and their boundaries for a 1-planet model. The min and max values for the systematic velocity parameter were the mean value of the raw radial velocities for that data file minus and plus $5000 ms^{-1}$ respectively.}
\label{tab:prior1p}
\smallskip
\scriptsize

\begin{center}
\begin{tabular}{cccccc}
\hline
\noalign{\smallskip}
 Parameter & Prior  & Mathematical form & Min & Max \\ 
\hline\hline
$V(ms^{-1})$ & Uniform  & $\frac{1}{V_{max}-V_{min}}$ & - & - \\ 
\noalign{\smallskip}
 $T_1(days)$& Jeffreys & $\frac{1}{T_1\,\ln\Big(\frac{T_{1\,max}}{T_{1\,min}}\Big)}$ & 0.2 & 15000 \\ 
 \noalign{\smallskip}
$K_{1}(ms^{-1})$ & Mod. Jeffreys & $\frac{(K_1+K_{1\,0})^{-1}}{\ln\big(\frac{K_{1\,0}+K_{1\,max}}{K_{1\,0}}\big)}$ & 0.0 &2000 \\ 
 $e_1$ & Uniform & 1 & 0 & 1 \\ 
 $\omega_1$ & Uniform & $\frac{1}{2\pi}$ & 0 & $2\pi$ \\ 
 $\chi_1$ & Uniform & 1 & 0 & 1 \\ 
$s(ms^{-1})$ & Mod. Jeffreys & $\frac{(s+s_{0})^{-1}}{\ln\big(\frac{s_{0}+s_{max}}{s_{0}}\big)}$ & 0 & $2000$\\
\hline
\noalign{\smallskip}
\end{tabular}
\end{center}
\end{table}

\begin{table}
\caption[Choice of priors for the 2-planet model]
{The assumed orbital parameter prior distributions and their boundaries for a 2-planet model. The boundaries for $V$ were as detailed previously.}
\label{tab:prior2p}
\smallskip
\scriptsize

\begin{center}
\begin{tabular}{cccccc}
\hline
\noalign{\smallskip}
 Para. & Prior  & Mathematical form & Min & Max \\ 
\hline\hline
$V(ms^{-1})$ & Uniform  & $\frac{1}{V_{max}-V_{min}}$ & - & - \\ 
\noalign{\smallskip}
 $T_1(days)$& Jeffreys & $\frac{1}{T_1\,\ln\Big(\frac{T_{1\,max}}{T_{1\,min}}\Big)}$ & 0.2 & 15000 \\ 
 \noalign{\smallskip}
$K_{1}(ms^{-1})$ & Mod. Jeffreys & $\frac{(K_1+K_{1\,0})^{-1}}{\ln\big(\frac{K_{1\,0}+K_{1\,max}}{K_{1\,0}}\big)}$ & 0.0 &2000 \\ 
 $e_1$ & Uniform & 1 & 0 & 1 \\ 
 $\omega_1$ & Uniform & $\frac{1}{2\pi}$ & 0 & $2\pi$ \\ 
 $\chi_1$ & Uniform & 1 & 0 & 1 \\ 
 $T_2(days)$& Jeffreys & $\frac{1}{T_2\,\ln\Big(\frac{T_{2\,max}}{T_{2\,min}}\Big)}$ & 0.2 & 15000 \\ 
 $K_{2}(ms^{-1})$ & Mod. Jeffreys &  $\frac{(K_2+K_{2\,0})^{-1}}{\ln\big(\frac{K_{2\,0}+K_{2\,max}}{K_{2\,0}}\big)}$ & 0.0 &2000 \\ 
 $e_2$ & Uniform & 1 & 0 & 1 \\ 
$\omega_2$ & Uniform & $\frac{1}{2\pi}$ & 0 & $2\pi$ \\ 
$\chi_2$ & Uniform & 1 & 0 & 1 \\ 
$s(ms^{-1})$ & Mod. Jeffreys & $\frac{(s+s_{0})^{-1}}{\ln\big(\frac{s_{0}+s_{max}}{s_{0}}\big)}$ & 0 & $2000$\\
\hline
\noalign{\smallskip}
\end{tabular}
\end{center}
\end{table}

\begin{table}
\caption[Systems analysed with different priors]
{Radial velocity data sets analysed with different period prior boundaries, for reasons explained in Section~\ref{subsec:ambig}. The initial value is set to the published value of the period, the maximum value is the initial value plus twice the published error, and the minimum is the initial value minus twice the published error. This approach was generally necessary for those systems (e.g. WASP and XO data sets) where the number of datapoints available at the time of selecting the data was low, thus requiring tighter priors to adequately constrain the solution.}
\label{tab:diff_prior}
\smallskip
\scriptsize

\begin{center}
\begin{tabular}{rlll}
\hline
\noalign{\smallskip}
\multicolumn{1}{c}{System} & 
\multicolumn{1}{c}{Initial period value} & 
\multicolumn{1}{c}{Min period} & 
\multicolumn{1}{c}{Max period} \\
\hline\hline
epsilon Eri & 2500 & 1800 & 3200 \\
gamma Cep & 906 & 899.84 & 912.1 \\
GJ849 & 1900 & 1400 & 2400 \\
GJ86 & 15.7649 & 15.76412 & 15.76568 \\
HAT-P-9 & 3.92289 & 3.92281 & 3.92297 \\
HD118203 & 6.1335 & 6.1323 & 6.1347 \\
HD12661 & 262.71 & 262.54 & 262.88 \\
HD128311 & 924 & 913.4 & 934.6 \\
HD131664 & 1950 & 1868 & 2032 \\
HD142 & 350 & 342.8 & 357.2 \\
HD149143 & 4.072 & 4.058 & 4.086 \\
HD162020 & 8.42820 & 8.428088 & 8.428312 \\
HD168443 & 58.1121 & 58.111142 & 58.113058 \\
HD169830 & 225.6 & 225.16 & 226.04 \\
HD183263 & 627 & 624.8 & 629.2 \\
HD187123 & 3.096583 & 3.09656732 & 3.09659868 \\
HD189733 & 2.2185757 & 2.2185754 & 2.2185760 \\
HD190360 & 2920 & 2862.2 & 2977.8 \\
HD196885 & 1330 & 1300 & 1360 \\
HD202206 & 255.87 & 255.75 & 255.99 \\
HD20868 & 380.85 & 380.67 & 381.03 \\
HD209458 & 3.5247486 & 3.52474784 & 3.52474936 \\
HD217107 & 7.12682 & 7.1267318 & 7.1268882 \\
HD219828 & 3.833 & 3.807 & 3.859 \\
HD28185 & 379 & 375 & 383 \\
HD330075 & 3.38773 & 3.38757 & 3.38789 \\
HD33636 & 2128 & 2111.6 & 2144.4 \\
HD38529 & 2146 & 2134.98 & 2157.02 \\
HD46375 & 3.02357 & 3.02344 & 3.0237 \\
HD47536 & 430 & 0.2 & 860 \\
HD50499 & 2460 & 2384.2 & 2535.8 \\
HD5319 & 670 & 636 & 704 \\
HD68988 & 6.2771 & 6.27668 & 6.27752 \\
HD73267 & 1260 & 1246 & 1274 \\
HD74156 & 2520 & 2490 & 2550 \\
HD75289 & 3.50927 & 3.509172 & 3.509398 \\
HD80606 & 111.4367 & 111.4359 & 111.4375 \\
HD86081 & 2.1375 & 2.1371 & 2.1379 \\
HD89307 & 2170 & 2094 & 2246 \\
HIP 75458 & 511.10 & 510.922 & 511.278 \\
tau Boo & 3.31246 & 3.312432 & 3.312488 \\
TrES-3 & 1.31 & 0.2 & 2.62 \\
WASP-2 & 2.152226 & 2.152218 & 2.152234 \\
WASP-3 & 1.846834 & 1.84683 & 1.846838 \\
XO-1 & 3.94153 & 3.941476 & 3.941584 \\
XO-2 & 2.615838 & 2.615822 & 2.615854 \\
XO-4 & 4.125083 & 4.125075 & 4.125091 \\
\hline
\end{tabular}
\end{center}
\end{table}

\subsection{Ambiguous systems}
\label{subsec:ambig}
For some of the systems analysed there is a clear trend in the radial velocities indicating the possibility of a second planet, but the data are not informative enough to properly constrain the orbital parameters of such an object. Plotting the resulting radial velocity curve and judging by eye can help to assess and distinguish between the 1- and 2-planet fits and evaluate the validity of the orbital solution, though such poorly-constrained orbits will lead to large error bars on the estimates of the orbital parameters. 

This did not always even lead to a clear classification though, and many systems were then re-run with tight priors on the period, given in Table~\ref{tab:diff_prior}. Some of these ambiguities were caused by data which were poor, or less accurate due to age, or too noisy due to stellar jitter. Others were simply due to the correlation between $\omega$ and $\chi$, or data not good enough to constrain these two parameters. This resulted in near-uniform posteriors for $\omega$ and $\chi$, and hence fits that match few of the measured datapoints as a result of being shifted in time. Estimates for masses and semi-major axes derived from these results are still valid however (providing reliable estimates for $T$, $K$ and $e$ are obtained, which was generally the case), as these values have no dependence on mean anomaly at epoch and the time evolution of the Keplerian orbit. 

\begin{figure}
\caption{The resulting radial velocity curve fits using the derived orbital parameters for HD89307. Imposing a 2-planet model on data with only one planet can have the result shown here, where the 2-planet fit is the same as the 1-planet fit with a superimposed artificial small-amplitude periodic variation.}
\label{multifig:3bclass}
\smallskip
\scriptsize

\begin{centering}
\subfigure[1-planet radial velocity plot of HD89307.]{
\includegraphics[width=0.9\columnwidth]{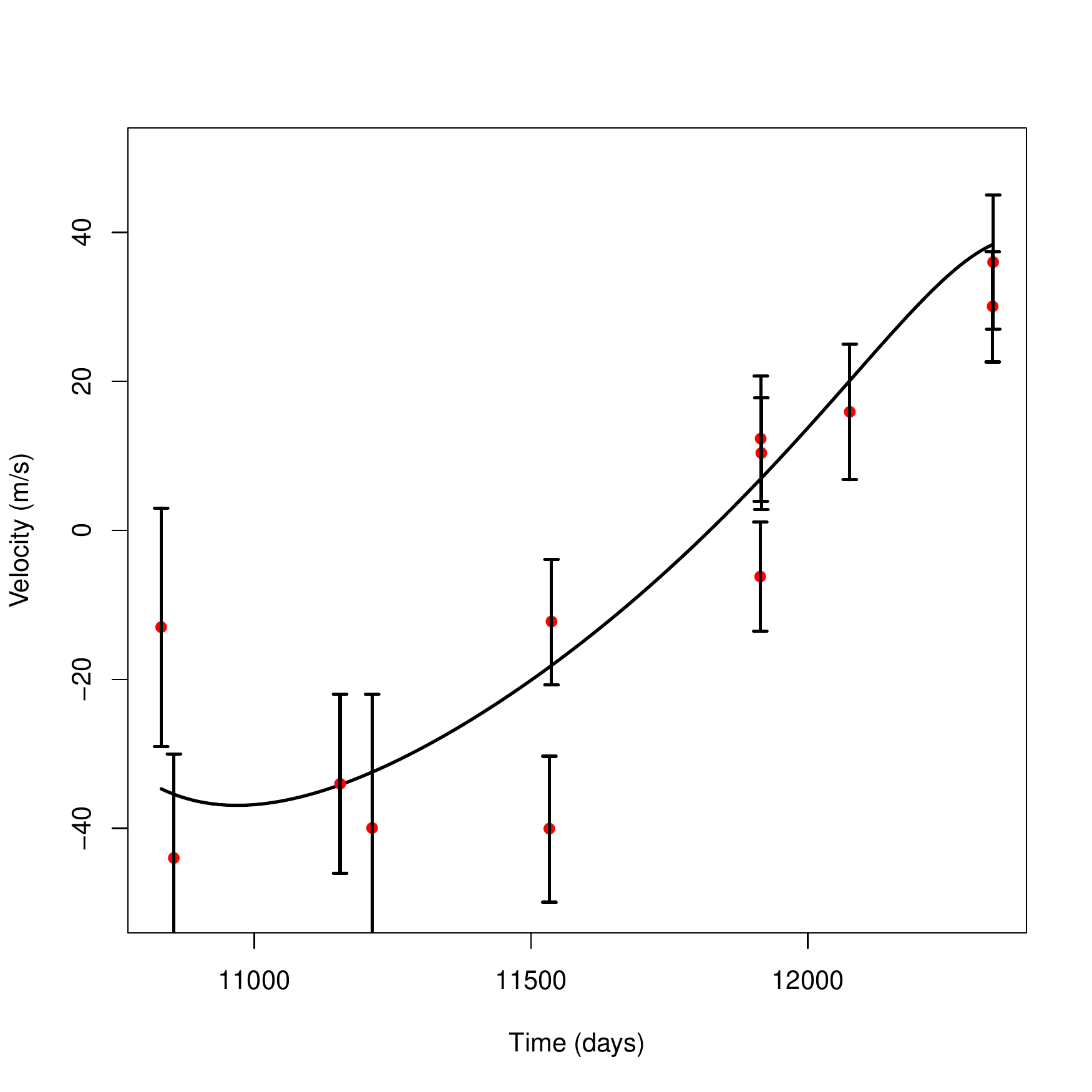}
\label{fig:HD89307_1p_orb}
}

\subfigure[2-planet radial velocity plot of HD89307.]{
\includegraphics[width=0.9\columnwidth]{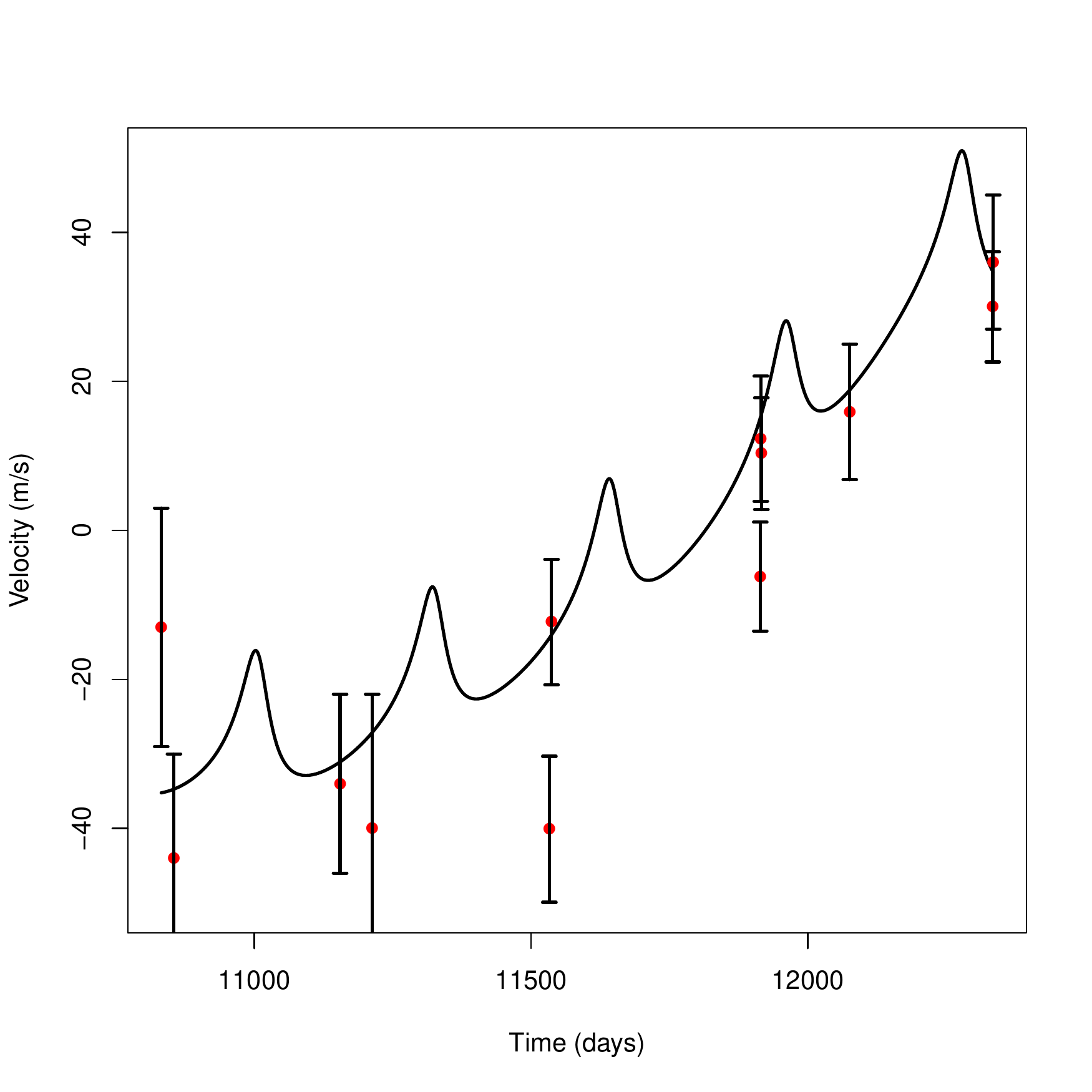}
\label{fig:HD89307_2p_orb}
}

\end{centering}
\end{figure}

The class 3 (both 1- and 2-planet fits equally good, or equally bad) systems, as introduced in Section~\ref{sec:model}, are those which were considered to be somewhat ambiguous even after being analysed with tighter priors. This category was subdivided further - in some cases these are distinct radial velocity solutions which provide plausible fits for both 1- and 2-planet models, and are classified as `3a'. However, there are also systems where the `second planet' fit just produces small-amplitude variations on the 1-planet solution (see Figure~\ref{multifig:3bclass} for an example), or where the 1- and 2-planet fits are identical but the `second planet' posteriors are peaks at very small values for $T$ and $K$, and uniform for $e$, $\omega$ and $\chi$ (i.e. there is no single solution for a second planet from these data). From this an `Occam's Razor' approach could be taken and the assumption made that the correct model for most of these `3b' class systems is in fact the single planet one. In a few cases though there may truly be a second planet present, and the data used are simply not good enough to change the likelihoods of the parameters from the initial `no-knowledge' (uniform prior) situation. So for all class 3 systems, better (or at least more up-to-date) data and more complete analyses (such as using the log likelihood ratio in more detail to narrow down the classification) are required to accurately determine the correct orbital solution.

\section{Comparison of orbital parameters}
\label{sec:param_comparison}
\begin{figure*}
\caption{Orbital parameter values taken from \textit{http://www.exoplanet.eu}, plotted against values yielded using \textsc{exofit}. Plotted systems are only those where \textsc{exofit} gave unambiguous (either class 1 or 2) results.}
\label{multifig:lit_comp}

\begin{centering}
\subfigure[Period (in days)]{
\includegraphics[width=0.9\columnwidth]{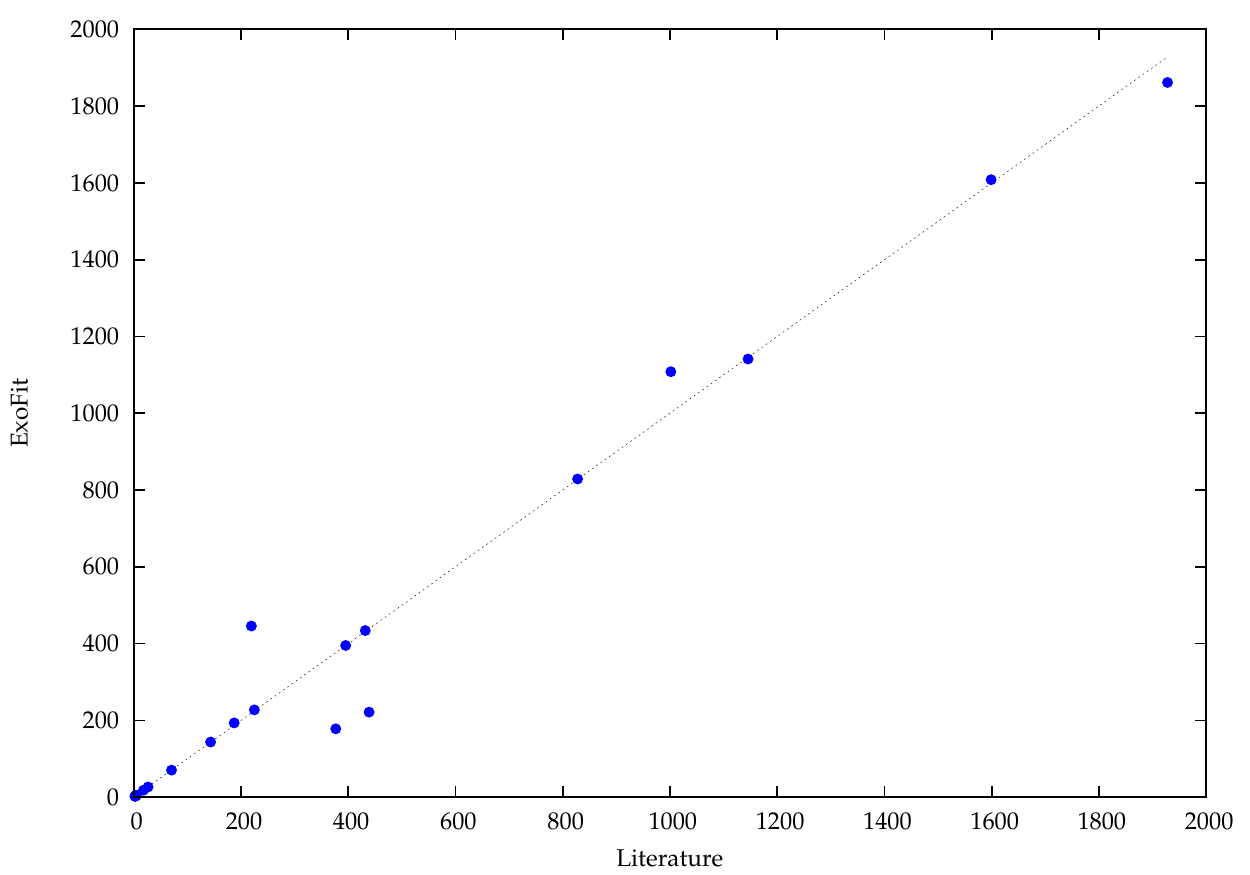}
\label{fig:period_comp}
}
\subfigure[Semi-major axis (in AU)]{
\includegraphics[width=0.9\columnwidth]{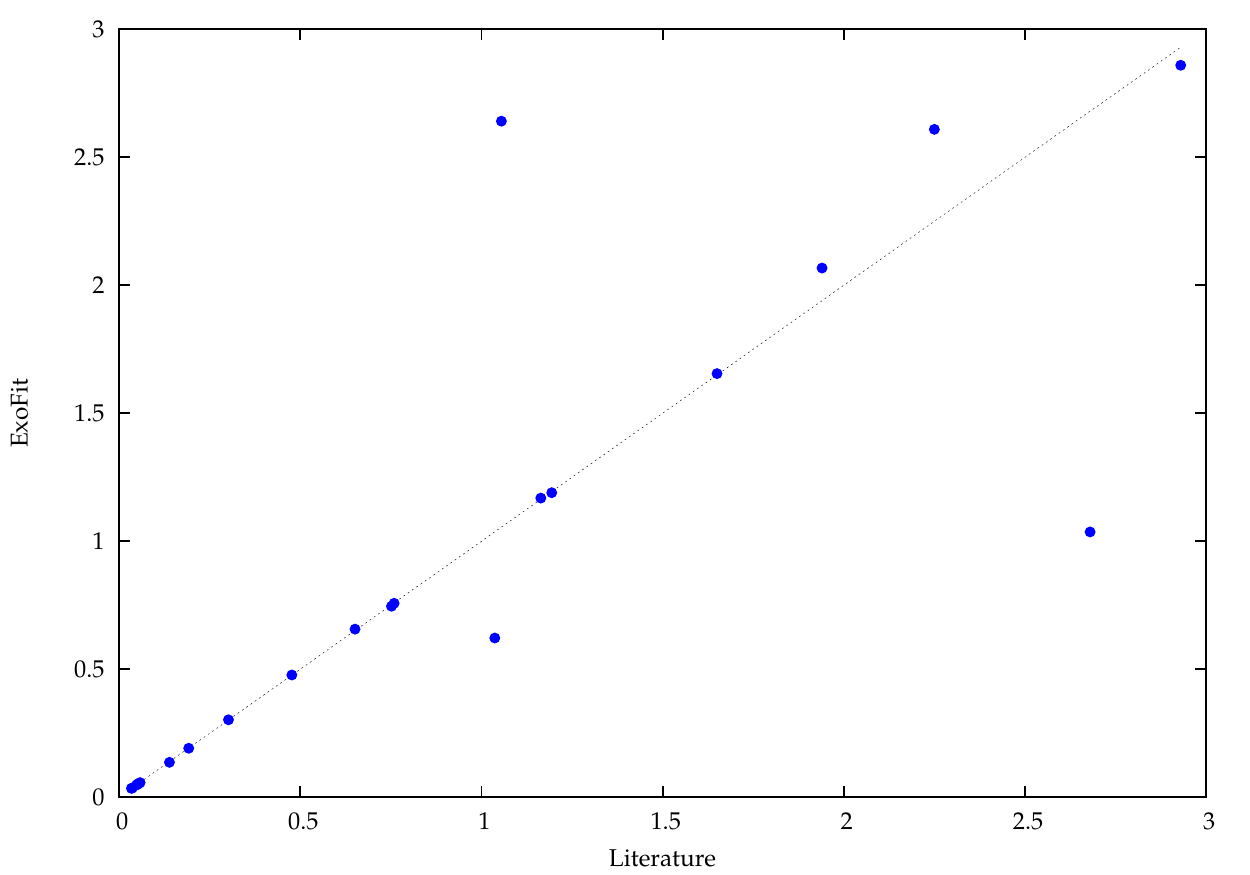}
\label{fig:axis_comp}
}
\subfigure[Mass (in M$_{\textnormal{Jup}}$)]{
\includegraphics[width=0.9\columnwidth]{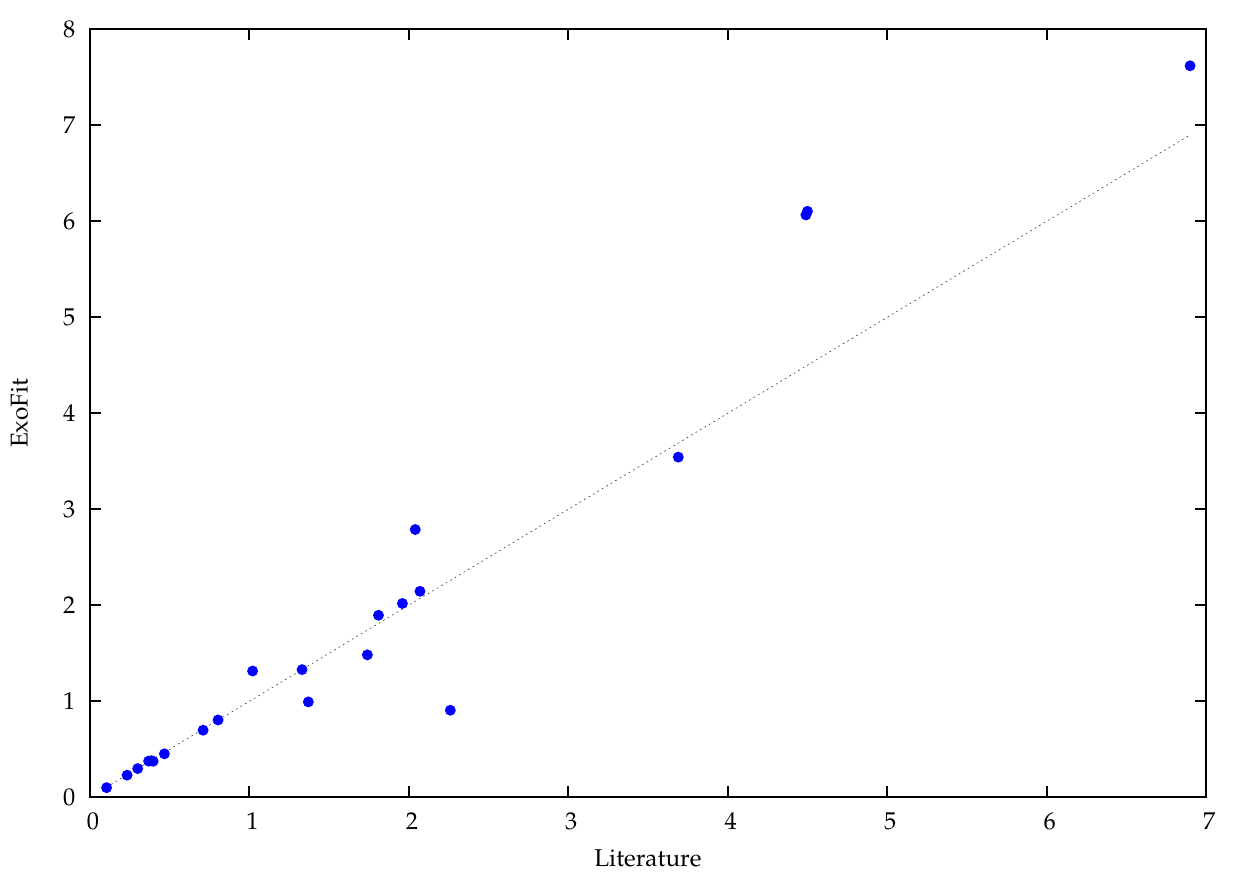}
\label{fig:mass_comp}
}
\subfigure[Eccentricity]{
\includegraphics[width=0.9\columnwidth]{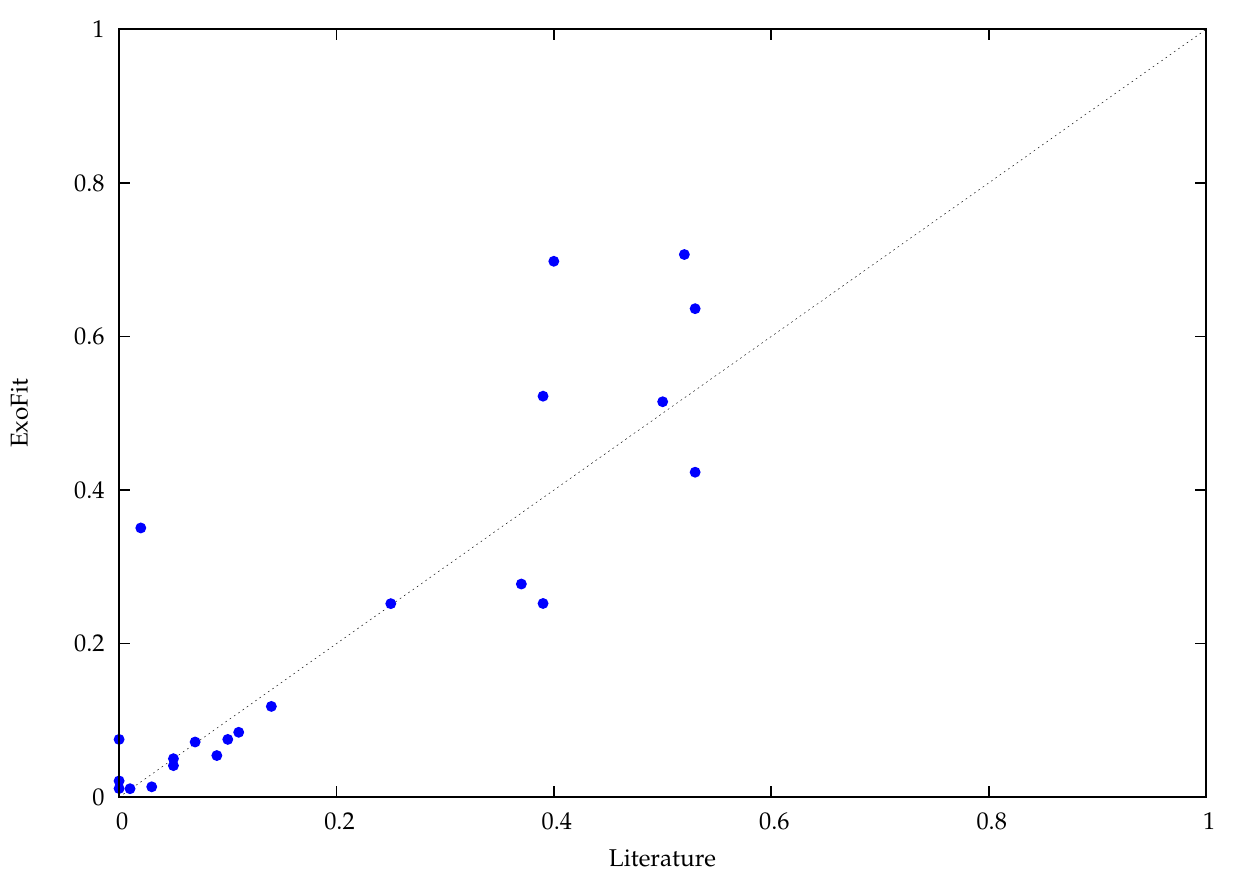}
\label{fig:ecc_comp}
}
\end{centering}
\end{figure*}

Figure~\ref{multifig:lit_comp} shows values for specified orbital elements from the literature against values yielded using \textsc{exofit} for each system. The mass, semi-major axis and period values all exhibit good correlations in general between the independently derived values and those in the published literature - this is unsurprising for the period as it has not been derived from other quantities. The eccentricity however shows a greater spread than expected - whereas our uniformly derived catalogue is consistent in taking the median, the published values use, in general, many different statistical measures to extract the final parameter values from varying analysis techniques. This highlights the value of using a consistent technique to build up reliable databases of orbital parameters. 

As mass and semi-major axis values are themselves derived from period and eccentricity, any inaccuracies in algorithms used will propagate, and also present discrepancies in the values yielded using \textsc{exofit} and are likely to amplify outliers in these plots. These outliers will be investigated in the future in order to assess the validity of the solutions. 

There are some discrepancies in the global distribution of parameter values between this catalogue and the published literature, especially for the eccentricity parameter. This may be partly due to poor or out-dated data, and is almost certainly affected by the ubiquitous effects of certain parameter correlations (as explained in Section~\ref{sec:discussion}). These should be analysed in more detail in the future, and techniques developed to explore the parameter space more efficiently and minimize or eradicate such dependencies. 

\section{Discussion}
\label{sec:discussion}
\begin{figure*}
\caption{The log likelihood ratio for each planet assigned to visual class 1 or 2 is shown in \subref{fig:llr1}. \subref{fig:llr2} shows the same data on a smaller scale, around the threshold at $R\,=\,0$. Class 2 systems (open circles) are all below the chi-square ambiguity threshold, and class 1 systems (filled circles) are all above, with the single exception of HD8574 (shown as a red triangle), class 1 but located just below the threshold with a value of $R\,=\,-0.03$.}
\label{multifig:llr}
\smallskip
\scriptsize

\begin{centering}
\mbox{\subfigure[Log likelihood ratio with planet for all class 1 and 2 systems.]{
\includegraphics[width=0.9\columnwidth]{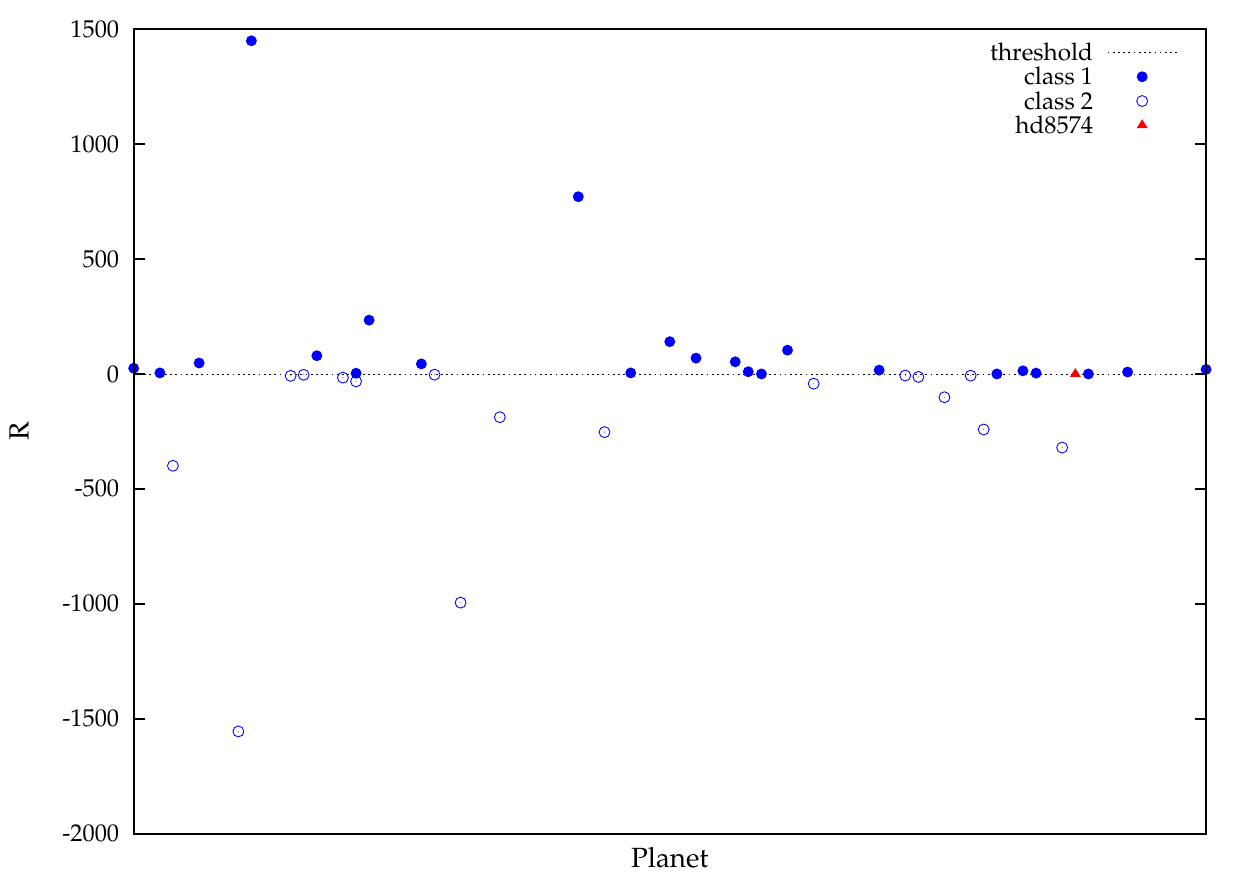}
\label{fig:llr1}
}\quad

\subfigure[Log likelihood ratio with planet for all class 1 and 2 systems, in the range $-20\,\leq\,R\,\leq\,+20$.]{
\includegraphics[width=0.9\columnwidth]{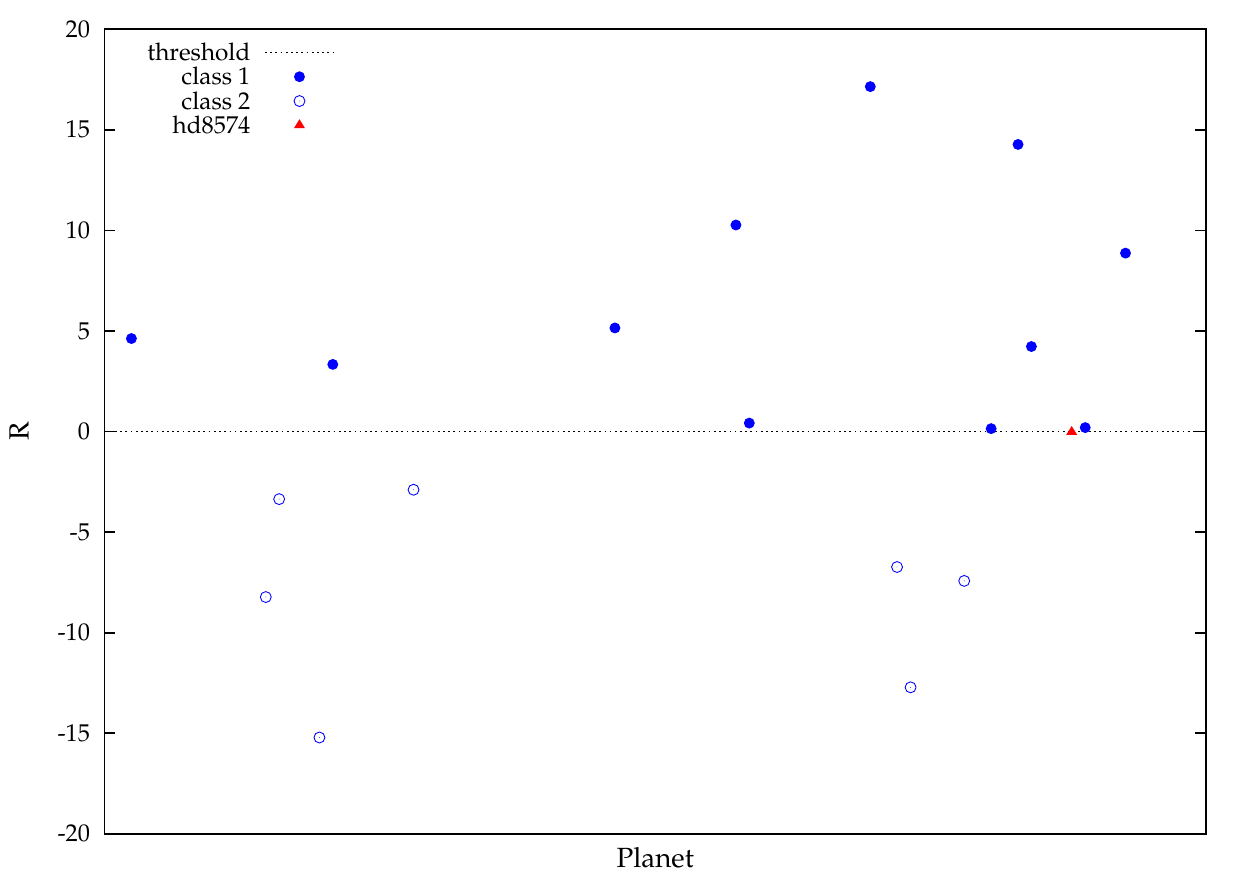}
\label{fig:llr2}
}
}

\end{centering}
\end{figure*}

The primary objective of this article is to analyse radial velocity data sets uniformly, using a single platform for the data analysis. \cite{butler2006} produced a catalogue of extrasolar planets using traditional methods (using periodograms and Levenberg-Marquardt minimisation). We have analysed a selection of radial velocity data sets using a Bayesian parameter estimation program for extrasolar planets. However, a model selection criteria is required for completion of the statistical inference process, and for this purpose, as described in Section~\ref{sec:model}, a chi-square statistic was employed as well as a visual flagging technique. Inconclusive results are obtained for a few data sets, but from those analysed here it can be seen that both model selection methods perform well, agreeing in $99\%$ of cases, as demonstrated in Figure~\ref{multifig:llr}.

\begin{figure}
\caption{Examples of bimodal and ambiguous posterior densities, obtained from HD49674 and HD190228.}\label{multifig:ambiguous}
\smallskip
\scriptsize

\begin{centering}
\subfigure[Posterior densities for a 1-planet fit to the HD49674 data, exhibiting some bimodality in the $\omega$ and $\chi$ orbital parameter values.]{
\includegraphics[width=0.9\columnwidth]{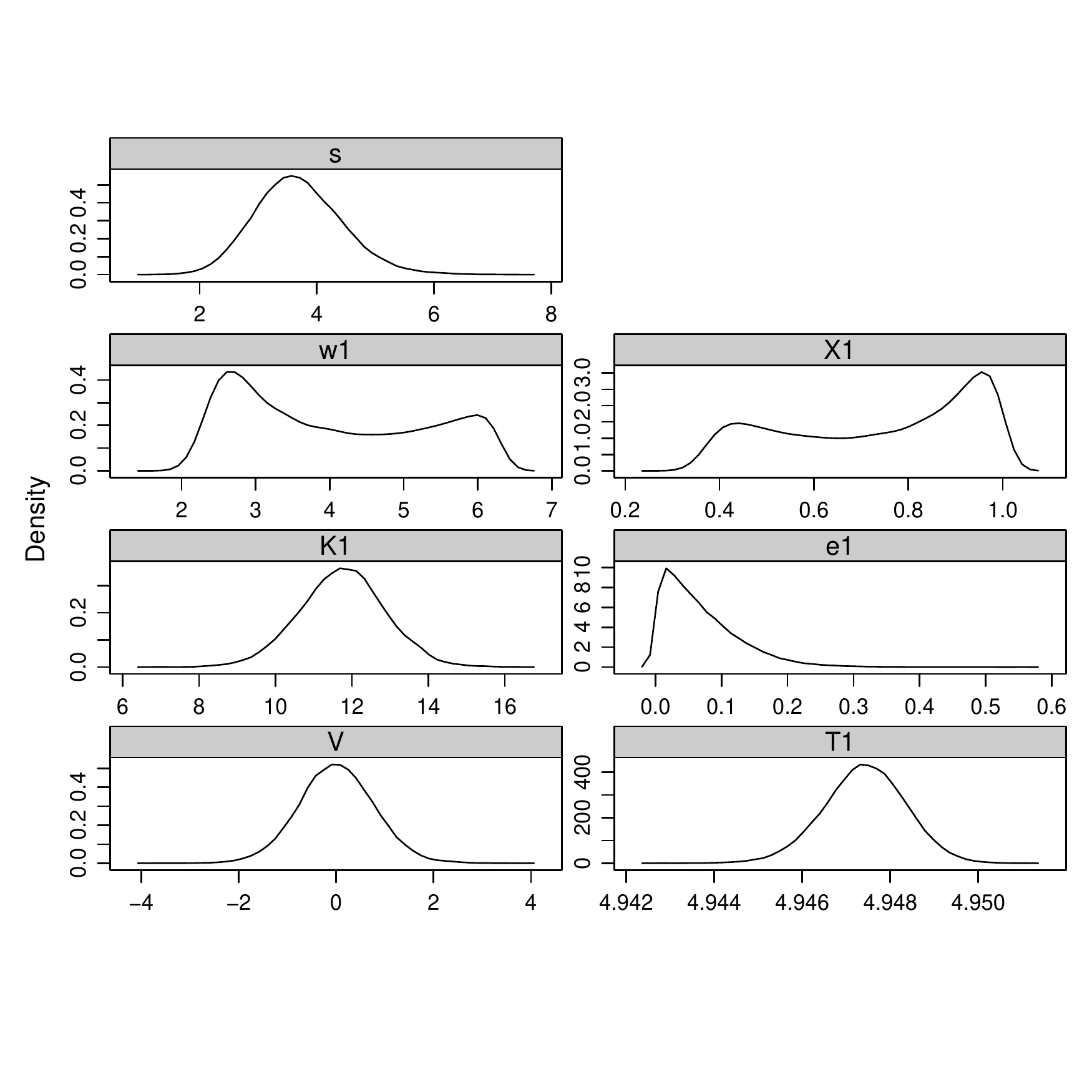}
}

\subfigure[Posterior densities for a 2-planet fit to the HD190228 data, showing ambiguity in the estimates for the $\omega$, $\chi$, and $e$ values for one of the planets.]{
\includegraphics[width=0.9\columnwidth]{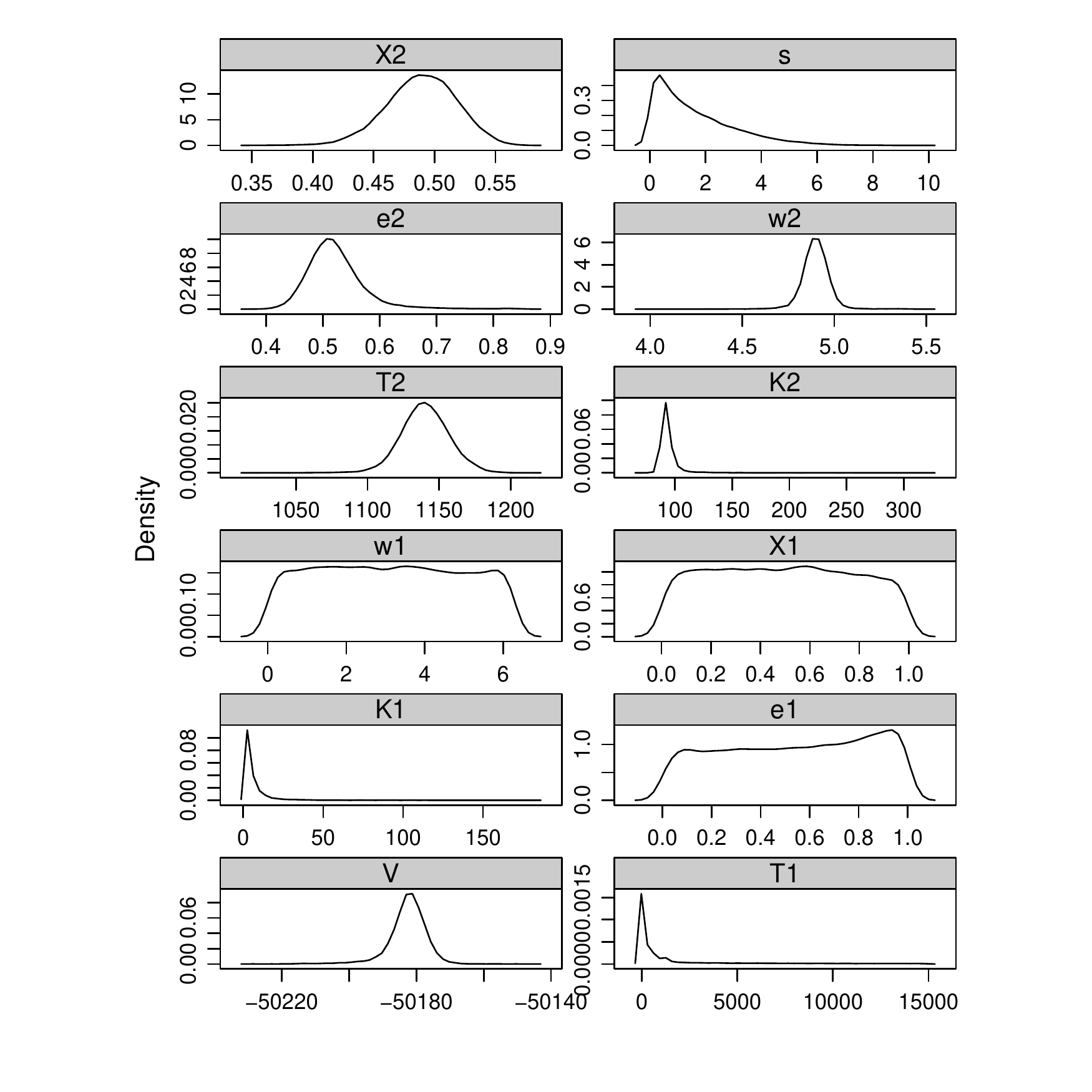}
}

\end{centering}
\end{figure}

Investigating further, we find that the chi-square values depend on the point estimates of the orbital parameters used to construct the predicted radial velocity curve. If the posterior distribution is unimodal such an approach will work flawlessly. However, posterior distributions of the orbital parameters exhibit multi-modality on many occasions. For example the parameters $\omega$ and $\chi$ are extremely correlated and their posterior distributions are bimodal for many data sets (an example of this is shown in Figure~\ref{multifig:ambiguous}), especially for planets with $e\,\approx\,0$. This problem has also been noted by \cite{gregory2007}, who proposed re-parameterisation of the problem as a possible way of dealing with this situation.

Many data sets contain planetary signals whose period is greater than the span of the observations, and so obtaining constraints on the orbital parameters of these objects is an extremely difficult task. There are several data sets where it was possible to obtain estimates for the orbital parameters for one of the planets, but then the second signal could not be constrained due to weak signal-to-noise. In most cases these signals appear to be a linear or quadratic trend in the radial velocity data. Therefore, it becomes extremely difficult to classify these objects as planets, and this is one of the reasons a visual flagging method was employed. One example of this is shown in Figure~\ref{multifig:ambiguous}, the results of the 2-planet fit to the data of the system HD190228. The strongest signal is picked up and well-constrained, as can be seen from the error bars in Table~\ref{tab:catalogue2p}, and in addition the values for this planet match well those from the fit for a single planet. Thus we can be reasonably sure of the parameters of the first planet, but those of the second, shown as HD190228b in Table~\ref{tab:catalogue2p}, are significantly less secure. 

Additionally, sharp prior boundaries were used on the orbital period for several data sets. In these cases we have found that either the planetary signal is very weak or the signal from a systematic trend from an additional companion in the radial velocities masks the weaker planetary signal. We have also found that the aliasing effects (see e.g. ~\citet{Dawson2010}) in observations can produce additional peaks in the posterior distributions, necessitating the use of the sharp prior on the period. 

In summary, a brief overview of the Bayesian theory has been given here, along with a description of the MCMC approach used in order to estimate the orbital parameters of extrasolar planets, more details of which can be found in~\cite{Balan2009}. A new catalogue of extrasolar planets is presented from the re-analysis of published radial velocity data sets, giving both 1- and 2-planet orbital solutions for 94 systems derived on a uniform basis. An attempt is made to distinguish between the solutions for each system by using both a visual categorisation method and a standard reduced chi-square technique, giving good agreement in $99\%$ of the cases presented here. Improvements in this `model selection' area of the analysis may be made by taking into account Bayesian Evidence, as seen in~\cite{gregory2007} and \cite{feroz2011a,feroz2011b}; more rigorous approaches such as these are outside the scope of this `proof of concept' study, but may be looked into in the future. Other further work will include updating this catalogue to incorporate the most up-to-date data, as well as extending \textsc{exofit} to be able to use transit and microlensing results, to search for an arbitrary number of planets, and to look into the possibility of accounting for interactions between planetary bodies.

\section{Acknowledgments}
\label{sec:acknowledge}
The authors would like to thank Paul Gorman, Robert Michaelides, Lisa Menahem, Andrew Strang, Robert C. Clouth, and Milroy Travasso for their help in analysing planetary data sets. MH is supported by an Impact/Perren studentship, STB acknowledges support from the Isaac Newton Studentship and a studentship from the Astrophysics Group, Cavendish Laboratory, and OL acknowledges a Royal Society Wolfson Research Merit Award. The authors acknowledge the use of the UCL Legion High Performance Computing Facility, and associated support services, in the completion of this work. This research has made use of the Exoplanet Orbit Database and the Exoplanet Data Explorer at \textit{http://exoplanets.org}.

The catalogue will be made available for public viewing at \textit{http://www.ucl.ac.uk/exoplanets/exocat}.

\bibliographystyle{mn2e}
\bibliography{hollis_bib}


\begin{table*}
\caption{Table of the orbital parameters for a 1-planet fit, both directly output from \textsc{exofit} and thence derived. The values of the parameters $T$, $K$, $e$ and $s$ (generated from \textsc{exofit}) are the medians of the parameter posterior distributions, with the associated $68.3\%$ confidence regions. The other parameters were calculated using these values and stellar masses taken from the published literature. Note that some parameters are extremely well-constrained, hence the errors on the parameter estimates are so small as to appear to be zero to the two decimal places shown in this table. A full table in machine-readable format will be provided on the website, and the reader is directed there if such data are required.}
\label{tab:catalogue1p}
\smallskip
\scriptsize

	\begin{center}
		\begin{tabular}{rcllllll}
			\hline\noalign{\smallskip}
\multicolumn{1}{c}{\textbf{planet}} & 
\multicolumn{1}{c}{\textbf{$m_{*} (M_{sol})$}} & 
\multicolumn{1}{c}{\textbf{$T (days)$}} & 
\multicolumn{1}{c}{\textbf{$K (ms^{-1})$}} & 
\multicolumn{1}{c}{\textbf{$e$}} & 
\multicolumn{1}{c}{\textbf{$s$}} & 
\multicolumn{1}{c}{\textbf{$m_p (M_{Jup})$}} & 
\multicolumn{1}{c}{\textbf{$a (AU)$}} \\
			\hline\hline

BD-17 63 b	&	0.74	&	$	655.49	^{+	0.59	}_{-	0.62	}$	&	$	172.44	^{+	0.62	}_{-	1.61	}$	&	$	0.54	^{+	0.01	}_{-	0.01	}$	&	$	4.59	^{+	0.95	}_{-	0.73	}$	&	$	5.06	^{+	0.05	}_{-	0.05	}$	&	$	1.34	^{+	0.00	}_{-	0.00	}$	\\
ChaHa8 b	&	0.10	&	$	304.59	^{+	1.81	}_{-	1.79	}$	&	$	1221.89	^{+	186.78	}_{-	128.02	}$	&	$	0.15	^{+	0.15	}_{-	0.10	}$	&	$	32.67	^{+	130.79	}_{-	30.48	}$	&	$	8.60	^{+	1.12	}_{-	0.89	}$	&	$	0.41	^{+	0.00	}_{-	0.00	}$	\\
epsilon Eri b	&	0.86	&	$	2503.68	^{+	57.36	}_{-	52.69	}$	&	$	17.83	^{+	1.93	}_{-	1.81	}$	&	$	0.16	^{+	0.16	}_{-	0.11	}$	&	$	9.44	^{+	0.91	}_{-	0.82	}$	&	$	1.05	^{+	0.10	}_{-	0.10	}$	&	$	3.43	^{+	0.05	}_{-	0.05	}$	\\
epsilon Tau b	&	2.70	&	$	597.53	^{+	12.02	}_{-	11.52	}$	&	$	96.16	^{+	3.93	}_{-	3.85	}$	&	$	0.13	^{+	0.04	}_{-	0.04	}$	&	$	8.85	^{+	2.48	}_{-	1.95	}$	&	$	7.66	^{+	0.30	}_{-	0.30	}$	&	$	1.93	^{+	0.03	}_{-	0.02	}$	\\
gamma Cep b	&	1.59	&	$	905.03	^{+	4.52	}_{-	3.68	}$	&	$	317.34	^{+	77.57	}_{-	71.25	}$	&	$	0.51	^{+	0.14	}_{-	0.16	}$	&	$	225.74	^{+	34.05	}_{-	26.78	}$	&	$	17.44	^{+	4.02	}_{-	4.15	}$	&	$	2.14	^{+	0.01	}_{-	0.01	}$	\\
GJ3021 b	&	0.90	&	$	133.70	^{+	0.20	}_{-	0.20	}$	&	$	167.02	^{+	3.87	}_{-	3.95	}$	&	$	0.51	^{+	0.02	}_{-	0.02	}$	&	$	15.86	^{+	2.34	}_{-	2.05	}$	&	$	3.36	^{+	0.08	}_{-	0.08	}$	&	$	0.49	^{+	0.00	}_{-	0.00	}$	\\
GJ317 b	&	0.24	&	$	672.33	^{+	8.26	}_{-	7.27	}$	&	$	90.96	^{+	46.14	}_{-	12.10	}$	&	$	0.45	^{+	0.20	}_{-	0.10	}$	&	$	15.63	^{+	4.38	}_{-	3.16	}$	&	$	1.36	^{+	0.40	}_{-	0.16	}$	&	$	0.93	^{+	0.01	}_{-	0.01	}$	\\
GJ674 b	&	0.35	&	$	4.69	^{+	0.00	}_{-	0.00	}$	&	$	9.50	^{+	0.99	}_{-	1.02	}$	&	$	0.11	^{+	0.10	}_{-	0.08	}$	&	$	3.55	^{+	0.58	}_{-	0.46	}$	&	$	0.04	^{+	0.00	}_{-	0.00	}$	&	$	0.04	^{+	0.00	}_{-	0.00	}$	\\
GJ849 b	&	0.49	&	$	2014.09	^{+	60.32	}_{-	61.27	}$	&	$	26.68	^{+	9.48	}_{-	4.46	}$	&	$	0.68	^{+	0.10	}_{-	0.09	}$	&	$	7.14	^{+	1.40	}_{-	1.08	}$	&	$	0.77	^{+	0.19	}_{-	0.11	}$	&	$	2.46	^{+	0.05	}_{-	0.05	}$	\\
GJ86 b	&	0.80	&	$	15.77	^{+	0.00	}_{-	0.00	}$	&	$	431.19	^{+	61.11	}_{-	59.29	}$	&	$	0.23	^{+	0.11	}_{-	0.12	}$	&	$	204.72	^{+	26.76	}_{-	21.68	}$	&	$	4.44	^{+	0.59	}_{-	0.59	}$	&	$	0.11	^{+	0.00	}_{-	0.00	}$	\\
HAT-P-6 b	&	1.29	&	$	3.85	^{+	0.00	}_{-	0.00	}$	&	$	115.69	^{+	3.99	}_{-	4.17	}$	&	$	0.04	^{+	0.04	}_{-	0.03	}$	&	$	8.73	^{+	3.25	}_{-	2.46	}$	&	$	1.06	^{+	0.04	}_{-	0.04	}$	&	$	0.05	^{+	0.00	}_{-	0.00	}$	\\
HAT-P-8 b	&	1.28	&	$	3.09	^{+	0.00	}_{-	0.00	}$	&	$	162.59	^{+	7.36	}_{-	6.46	}$	&	$	0.05	^{+	0.05	}_{-	0.03	}$	&	$	6.56	^{+	4.14	}_{-	3.02	}$	&	$	1.37	^{+	0.06	}_{-	0.05	}$	&	$	0.05	^{+	0.00	}_{-	0.00	}$	\\
HAT-P-9 b	&	1.30	&	$	3.92	^{+	0.00	}_{-	0.00	}$	&	$	84.50	^{+	10.56	}_{-	9.37	}$	&	$	0.12	^{+	0.14	}_{-	0.09	}$	&	$	4.09	^{+	9.61	}_{-	3.40	}$	&	$	0.77	^{+	0.09	}_{-	0.09	}$	&	$	0.05	^{+	0.00	}_{-	0.00	}$	\\
HD101930 b	&	0.74	&	$	70.58	^{+	0.40	}_{-	0.37	}$	&	$	17.99	^{+	0.89	}_{-	0.91	}$	&	$	0.08	^{+	0.05	}_{-	0.05	}$	&	$	1.92	^{+	0.65	}_{-	0.46	}$	&	$	0.30	^{+	0.01	}_{-	0.02	}$	&	$	0.30	^{+	0.00	}_{-	0.00	}$	\\
HD108874 b	&	0.95	&	$	395.16	^{+	5.60	}_{-	4.43	}$	&	$	34.93	^{+	3.82	}_{-	3.57	}$	&	$	0.05	^{+	0.08	}_{-	0.04	}$	&	$	13.00	^{+	1.65	}_{-	1.38	}$	&	$	1.21	^{+	0.13	}_{-	0.12	}$	&	$	1.04	^{+	0.01	}_{-	0.01	}$	\\
HD11506 b	&	1.19	&	$	1456.01	^{+	136.42	}_{-	85.10	}$	&	$	81.49	^{+	13.56	}_{-	4.62	}$	&	$	0.37	^{+	0.16	}_{-	0.10	}$	&	$	10.60	^{+	2.06	}_{-	1.60	}$	&	$	4.76	^{+	0.46	}_{-	0.23	}$	&	$	2.66	^{+	0.16	}_{-	0.10	}$	\\
HD118203 b	&	1.23	&	$	6.13	^{+	0.00	}_{-	0.00	}$	&	$	213.96	^{+	6.49	}_{-	6.40	}$	&	$	0.30	^{+	0.03	}_{-	0.03	}$	&	$	22.83	^{+	3.88	}_{-	3.28	}$	&	$	2.11	^{+	0.06	}_{-	0.06	}$	&	$	0.07	^{+	0.00	}_{-	0.00	}$	\\
HD12661 b	&	1.14	&	$	262.75	^{+	0.09	}_{-	0.13	}$	&	$	77.37	^{+	2.52	}_{-	2.55	}$	&	$	0.27	^{+	0.03	}_{-	0.03	}$	&	$	17.60	^{+	1.38	}_{-	1.24	}$	&	$	2.56	^{+	0.08	}_{-	0.09	}$	&	$	0.84	^{+	0.00	}_{-	0.00	}$	\\
HD128311 b	&	0.83	&	$	921.18	^{+	6.65	}_{-	5.13	}$	&	$	93.88	^{+	7.75	}_{-	7.27	}$	&	$	0.46	^{+	0.05	}_{-	0.05	}$	&	$	30.34	^{+	2.80	}_{-	2.42	}$	&	$	3.51	^{+	0.24	}_{-	0.24	}$	&	$	1.74	^{+	0.01	}_{-	0.01	}$	\\
HD131664 b	&	1.10	&	$	1976.18	^{+	32.94	}_{-	41.05	}$	&	$	356.10	^{+	24.90	}_{-	18.59	}$	&	$	0.64	^{+	0.02	}_{-	0.02	}$	&	$	5.11	^{+	0.79	}_{-	0.66	}$	&	$	18.03	^{+	0.85	}_{-	0.65	}$	&	$	3.18	^{+	0.04	}_{-	0.04	}$	\\
HD132406 b	&	1.09	&	$	1172.21	^{+	75.55	}_{-	49.55	}$	&	$	122.19	^{+	157.18	}_{-	32.90	}$	&	$	0.34	^{+	0.28	}_{-	0.19	}$	&	$	17.04	^{+	4.70	}_{-	3.61	}$	&	$	6.31	^{+	5.80	}_{-	1.47	}$	&	$	2.24	^{+	0.10	}_{-	0.06	}$	\\
HD142 b	&	1.23	&	$	344.05	^{+	2.12	}_{-	0.93	}$	&	$	32.00	^{+	7.12	}_{-	6.14	}$	&	$	0.19	^{+	0.16	}_{-	0.13	}$	&	$	20.29	^{+	2.53	}_{-	2.15	}$	&	$	1.24	^{+	0.24	}_{-	0.23	}$	&	$	1.03	^{+	0.00	}_{-	0.00	}$	\\
HD142022 b	&	0.90	&	$	1861.69	^{+	14.86	}_{-	13.47	}$	&	$	140.10	^{+	112.02	}_{-	39.74	}$	&	$	0.64	^{+	0.12	}_{-	0.09	}$	&	$	3.00	^{+	1.67	}_{-	1.29	}$	&	$	6.10	^{+	3.21	}_{-	1.36	}$	&	$	2.86	^{+	0.02	}_{-	0.01	}$	\\
HD149143 b	&	1.20	&	$	4.07	^{+	0.00	}_{-	0.00	}$	&	$	149.71	^{+	1.67	}_{-	1.61	}$	&	$	0.01	^{+	0.01	}_{-	0.01	}$	&	$	1.21	^{+	1.69	}_{-	0.92	}$	&	$	1.33	^{+	0.01	}_{-	0.01	}$	&	$	0.05	^{+	0.00	}_{-	0.00	}$	\\
HD154345 b	&	0.89	&	$	3332.50	^{+	84.05	}_{-	74.54	}$	&	$	14.10	^{+	0.84	}_{-	0.85	}$	&	$	0.05	^{+	0.05	}_{-	0.04	}$	&	$	2.84	^{+	0.37	}_{-	0.32	}$	&	$	0.96	^{+	0.06	}_{-	0.06	}$	&	$	4.20	^{+	0.07	}_{-	0.06	}$	\\
HD155358 b	&	0.87	&	$	194.26	^{+	0.88	}_{-	0.80	}$	&	$	31.86	^{+	1.98	}_{-	1.97	}$	&	$	0.21	^{+	0.06	}_{-	0.06	}$	&	$	9.69	^{+	1.03	}_{-	0.89	}$	&	$	0.81	^{+	0.05	}_{-	0.05	}$	&	$	0.63	^{+	0.00	}_{-	0.00	}$	\\
HD162020 b	&	0.80	&	$	8.43	^{+	0.00	}_{-	0.00	}$	&	$	1808.97	^{+	5.15	}_{-	5.13	}$	&	$	0.28	^{+	0.00	}_{-	0.00	}$	&	$	11.13	^{+	2.65	}_{-	2.39	}$	&	$	15.01	^{+	0.04	}_{-	0.04	}$	&	$	0.08	^{+	0.00	}_{-	0.00	}$	\\
HD168443 b	&	1.01	&	$	58.11	^{+	0.00	}_{-	0.00	}$	&	$	510.46	^{+	252.18	}_{-	117.22	}$	&	$	0.52	^{+	0.20	}_{-	0.18	}$	&	$	220.77	^{+	46.40	}_{-	34.29	}$	&	$	8.28	^{+	2.91	}_{-	1.94	}$	&	$	0.29	^{+	0.00	}_{-	0.00	}$	\\
HD169830 b	&	1.41	&	$	225.62	^{+	0.29	}_{-	0.31	}$	&	$	83.07	^{+	3.05	}_{-	3.09	}$	&	$	0.37	^{+	0.03	}_{-	0.03	}$	&	$	1.52	^{+	2.40	}_{-	1.18	}$	&	$	2.91	^{+	0.10	}_{-	0.10	}$	&	$	0.81	^{+	0.00	}_{-	0.00	}$	\\
HD171028 b	&	0.99	&	$	545.13	^{+	10.10	}_{-	12.19	}$	&	$	59.75	^{+	3.04	}_{-	2.05	}$	&	$	0.59	^{+	0.02	}_{-	0.02	}$	&	$	2.55	^{+	0.70	}_{-	0.51	}$	&	$	1.92	^{+	0.13	}_{-	0.10	}$	&	$	1.30	^{+	0.02	}_{-	0.02	}$	\\
HD183263 b	&	1.12	&	$	627.80	^{+	1.03	}_{-	1.64	}$	&	$	89.99	^{+	13.01	}_{-	11.63	}$	&	$	0.42	^{+	0.08	}_{-	0.09	}$	&	$	26.59	^{+	3.48	}_{-	2.91	}$	&	$	3.72	^{+	0.42	}_{-	0.41	}$	&	$	1.49	^{+	0.00	}_{-	0.00	}$	\\
HD185269 b	&	1.30	&	$	6.84	^{+	0.00	}_{-	0.00	}$	&	$	89.57	^{+	4.12	}_{-	4.02	}$	&	$	0.28	^{+	0.03	}_{-	0.04	}$	&	$	7.72	^{+	1.92	}_{-	1.61	}$	&	$	0.96	^{+	0.04	}_{-	0.04	}$	&	$	0.08	^{+	0.00	}_{-	0.00	}$	\\
HD187123 b	&	1.04	&	$	3.10	^{+	0.00	}_{-	0.00	}$	&	$	65.68	^{+	3.34	}_{-	3.35	}$	&	$	0.05	^{+	0.06	}_{-	0.04	}$	&	$	18.33	^{+	1.83	}_{-	1.58	}$	&	$	0.48	^{+	0.02	}_{-	0.03	}$	&	$	0.04	^{+	0.00	}_{-	0.00	}$	\\
HD189733 b	&	0.81	&	$	2.22	^{+	0.00	}_{-	0.00	}$	&	$	204.58	^{+	5.15	}_{-	5.10	}$	&	$	0.01	^{+	0.01	}_{-	0.01	}$	&	$	15.65	^{+	1.33	}_{-	1.17	}$	&	$	1.14	^{+	0.03	}_{-	0.03	}$	&	$	0.03	^{+	0.00	}_{-	0.00	}$	\\
HD190228 b	&	1.82	&	$	1141.21	^{+	15.40	}_{-	14.66	}$	&	$	92.26	^{+	4.58	}_{-	3.48	}$	&	$	0.52	^{+	0.04	}_{-	0.04	}$	&	$	1.23	^{+	1.87	}_{-	0.94	}$	&	$	6.07	^{+	0.17	}_{-	0.15	}$	&	$	2.61	^{+	0.02	}_{-	0.02	}$	\\
HD190360 b	&	0.98	&	$	2925.83	^{+	36.05	}_{-	41.53	}$	&	$	19.38	^{+	2.67	}_{-	2.28	}$	&	$	0.33	^{+	0.11	}_{-	0.10	}$	&	$	5.92	^{+	1.47	}_{-	1.47	}$	&	$	1.27	^{+	0.14	}_{-	0.13	}$	&	$	3.98	^{+	0.03	}_{-	0.04	}$	\\
HD190647 b	&	1.10	&	$	1038.09	^{+	5.27	}_{-	5.38	}$	&	$	36.78	^{+	1.19	}_{-	1.17	}$	&	$	0.17	^{+	0.02	}_{-	0.02	}$	&	$	0.97	^{+	0.70	}_{-	0.65	}$	&	$	1.92	^{+	0.06	}_{-	0.06	}$	&	$	2.07	^{+	0.01	}_{-	0.01	}$	\\
HD195019 b	&	1.02	&	$	18.20	^{+	0.00	}_{-	0.00	}$	&	$	270.12	^{+	1.54	}_{-	1.55	}$	&	$	0.01	^{+	0.01	}_{-	0.01	}$	&	$	10.42	^{+	1.19	}_{-	1.14	}$	&	$	3.54	^{+	0.02	}_{-	0.02	}$	&	$	0.14	^{+	0.00	}_{-	0.00	}$	\\
HD202206 b	&	1.07	&	$	255.90	^{+	0.06	}_{-	0.09	}$	&	$	585.94	^{+	6.24	}_{-	6.13	}$	&	$	0.42	^{+	0.01	}_{-	0.01	}$	&	$	30.18	^{+	2.61	}_{-	2.31	}$	&	$	17.41	^{+	0.16	}_{-	0.16	}$	&	$	0.81	^{+	0.00	}_{-	0.00	}$	\\
HD20868 b	&	0.78	&	$	380.79	^{+	0.13	}_{-	0.09	}$	&	$	97.02	^{+	7.97	}_{-	7.95	}$	&	$	0.61	^{+	0.04	}_{-	0.04	}$	&	$	32.81	^{+	4.04	}_{-	3.35	}$	&	$	2.31	^{+	0.18	}_{-	0.18	}$	&	$	0.95	^{+	0.00	}_{-	0.00	}$	\\
HD209458 b	&	1.13	&	$	3.52	^{+	0.00	}_{-	0.00	}$	&	$	84.33	^{+	0.87	}_{-	0.87	}$	&	$	0.01	^{+	0.01	}_{-	0.01	}$	&	$	3.34	^{+	0.69	}_{-	0.67	}$	&	$	0.69	^{+	0.01	}_{-	0.01	}$	&	$	0.05	^{+	0.00	}_{-	0.00	}$	\\
HD212301 b	&	1.05	&	$	2.27	^{+	0.00	}_{-	0.00	}$	&	$	56.31	^{+	5.83	}_{-	5.94	}$	&	$	0.08	^{+	0.08	}_{-	0.05	}$	&	$	17.70	^{+	3.65	}_{-	2.73	}$	&	$	0.37	^{+	0.04	}_{-	0.04	}$	&	$	0.03	^{+	0.00	}_{-	0.00	}$	\\
HD217107 b	&	1.11	&	$	7.13	^{+	0.00	}_{-	0.00	}$	&	$	140.71	^{+	2.35	}_{-	2.41	}$	&	$	0.15	^{+	0.02	}_{-	0.02	}$	&	$	22.70	^{+	1.23	}_{-	1.12	}$	&	$	1.41	^{+	0.02	}_{-	0.02	}$	&	$	0.08	^{+	0.00	}_{-	0.00	}$	\\
HD219828 b	&	1.24	&	$	3.84	^{+	0.01	}_{-	0.02	}$	&	$	3.11	^{+	7.83	}_{-	2.55	}$	&	$	0.58	^{+	0.32	}_{-	0.39	}$	&	$	15.98	^{+	2.97	}_{-	2.27	}$	&	$	0.02	^{+	0.05	}_{-	0.02	}$	&	$	0.05	^{+	0.00	}_{-	0.00	}$	\\
HD221287 b	&	1.30	&	$	458.77	^{+	8.13	}_{-	6.20	}$	&	$	69.77	^{+	8.66	}_{-	6.77	}$	&	$	0.11	^{+	0.10	}_{-	0.07	}$	&	$	10.16	^{+	1.97	}_{-	1.50	}$	&	$	3.14	^{+	0.38	}_{-	0.32	}$	&	$	1.27	^{+	0.01	}_{-	0.01	}$	\\
HD224693 b	&	1.30	&	$	26.73	^{+	0.03	}_{-	0.03	}$	&	$	39.92	^{+	1.52	}_{-	1.53	}$	&	$	0.04	^{+	0.04	}_{-	0.03	}$	&	$	1.92	^{+	1.07	}_{-	1.10	}$	&	$	0.70	^{+	0.03	}_{-	0.03	}$	&	$	0.19	^{+	0.00	}_{-	0.00	}$	\\
HD23127 b	&	1.13	&	$	1226.63	^{+	21.59	}_{-	21.71	}$	&	$	27.75	^{+	3.08	}_{-	2.84	}$	&	$	0.44	^{+	0.09	}_{-	0.10	}$	&	$	10.89	^{+	2.03	}_{-	1.67	}$	&	$	1.42	^{+	0.17	}_{-	0.16	}$	&	$	2.34	^{+	0.03	}_{-	0.03	}$	\\
HD2638 b	&	0.93	&	$	3.44	^{+	0.00	}_{-	0.00	}$	&	$	67.59	^{+	1.06	}_{-	1.02	}$	&	$	0.01	^{+	0.01	}_{-	0.01	}$	&	$	3.31	^{+	0.70	}_{-	0.57	}$	&	$	0.48	^{+	0.01	}_{-	0.01	}$	&	$	0.04	^{+	0.00	}_{-	0.00	}$	\\
HD27442 b	&	1.20	&	$	415.32	^{+	6.25	}_{-	5.74	}$	&	$	32.48	^{+	1.79	}_{-	1.76	}$	&	$	0.07	^{+	0.06	}_{-	0.04	}$	&	$	2.98	^{+	1.41	}_{-	1.13	}$	&	$	1.34	^{+	0.07	}_{-	0.07	}$	&	$	1.16	^{+	0.01	}_{-	0.01	}$	\\
HD27894 b	&	0.75	&	$	18.01	^{+	0.02	}_{-	0.01	}$	&	$	57.01	^{+	1.61	}_{-	1.66	}$	&	$	0.04	^{+	0.03	}_{-	0.02	}$	&	$	4.39	^{+	1.08	}_{-	0.83	}$	&	$	0.61	^{+	0.02	}_{-	0.02	}$	&	$	0.12	^{+	0.00	}_{-	0.00	}$	\\
HD28185 b	&	0.99	&	$	381.81	^{+	0.83	}_{-	1.32	}$	&	$	174.72	^{+	12.09	}_{-	7.75	}$	&	$	0.05	^{+	0.02	}_{-	0.02	}$	&	$	7.82	^{+	1.76	}_{-	1.53	}$	&	$	6.19	^{+	0.43	}_{-	0.28	}$	&	$	1.03	^{+	0.00	}_{-	0.00	}$	\\
HD285968 b	&	0.49	&	$	10.23	^{+	0.00	}_{-	0.00	}$	&	$	11.88	^{+	2.24	}_{-	1.79	}$	&	$	0.25	^{+	0.20	}_{-	0.17	}$	&	$	2.47	^{+	1.77	}_{-	1.74	}$	&	$	0.08	^{+	0.01	}_{-	0.01	}$	&	$	0.07	^{+	0.00	}_{-	0.00	}$	\\
HD330075 b	&	0.70	&	$	3.39	^{+	0.00	}_{-	0.00	}$	&	$	107.34	^{+	1.00	}_{-	1.03	}$	&	$	0.01	^{+	0.01	}_{-	0.00	}$	&	$	2.02	^{+	0.86	}_{-	0.74	}$	&	$	0.63	^{+	0.01	}_{-	0.01	}$	&	$	0.04	^{+	0.00	}_{-	0.00	}$	\\
HD33636 b	&	1.02	&	$	2127.74	^{+	11.40	}_{-	11.04	}$	&	$	389.98	^{+	156.81	}_{-	152.41	}$	&	$	0.90	^{+	0.03	}_{-	0.11	}$	&	$	0.57	^{+	0.69	}_{-	0.42	}$	&	$	11.02	^{+	2.19	}_{-	1.89	}$	&	$	3.26	^{+	0.01	}_{-	0.01	}$	\\
HD3651 b	&	0.88	&	$	60.36	^{+	0.04	}_{-	0.05	}$	&	$	9.60	^{+	1.91	}_{-	1.60	}$	&	$	0.54	^{+	0.15	}_{-	0.16	}$	&	$	7.73	^{+	0.68	}_{-	0.61	}$	&	$	0.14	^{+	0.02	}_{-	0.02	}$	&	$	0.29	^{+	0.00	}_{-	0.00	}$	\\
HD38529 b	&	1.48	&	$	2143.62	^{+	8.24	}_{-	6.24	}$	&	$	177.12	^{+	6.26	}_{-	6.00	}$	&	$	0.35	^{+	0.03	}_{-	0.03	}$	&	$	40.24	^{+	2.46	}_{-	2.22	}$	&	$	13.65	^{+	0.42	}_{-	0.42	}$	&	$	3.71	^{+	0.01	}_{-	0.01	}$	\\

\multicolumn{8}{r}{\tiny{\textit{continued on next page...}}}
		\end{tabular}
	\end{center}
\end{table*}

\begin{table*}
\scriptsize
\smallskip
	\begin{center}
		\begin{tabular}{rcllllll}
			\hline\noalign{\smallskip}
\multicolumn{1}{c}{\textbf{planet}} & 
\multicolumn{1}{c}{\textbf{$m_{*} (M_{sol})$}} & 
\multicolumn{1}{c}{\textbf{$T (days)$}} & 
\multicolumn{1}{c}{\textbf{$K (ms^{-1})$}} & 
\multicolumn{1}{c}{\textbf{$e$}} & 
\multicolumn{1}{c}{\textbf{$s$}} & 
\multicolumn{1}{c}{\textbf{$m_p (M_{Jup})$}} & 
\multicolumn{1}{c}{\textbf{$a (AU)$}} \\
			\hline\hline

HD4203 b	&	1.13	&	$	434.21	^{+	2.65	}_{-	1.97	}$	&	$	74.24	^{+	88.38	}_{-	18.24	}$	&	$	0.71	^{+	0.14	}_{-	0.09	}$	&	$	5.63	^{+	1.49	}_{-	1.18	}$	&	$	2.15	^{+	1.42	}_{-	0.39	}$	&	$	1.17	^{+	0.00	}_{-	0.00	}$	\\
HD4208 b	&	0.88	&	$	828.90	^{+	8.03	}_{-	7.83	}$	&	$	19.01	^{+	0.68	}_{-	0.70	}$	&	$	0.05	^{+	0.04	}_{-	0.03	}$	&	$	1.26	^{+	0.86	}_{-	0.85	}$	&	$	0.81	^{+	0.03	}_{-	0.03	}$	&	$	1.65	^{+	0.01	}_{-	0.01	}$	\\
HD43691 b	&	1.38	&	$	36.93	^{+	0.04	}_{-	0.04	}$	&	$	123.78	^{+	4.76	}_{-	4.65	}$	&	$	0.11	^{+	0.05	}_{-	0.05	}$	&	$	14.47	^{+	3.07	}_{-	2.51	}$	&	$	2.50	^{+	0.10	}_{-	0.10	}$	&	$	0.24	^{+	0.00	}_{-	0.00	}$	\\
HD43848 b	&	0.93	&	$	2390.59	^{+	121.57	}_{-	85.08	}$	&	$	888.17	^{+	447.92	}_{-	286.53	}$	&	$	0.81	^{+	0.06	}_{-	0.08	}$	&	$	5.58	^{+	4.19	}_{-	2.90	}$	&	$	33.07	^{+	9.51	}_{-	7.29	}$	&	$	3.42	^{+	0.11	}_{-	0.08	}$	\\
HD46375 b	&	0.93	&	$	3.02	^{+	0.00	}_{-	0.00	}$	&	$	33.67	^{+	0.79	}_{-	0.80	}$	&	$	0.06	^{+	0.03	}_{-	0.03	}$	&	$	3.30	^{+	0.59	}_{-	0.52	}$	&	$	0.23	^{+	0.01	}_{-	0.01	}$	&	$	0.04	^{+	0.00	}_{-	0.00	}$	\\
HD47536 b	&	0.94	&	$	717.27	^{+	15.06	}_{-	13.33	}$	&	$	108.71	^{+	12.96	}_{-	12.35	}$	&	$	0.15	^{+	0.10	}_{-	0.09	}$	&	$	6.65	^{+	9.40	}_{-	5.62	}$	&	$	4.53	^{+	0.57	}_{-	0.56	}$	&	$	1.54	^{+	0.02	}_{-	0.02	}$	\\
HD49674 b	&	1.01	&	$	4.95	^{+	0.00	}_{-	0.00	}$	&	$	11.76	^{+	1.09	}_{-	1.13	}$	&	$	0.05	^{+	0.07	}_{-	0.04	}$	&	$	3.65	^{+	0.78	}_{-	0.69	}$	&	$	0.10	^{+	0.01	}_{-	0.01	}$	&	$	0.06	^{+	0.00	}_{-	0.00	}$	\\
HD50499 b	&	1.28	&	$	2460.46	^{+	51.66	}_{-	51.79	}$	&	$	25.77	^{+	5.54	}_{-	4.56	}$	&	$	0.29	^{+	0.21	}_{-	0.19	}$	&	$	13.38	^{+	2.08	}_{-	1.71	}$	&	$	1.91	^{+	0.33	}_{-	0.32	}$	&	$	3.87	^{+	0.05	}_{-	0.05	}$	\\
HD5319 b	&	1.60	&	$	684.51	^{+	12.52	}_{-	16.78	}$	&	$	36.23	^{+	17.09	}_{-	5.87	}$	&	$	0.10	^{+	0.20	}_{-	0.07	}$	&	$	10.44	^{+	1.73	}_{-	1.37	}$	&	$	2.13	^{+	0.91	}_{-	0.35	}$	&	$	1.78	^{+	0.02	}_{-	0.03	}$	\\
HD63454 b	&	0.80	&	$	2.82	^{+	0.00	}_{-	0.00	}$	&	$	63.32	^{+	1.76	}_{-	1.81	}$	&	$	0.02	^{+	0.03	}_{-	0.02	}$	&	$	5.80	^{+	1.25	}_{-	0.98	}$	&	$	0.38	^{+	0.01	}_{-	0.01	}$	&	$	0.04	^{+	0.00	}_{-	0.00	}$	\\
HD68988	&	1.12	&	$	6.28	^{+	0.00	}_{-	0.00	}$	&	$	183.95	^{+	13.27	}_{-	13.39	}$	&	$	0.09	^{+	0.06	}_{-	0.06	}$	&	$	42.29	^{+	7.33	}_{-	5.71	}$	&	$	1.79	^{+	0.12	}_{-	0.13	}$	&	$	0.07	^{+	0.00	}_{-	0.00	}$	\\
HD73267 b	&	0.89	&	$	1259.62	^{+	6.47	}_{-	6.80	}$	&	$	64.28	^{+	0.44	}_{-	0.46	}$	&	$	0.26	^{+	0.01	}_{-	0.01	}$	&	$	0.72	^{+	0.50	}_{-	0.48	}$	&	$	3.06	^{+	0.02	}_{-	0.02	}$	&	$	2.20	^{+	0.01	}_{-	0.01	}$	\\
HD73526 b	&	1.01	&	$	193.32	^{+	0.75	}_{-	1.05	}$	&	$	114.79	^{+	22.33	}_{-	18.22	}$	&	$	0.57	^{+	0.07	}_{-	0.09	}$	&	$	25.52	^{+	4.75	}_{-	3.69	}$	&	$	2.70	^{+	0.40	}_{-	0.35	}$	&	$	0.66	^{+	0.00	}_{-	0.00	}$	\\
HD74156 b	&	1.24	&	$	2519.62	^{+	20.55	}_{-	20.33	}$	&	$	120.88	^{+	139.67	}_{-	42.57	}$	&	$	0.89	^{+	0.08	}_{-	0.13	}$	&	$	54.09	^{+	4.69	}_{-	4.09	}$	&	$	4.31	^{+	1.34	}_{-	0.68	}$	&	$	3.89	^{+	0.02	}_{-	0.02	}$	\\
HD75289 b	&	1.19	&	$	3.51	^{+	0.00	}_{-	0.00	}$	&	$	53.90	^{+	1.31	}_{-	1.31	}$	&	$	0.01	^{+	0.02	}_{-	0.01	}$	&	$	0.73	^{+	1.04	}_{-	0.54	}$	&	$	0.45	^{+	0.01	}_{-	0.01	}$	&	$	0.05	^{+	0.00	}_{-	0.00	}$	\\
HD75898 b	&	1.30	&	$	421.64	^{+	7.49	}_{-	7.35	}$	&	$	74.99	^{+	3.83	}_{-	3.66	}$	&	$	0.05	^{+	0.07	}_{-	0.04	}$	&	$	9.56	^{+	2.16	}_{-	1.59	}$	&	$	3.29	^{+	0.17	}_{-	0.16	}$	&	$	1.20	^{+	0.01	}_{-	0.01	}$	\\
HD76700 b	&	1.13	&	$	3.97	^{+	0.00	}_{-	0.00	}$	&	$	27.30	^{+	1.35	}_{-	1.29	}$	&	$	0.08	^{+	0.06	}_{-	0.05	}$	&	$	2.79	^{+	1.46	}_{-	1.58	}$	&	$	0.23	^{+	0.01	}_{-	0.01	}$	&	$	0.05	^{+	0.00	}_{-	0.00	}$	\\
HD80606 b	&	0.96	&	$	111.44	^{+	0.00	}_{-	0.00	}$	&	$	560.69	^{+	187.15	}_{-	113.26	}$	&	$	0.95	^{+	0.01	}_{-	0.02	}$	&	$	12.47	^{+	2.65	}_{-	2.53	}$	&	$	3.84	^{+	0.51	}_{-	0.30	}$	&	$	0.45	^{+	0.00	}_{-	0.00	}$	\\
HD81040 b	&	0.96	&	$	1108.32	^{+	7.45	}_{-	4.65	}$	&	$	169.71	^{+	10.03	}_{-	10.03	}$	&	$	0.42	^{+	0.07	}_{-	0.14	}$	&	$	25.51	^{+	6.90	}_{-	5.16	}$	&	$	7.62	^{+	0.56	}_{-	0.47	}$	&	$	2.07	^{+	0.01	}_{-	0.01	}$	\\
HD82943  b	&	1.13	&	$	4030.24	^{+	7058.52	}_{-	3812.23	}$	&	$	88.77	^{+	347.92	}_{-	55.10	}$	&	$	0.49	^{+	0.30	}_{-	0.33	}$	&	$	34.40	^{+	8.27	}_{-	6.74	}$	&	$	5.58	^{+	28.20	}_{-	4.69	}$	&	$	5.16	^{+	4.97	}_{-	4.42	}$	\\
HD8574 b	&	1.12	&	$	227.45	^{+	0.82	}_{-	0.79	}$	&	$	64.92	^{+	4.36	}_{-	4.37	}$	&	$	0.28	^{+	0.05	}_{-	0.05	}$	&	$	9.70	^{+	2.53	}_{-	2.39	}$	&	$	2.02	^{+	0.13	}_{-	0.13	}$	&	$	0.76	^{+	0.00	}_{-	0.00	}$	\\
HD86081 b	&	1.21	&	$	2.14	^{+	0.00	}_{-	0.00	}$	&	$	207.49	^{+	0.86	}_{-	0.84	}$	&	$	0.01	^{+	0.01	}_{-	0.01	}$	&	$	0.76	^{+	0.92	}_{-	0.56	}$	&	$	1.49	^{+	0.01	}_{-	0.01	}$	&	$	0.03	^{+	0.00	}_{-	0.00	}$	\\
HD89307	&	0.99	&	$	2174.23	^{+	49.69	}_{-	53.83	}$	&	$	38.05	^{+	26.97	}_{-	7.05	}$	&	$	0.29	^{+	0.36	}_{-	0.20	}$	&	$	3.31	^{+	5.42	}_{-	2.68	}$	&	$	2.29	^{+	0.83	}_{-	0.41	}$	&	$	3.27	^{+	0.05	}_{-	0.05	}$	\\
HD93083 b	&	0.70	&	$	143.99	^{+	1.54	}_{-	1.47	}$	&	$	18.68	^{+	1.26	}_{-	1.20	}$	&	$	0.12	^{+	0.07	}_{-	0.06	}$	&	$	2.28	^{+	0.75	}_{-	0.54	}$	&	$	0.38	^{+	0.03	}_{-	0.03	}$	&	$	0.48	^{+	0.00	}_{-	0.00	}$	\\
HR 810 b	&	1.11	&	$	302.94	^{+	2.27	}_{-	2.18	}$	&	$	57.17	^{+	5.60	}_{-	5.43	}$	&	$	0.12	^{+	0.10	}_{-	0.08	}$	&	$	17.92	^{+	3.70	}_{-	2.91	}$	&	$	2.00	^{+	0.20	}_{-	0.20	}$	&	$	0.91	^{+	0.00	}_{-	0.00	}$	\\
kappa CrB b	&	1.80	&	$	1218.78	^{+	35.22	}_{-	28.41	}$	&	$	24.06	^{+	1.39	}_{-	1.41	}$	&	$	0.12	^{+	0.08	}_{-	0.07	}$	&	$	4.73	^{+	0.86	}_{-	0.75	}$	&	$	1.86	^{+	0.11	}_{-	0.11	}$	&	$	2.72	^{+	0.05	}_{-	0.04	}$	\\
NGC 2423 3 b	&	2.40	&	$	713.95	^{+	4.99	}_{-	5.20	}$	&	$	133.14	^{+	7.95	}_{-	7.28	}$	&	$	0.18	^{+	0.06	}_{-	0.06	}$	&	$	18.05	^{+	2.72	}_{-	2.32	}$	&	$	10.32	^{+	0.56	}_{-	0.55	}$	&	$	2.09	^{+	0.01	}_{-	0.01	}$	\\
NGC 4349 127 b	&	3.90	&	$	676.93	^{+	4.37	}_{-	4.48	}$	&	$	189.46	^{+	10.22	}_{-	9.44	}$	&	$	0.19	^{+	0.05	}_{-	0.05	}$	&	$	14.59	^{+	3.43	}_{-	2.50	}$	&	$	19.89	^{+	1.06	}_{-	1.02	}$	&	$	2.38	^{+	0.01	}_{-	0.01	}$	\\
tau Boo b	&	1.34	&	$	3.31	^{+	10.38	}_{-	10.09	}$	&	$	469.05	^{+	14.76	}_{-	14.87	}$	&	$	0.07	^{+	0.03	}_{-	0.04	}$	&	$	96.65	^{+	8.42	}_{-	7.39	}$	&	$	4.17	^{+	0.13	}_{-	0.13	}$	&	$	0.05	^{+	0.00	}_{-	0.00	}$	\\
TrES-3 b	&	0.92	&	$	1.34	^{+	0.04	}_{-	0.04	}$	&	$	352.25	^{+	17.08	}_{-	21.08	}$	&	$	0.06	^{+	0.09	}_{-	0.04	}$	&	$	10.70	^{+	24.41	}_{-	9.07	}$	&	$	1.80	^{+	0.09	}_{-	0.12	}$	&	$	0.02	^{+	0.00	}_{-	0.00	}$	\\
WASP-2 b	&	0.88	&	$	2.15	^{+	0.00	}_{-	0.00	}$	&	$	159.43	^{+	8.00	}_{-	6.39	}$	&	$	0.25	^{+	0.10	}_{-	0.11	}$	&	$	2.32	^{+	5.14	}_{-	1.87	}$	&	$	0.90	^{+	0.04	}_{-	0.03	}$	&	$	0.03	^{+	0.00	}_{-	0.00	}$	\\
WASP-3 b	&	1.22	&	$	1.85	^{+	0.00	}_{-	0.00	}$	&	$	250.29	^{+	11.08	}_{-	11.16	}$	&	$	0.07	^{+	0.05	}_{-	0.04	}$	&	$	3.85	^{+	11.11	}_{-	3.21	}$	&	$	1.72	^{+	0.08	}_{-	0.08	}$	&	$	0.03	^{+	0.00	}_{-	0.00	}$	\\
WASP-4 b	&	0.91	&	$	1.34	^{+	0.00	}_{-	0.00	}$	&	$	238.69	^{+	10.83	}_{-	11.83	}$	&	$	0.03	^{+	0.04	}_{-	0.02	}$	&	$	11.37	^{+	11.28	}_{-	9.52	}$	&	$	1.21	^{+	0.06	}_{-	0.06	}$	&	$	0.02	^{+	0.00	}_{-	0.00	}$	\\
WASP-5 b	&	1.01	&	$	1.63	^{+	0.00	}_{-	0.00	}$	&	$	282.92	^{+	9.11	}_{-	9.08	}$	&	$	0.07	^{+	0.04	}_{-	0.03	}$	&	$	2.77	^{+	6.48	}_{-	2.23	}$	&	$	1.64	^{+	0.05	}_{-	0.05	}$	&	$	0.03	^{+	0.00	}_{-	0.00	}$	\\
XO-1 b	&	1.03	&	$	3.94	^{+	0.00	}_{-	0.00	}$	&	$	119.02	^{+	12.73	}_{-	11.60	}$	&	$	0.10	^{+	0.11	}_{-	0.07	}$	&	$	3.09	^{+	8.03	}_{-	2.54	}$	&	$	0.94	^{+	0.10	}_{-	0.09	}$	&	$	0.05	^{+	0.00	}_{-	0.00	}$	\\
XO-2 b	&	0.97	&	$	2.62	^{+	0.00	}_{-	0.00	}$	&	$	86.00	^{+	10.51	}_{-	10.07	}$	&	$	0.19	^{+	0.15	}_{-	0.13	}$	&	$	3.58	^{+	9.40	}_{-	2.94	}$	&	$	0.56	^{+	0.06	}_{-	0.07	}$	&	$	0.04	^{+	0.00	}_{-	0.00	}$	\\
XO-3 b	&	1.41	&	$	3.19	^{+	0.00	}_{-	0.00	}$	&	$	1501.73	^{+	16.33	}_{-	16.28	}$	&	$	0.29	^{+	0.01	}_{-	0.01	}$	&	$	39.89	^{+	7.29	}_{-	6.39	}$	&	$	13.09	^{+	0.14	}_{-	0.14	}$	&	$	0.05	^{+	0.00	}_{-	0.00	}$	\\
XO-4 b	&	1.32	&	$	4.13	^{+	0.00	}_{-	0.00	}$	&	$	200.12	^{+	92.23	}_{-	34.80	}$	&	$	0.35	^{+	0.22	}_{-	0.20	}$	&	$	4.34	^{+	13.06	}_{-	3.64	}$	&	$	1.79	^{+	0.50	}_{-	0.27	}$	&	$	0.06	^{+	0.00	}_{-	0.00	}$	\\

\hline
		\end{tabular}
	\end{center}
\end{table*}


\begin{table*}
\caption{Table of the orbital parameters for a 2-planet fit, both directly output from \textsc{exofit} and thence derived. The values of the parameters $T$, $K$, $e$ and $s$ (generated from \textsc{exofit}) are the medians of the parameter posterior distributions, with the associated $68.3\%$ confidence regions. The other parameters were calculated using these values and stellar masses taken from the published literature. Note that some parameters are extremely well-constrained, hence the errors on the parameter estimates are so small as to appear to be zero to the two decimal places shown in this table. A full table in machine-readable format will be provided on the website, and the reader is directed there if such data are required.}
\label{tab:catalogue2p}
\smallskip
\scriptsize

	\begin{center}
		\begin{tabular}{rcllllll}
			\hline\noalign{\smallskip}
\multicolumn{1}{c}{\textbf{planet}} & 
\multicolumn{1}{c}{\textbf{$m_{*} (M_{sol})$}} & 
\multicolumn{1}{c}{\textbf{$T (days)$}} & 
\multicolumn{1}{c}{\textbf{$K (ms^{-1})$}} & 
\multicolumn{1}{c}{\textbf{$e$}} & 
\multicolumn{1}{c}{\textbf{$s$}} & 
\multicolumn{1}{c}{\textbf{$m_p (M_{Jup})$}} & 
\multicolumn{1}{c}{\textbf{$a (AU)$}} \\
			\hline\hline

BD-17 63 b	&	0.74	&	$	1976.15	^{+	226.40	}_{-	182.28	}$	&	$	297.62	^{+	389.34	}_{-	93.47	}$	&	$	0.87	^{+	0.07	}_{-	0.07	}$	&	$	9.46	^{+	1.96	}_{-	1.47	}$	&	$	7.48	^{+	4.67	}_{-	1.37	}$	&	$	2.79	^{+	0.21	}_{-	0.17	}$	\\
c	&		&	$	2829.26	^{+	303.62	}_{-	183.09	}$	&	$	182.40	^{+	7.20	}_{-	5.52	}$	&	$	0.84	^{+	0.01	}_{-	0.01	}$	&								&	$	5.68	^{+	0.17	}_{-	0.14	}$	&	$	3.54	^{+	0.25	}_{-	0.15	}$	\\
ChaHa8 b	&	0.10	&	$	1184.29	^{+	619.30	}_{-	1170.75	}$	&	$	870.23	^{+	660.70	}_{-	855.84	}$	&	$	0.43	^{+	0.30	}_{-	0.29	}$	&	$	29.42	^{+	159.46	}_{-	27.44	}$	&	$	6.15	^{+	10.65	}_{-	6.11	}$	&	$	1.02	^{+	0.33	}_{-	0.97	}$	\\
c	&		&	$	504.54	^{+	1270.37	}_{-	493.52	}$	&	$	965.92	^{+	484.80	}_{-	952.10	}$	&	$	0.39	^{+	0.32	}_{-	0.27	}$	&								&	$	6.84	^{+	9.21	}_{-	6.80	}$	&	$	0.58	^{+	0.76	}_{-	0.53	}$	\\
epsilon Eri b	&	0.86	&	$	2443.29	^{+	45.64	}_{-	44.45	}$	&	$	13.91	^{+	1.51	}_{-	1.49	}$	&	$	0.05	^{+	0.07	}_{-	0.04	}$	&	$	7.72	^{+	0.84	}_{-	0.76	}$	&	$	0.83	^{+	0.09	}_{-	0.09	}$	&	$	3.37	^{+	0.04	}_{-	0.04	}$	\\
c	&		&	$	541.54	^{+	2.09	}_{-	1.05	}$	&	$	14.86	^{+	8.97	}_{-	4.11	}$	&	$	0.82	^{+	0.09	}_{-	0.10	}$	&								&	$	0.30	^{+	0.08	}_{-	0.06	}$	&	$	1.23	^{+	0.00	}_{-	0.00	}$	\\
epsilon Tau b	&	2.70	&	$	598.14	^{+	11.23	}_{-	10.56	}$	&	$	97.38	^{+	4.08	}_{-	3.98	}$	&	$	0.13	^{+	0.06	}_{-	0.05	}$	&	$	6.17	^{+	3.70	}_{-	5.02	}$	&	$	7.76	^{+	0.30	}_{-	0.31	}$	&	$	1.93	^{+	0.02	}_{-	0.02	}$	\\
c	&		&	$	71.16	^{+	1113.78	}_{-	70.55	}$	&	$	12.48	^{+	13.93	}_{-	10.77	}$	&	$	0.66	^{+	0.25	}_{-	0.37	}$	&								&	$	0.18	^{+	0.90	}_{-	0.14	}$	&	$	0.47	^{+	2.58	}_{-	0.45	}$	\\
gamma Cep b	&	1.59	&	$	905.43	^{+	4.49	}_{-	3.95	}$	&	$	37.97	^{+	4.81	}_{-	4.42	}$	&	$	0.04	^{+	0.07	}_{-	0.03	}$	&	$	1.51	^{+	2.27	}_{-	1.16	}$	&	$	2.46	^{+	0.31	}_{-	0.29	}$	&	$	2.14	^{+	0.01	}_{-	0.01	}$	\\
c	&		&	$	3648.72	^{+	2293.49	}_{-	661.35	}$	&	$	1761.05	^{+	176.92	}_{-	243.13	}$	&	$	0.75	^{+	0.06	}_{-	0.09	}$	&								&	$	117.32	^{+	51.84	}_{-	22.41	}$	&	$	5.41	^{+	2.08	}_{-	0.68	}$	\\
GJ3021 b	&	0.90	&	$	133.71	^{+	0.20	}_{-	0.21	}$	&	$	166.78	^{+	3.96	}_{-	3.91	}$	&	$	0.51	^{+	0.02	}_{-	0.02	}$	&	$	15.40	^{+	2.49	}_{-	2.32	}$	&	$	3.36	^{+	0.08	}_{-	0.08	}$	&	$	0.49	^{+	0.00	}_{-	0.00	}$	\\
c	&		&	$	33.23	^{+	3748.49	}_{-	32.08	}$	&	$	5.00	^{+	14.75	}_{-	4.26	}$	&	$	0.56	^{+	0.32	}_{-	0.38	}$	&								&	$	0.05	^{+	0.26	}_{-	0.05	}$	&	$	0.20	^{+	4.39	}_{-	0.17	}$	\\
GJ317 b	&	0.24	&	$	4602.69	^{+	3780.22	}_{-	1555.08	}$	&	$	33.00	^{+	22.89	}_{-	7.52	}$	&	$	0.35	^{+	0.23	}_{-	0.19	}$	&	$	4.51	^{+	4.39	}_{-	2.90	}$	&	$	0.95	^{+	0.86	}_{-	0.26	}$	&	$	3.37	^{+	1.65	}_{-	0.81	}$	\\
c	&		&	$	682.64	^{+	4.93	}_{-	4.68	}$	&	$	82.68	^{+	4.15	}_{-	3.87	}$	&	$	0.27	^{+	0.05	}_{-	0.06	}$	&								&	$	1.33	^{+	0.06	}_{-	0.06	}$	&	$	0.94	^{+	0.00	}_{-	0.00	}$	\\
GJ674 b	&	0.35	&	$	34.82	^{+	0.07	}_{-	0.08	}$	&	$	4.84	^{+	0.40	}_{-	0.42	}$	&	$	0.23	^{+	0.08	}_{-	0.08	}$	&	$	1.17	^{+	0.28	}_{-	0.22	}$	&	$	0.04	^{+	0.00	}_{-	0.00	}$	&	$	0.15	^{+	0.00	}_{-	0.00	}$	\\
c	&		&	$	4.69	^{+	0.00	}_{-	0.00	}$	&	$	8.59	^{+	0.39	}_{-	0.39	}$	&	$	0.14	^{+	0.04	}_{-	0.05	}$	&								&	$	0.03	^{+	0.00	}_{-	0.00	}$	&	$	0.04	^{+	0.00	}_{-	0.00	}$	\\
GJ849 b	&	0.49	&	$	1971.53	^{+	178.78	}_{-	104.48	}$	&	$	21.41	^{+	1.79	}_{-	1.68	}$	&	$	0.11	^{+	0.07	}_{-	0.07	}$	&	$	2.89	^{+	0.92	}_{-	0.85	}$	&	$	0.81	^{+	0.08	}_{-	0.07	}$	&	$	2.43	^{+	0.14	}_{-	0.09	}$	\\
c	&		&	$	9065.80	^{+	3863.21	}_{-	3507.96	}$	&	$	65.32	^{+	84.73	}_{-	38.59	}$	&	$	0.59	^{+	0.31	}_{-	0.37	}$	&								&	$	3.06	^{+	2.95	}_{-	1.70	}$	&	$	6.71	^{+	1.79	}_{-	1.87	}$	\\
GJ86 b	&	0.80	&	$	15.77	^{+	0.00	}_{-	0.00	}$	&	$	378.50	^{+	1.12	}_{-	1.11	}$	&	$	0.05	^{+	0.00	}_{-	0.00	}$	&	$	0.98	^{+	1.25	}_{-	0.73	}$	&	$	4.02	^{+	0.01	}_{-	0.01	}$	&	$	0.11	^{+	0.00	}_{-	0.00	}$	\\
c	&		&	$	6771.07	^{+	1223.25	}_{-	972.87	}$	&	$	1695.71	^{+	186.49	}_{-	272.94	}$	&	$	0.80	^{+	0.04	}_{-	0.03	}$	&								&	$	80.82	^{+	7.66	}_{-	7.35	}$	&	$	6.50	^{+	0.76	}_{-	0.64	}$	\\
HAT-P-6 b	&	1.29	&	$	87.07	^{+	4320.01	}_{-	85.42	}$	&	$	8.92	^{+	63.07	}_{-	7.84	}$	&	$	0.55	^{+	0.33	}_{-	0.38	}$	&	$	48.20	^{+	37.86	}_{-	35.69	}$	&	$	0.14	^{+	1.66	}_{-	0.12	}$	&	$	0.42	^{+	5.31	}_{-	0.39	}$	\\
c	&		&	$	3.85	^{+	1126.88	}_{-	0.00	}$	&	$	107.11	^{+	13.37	}_{-	101.50	}$	&	$	0.11	^{+	0.60	}_{-	0.09	}$	&								&	$	1.00	^{+	0.14	}_{-	0.91	}$	&	$	0.05	^{+	2.26	}_{-	0.00	}$	\\
HAT-P-8 b	&	1.28	&	$	24.39	^{+	2317.94	}_{-	23.15	}$	&	$	11.23	^{+	49.73	}_{-	9.61	}$	&	$	0.54	^{+	0.33	}_{-	0.37	}$	&	$	4.94	^{+	4.87	}_{-	3.95	}$	&	$	0.13	^{+	1.02	}_{-	0.11	}$	&	$	0.18	^{+	3.57	}_{-	0.15	}$	\\
c	&		&	$	3.08	^{+	0.00	}_{-	0.01	}$	&	$	160.51	^{+	8.09	}_{-	6.77	}$	&	$	0.05	^{+	0.06	}_{-	0.03	}$	&								&	$	1.35	^{+	0.07	}_{-	0.06	}$	&	$	0.05	^{+	0.00	}_{-	0.00	}$	\\
HAT-P-9 b	&	1.30	&	$	3.92	^{+	0.00	}_{-	0.00	}$	&	$	84.33	^{+	11.66	}_{-	9.75	}$	&	$	0.17	^{+	0.14	}_{-	0.10	}$	&	$	3.28	^{+	8.16	}_{-	2.70	}$	&	$	0.77	^{+	0.10	}_{-	0.09	}$	&	$	0.05	^{+	0.00	}_{-	0.00	}$	\\
c	&		&	$	138.54	^{+	4330.13	}_{-	136.58	}$	&	$	8.19	^{+	35.70	}_{-	7.18	}$	&	$	0.53	^{+	0.33	}_{-	0.37	}$	&								&	$	0.13	^{+	1.31	}_{-	0.12	}$	&	$	0.57	^{+	5.22	}_{-	0.54	}$	\\
HD101930 b	&	0.74	&	$	362.82	^{+	3814.57	}_{-	357.99	}$	&	$	2.85	^{+	7.90	}_{-	2.19	}$	&	$	0.55	^{+	0.33	}_{-	0.36	}$	&	$	1.62	^{+	0.71	}_{-	0.64	}$	&	$	0.05	^{+	0.26	}_{-	0.04	}$	&	$	0.90	^{+	3.69	}_{-	0.85	}$	\\
c	&		&	$	70.64	^{+	0.48	}_{-	0.44	}$	&	$	18.14	^{+	0.93	}_{-	0.92	}$	&	$	0.08	^{+	0.05	}_{-	0.05	}$	&								&	$	0.30	^{+	0.02	}_{-	0.02	}$	&	$	0.30	^{+	0.00	}_{-	0.00	}$	\\
HD108874 b	&	0.95	&	$	395.33	^{+	0.97	}_{-	0.94	}$	&	$	37.76	^{+	1.14	}_{-	1.14	}$	&	$	0.07	^{+	0.03	}_{-	0.03	}$	&	$	0.81	^{+	0.91	}_{-	0.60	}$	&	$	1.31	^{+	0.04	}_{-	0.04	}$	&	$	1.04	^{+	0.00	}_{-	0.00	}$	\\
c	&		&	$	1608.44	^{+	41.21	}_{-	33.38	}$	&	$	18.43	^{+	0.95	}_{-	0.90	}$	&	$	0.25	^{+	0.04	}_{-	0.04	}$	&								&	$	0.99	^{+	0.06	}_{-	0.05	}$	&	$	2.64	^{+	0.04	}_{-	0.04	}$	\\
HD11506 b	&	1.19	&	$	1337.48	^{+	247.15	}_{-	64.76	}$	&	$	65.38	^{+	18.91	}_{-	6.35	}$	&	$	0.30	^{+	0.16	}_{-	0.08	}$	&	$	4.08	^{+	1.63	}_{-	1.35	}$	&	$	3.78	^{+	1.15	}_{-	0.35	}$	&	$	2.52	^{+	0.30	}_{-	0.08	}$	\\
c	&		&	$	170.36	^{+	1.82	}_{-	89.93	}$	&	$	25.54	^{+	6.16	}_{-	7.51	}$	&	$	0.36	^{+	0.13	}_{-	0.17	}$	&								&	$	0.73	^{+	0.16	}_{-	0.32	}$	&	$	0.64	^{+	0.00	}_{-	0.25	}$	\\
HD118203 b	&	1.23	&	$	6.13	^{+	0.00	}_{-	0.00	}$	&	$	217.20	^{+	4.95	}_{-	4.82	}$	&	$	0.31	^{+	0.02	}_{-	0.02	}$	&	$	14.34	^{+	3.29	}_{-	2.96	}$	&	$	2.14	^{+	0.04	}_{-	0.04	}$	&	$	0.07	^{+	0.00	}_{-	0.00	}$	\\
c	&		&	$	6753.13	^{+	5243.74	}_{-	3967.75	}$	&	$	397.12	^{+	386.81	}_{-	294.44	}$	&	$	0.51	^{+	0.24	}_{-	0.33	}$	&								&	$	31.82	^{+	40.68	}_{-	24.02	}$	&	$	7.49	^{+	3.50	}_{-	3.34	}$	\\
HD12661 b	&	1.14	&	$	262.67	^{+	0.12	}_{-	0.09	}$	&	$	74.33	^{+	0.73	}_{-	0.72	}$	&	$	0.36	^{+	0.01	}_{-	0.01	}$	&	$	3.02	^{+	0.67	}_{-	0.64	}$	&	$	2.39	^{+	0.02	}_{-	0.02	}$	&	$	0.84	^{+	0.00	}_{-	0.00	}$	\\
c	&		&	$	1681.47	^{+	29.15	}_{-	26.29	}$	&	$	29.16	^{+	0.79	}_{-	0.83	}$	&	$	0.02	^{+	0.02	}_{-	0.01	}$	&								&	$	1.86	^{+	0.05	}_{-	0.05	}$	&	$	2.89	^{+	0.03	}_{-	0.03	}$	\\
HD128311 b	&	0.83	&	$	924.69	^{+	6.43	}_{-	6.83	}$	&	$	75.43	^{+	3.28	}_{-	3.15	}$	&	$	0.14	^{+	0.10	}_{-	0.10	}$	&	$	16.15	^{+	1.55	}_{-	1.37	}$	&	$	3.14	^{+	0.12	}_{-	0.12	}$	&	$	1.74	^{+	0.01	}_{-	0.01	}$	\\
c	&		&	$	458.32	^{+	3.00	}_{-	2.95	}$	&	$	53.71	^{+	5.30	}_{-	5.55	}$	&	$	0.33	^{+	0.06	}_{-	0.06	}$	&								&	$	1.69	^{+	0.18	}_{-	0.19	}$	&	$	1.09	^{+	0.00	}_{-	0.00	}$	\\
HD131664 b	&	1.10	&	$	1964.51	^{+	35.83	}_{-	45.84	}$	&	$	363.85	^{+	31.98	}_{-	23.96	}$	&	$	0.64	^{+	0.03	}_{-	0.03	}$	&	$	4.67	^{+	0.86	}_{-	0.74	}$	&	$	18.28	^{+	1.10	}_{-	0.86	}$	&	$	3.17	^{+	0.04	}_{-	0.05	}$	\\
c	&		&	$	561.36	^{+	1790.01	}_{-	558.23	}$	&	$	5.40	^{+	10.32	}_{-	4.29	}$	&	$	0.46	^{+	0.39	}_{-	0.31	}$	&								&	$	0.14	^{+	0.53	}_{-	0.13	}$	&	$	1.37	^{+	2.20	}_{-	1.33	}$	\\
HD132406 b	&	1.09	&	$	540.91	^{+	892.18	}_{-	326.23	}$	&	$	49.33	^{+	68.99	}_{-	27.88	}$	&	$	0.39	^{+	0.31	}_{-	0.24	}$	&	$	6.90	^{+	9.73	}_{-	5.67	}$	&	$	1.74	^{+	4.10	}_{-	1.19	}$	&	$	1.34	^{+	1.22	}_{-	0.62	}$	\\
c	&		&	$	1156.94	^{+	381.48	}_{-	815.09	}$	&	$	89.58	^{+	49.38	}_{-	63.98	}$	&	$	0.39	^{+	0.29	}_{-	0.25	}$	&								&	$	4.75	^{+	2.15	}_{-	3.97	}$	&	$	2.22	^{+	0.46	}_{-	1.23	}$	\\
HD142 b	&	1.23	&	$	349.57	^{+	3.40	}_{-	3.58	}$	&	$	29.66	^{+	4.66	}_{-	4.22	}$	&	$	0.20	^{+	0.11	}_{-	0.11	}$	&	$	8.74	^{+	1.54	}_{-	1.34	}$	&	$	1.15	^{+	0.16	}_{-	0.15	}$	&	$	1.04	^{+	0.01	}_{-	0.01	}$	\\
c	&		&	$	9822.95	^{+	3348.44	}_{-	3208.17	}$	&	$	53.42	^{+	53.99	}_{-	17.59	}$	&	$	0.18	^{+	0.20	}_{-	0.13	}$	&								&	$	6.27	^{+	7.33	}_{-	2.53	}$	&	$	9.62	^{+	2.08	}_{-	2.23	}$	\\
HD142022 b	&	0.90	&	$	244.04	^{+	6905.04	}_{-	242.28	}$	&	$	3.41	^{+	8.61	}_{-	2.65	}$	&	$	0.52	^{+	0.35	}_{-	0.36	}$	&	$	2.84	^{+	1.76	}_{-	1.47	}$	&	$	0.05	^{+	0.52	}_{-	0.05	}$	&	$	0.74	^{+	6.27	}_{-	0.71	}$	\\
c	&		&	$	1877.47	^{+	35.74	}_{-	20.20	}$	&	$	127.61	^{+	84.60	}_{-	35.05	}$	&	$	0.62	^{+	0.11	}_{-	0.10	}$	&								&	$	5.70	^{+	2.52	}_{-	1.21	}$	&	$	2.88	^{+	0.04	}_{-	0.02	}$	\\
HD149143 b	&	1.20	&	$	4.07	^{+	0.00	}_{-	0.00	}$	&	$	149.85	^{+	1.79	}_{-	1.66	}$	&	$	0.01	^{+	0.01	}_{-	0.00	}$	&	$	1.33	^{+	1.83	}_{-	1.02	}$	&	$	1.33	^{+	0.02	}_{-	0.01	}$	&	$	0.05	^{+	0.00	}_{-	0.00	}$	\\
c	&		&	$	248.47	^{+	3300.06	}_{-	245.40	}$	&	$	3.73	^{+	13.28	}_{-	3.00	}$	&	$	0.55	^{+	0.33	}_{-	0.37	}$	&								&	$	0.07	^{+	0.55	}_{-	0.06	}$	&	$	0.82	^{+	4.02	}_{-	0.78	}$	\\
HD154345 b	&	0.89	&	$	3216.13	^{+	150.82	}_{-	3170.93	}$	&	$	13.11	^{+	1.63	}_{-	10.59	}$	&	$	0.10	^{+	0.47	}_{-	0.08	}$	&	$	2.38	^{+	0.47	}_{-	0.41	}$	&	$	0.87	^{+	0.12	}_{-	0.84	}$	&	$	4.10	^{+	0.13	}_{-	3.86	}$	\\
c	&		&	$	2695.22	^{+	697.19	}_{-	2649.86	}$	&	$	6.21	^{+	8.30	}_{-	4.22	}$	&	$	0.15	^{+	0.55	}_{-	0.11	}$	&								&	$	0.15	^{+	0.83	}_{-	0.13	}$	&	$	3.65	^{+	0.60	}_{-	3.41	}$	\\
HD155358 b	&	0.87	&	$	193.24	^{+	0.98	}_{-	0.97	}$	&	$	32.88	^{+	1.51	}_{-	1.50	}$	&	$	0.13	^{+	0.04	}_{-	0.04	}$	&	$	5.59	^{+	0.72	}_{-	0.65	}$	&	$	0.84	^{+	0.04	}_{-	0.04	}$	&	$	0.62	^{+	0.00	}_{-	0.00	}$	\\
c	&		&	$	314.82	^{+	4.92	}_{-	3.95	}$	&	$	33.63	^{+	41.74	}_{-	12.77	}$	&	$	0.83	^{+	0.12	}_{-	0.17	}$	&								&	$	0.57	^{+	0.23	}_{-	0.11	}$	&	$	0.86	^{+	0.01	}_{-	0.01	}$	\\
HD162020 b	&	0.80	&	$	8.43	^{+	0.00	}_{-	0.00	}$	&	$	1808.83	^{+	5.13	}_{-	5.09	}$	&	$	0.28	^{+	0.00	}_{-	0.00	}$	&	$	10.45	^{+	2.84	}_{-	2.75	}$	&	$	15.00	^{+	0.04	}_{-	0.04	}$	&	$	0.08	^{+	0.00	}_{-	0.00	}$	\\
c	&		&	$	180.71	^{+	3271.59	}_{-	175.94	}$	&	$	6.73	^{+	21.75	}_{-	5.73	}$	&	$	0.57	^{+	0.33	}_{-	0.39	}$	&								&	$	0.10	^{+	0.41	}_{-	0.09	}$	&	$	0.58	^{+	3.57	}_{-	0.53	}$	\\
HD168443 b	&	1.01	&	$	58.11	^{+	0.00	}_{-	0.00	}$	&	$	529.04	^{+	82.29	}_{-	52.08	}$	&	$	0.57	^{+	0.04	}_{-	0.04	}$	&	$	7.80	^{+	3.15	}_{-	2.74	}$	&	$	8.37	^{+	0.93	}_{-	0.61	}$	&	$	0.29	^{+	0.00	}_{-	0.00	}$	\\
c	&		&	$	1755.58	^{+	6.65	}_{-	7.25	}$	&	$	302.53	^{+	3.86	}_{-	3.76	}$	&	$	0.23	^{+	0.02	}_{-	0.02	}$	&								&	$	17.61	^{+	0.21	}_{-	0.21	}$	&	$	2.86	^{+	0.01	}_{-	0.01	}$	\\

\multicolumn{8}{r}{\tiny{\textit{continued on next page...}}}
		\end{tabular}
	\end{center}
\end{table*}

\begin{table*}
\scriptsize
\smallskip
	\begin{center}
		\begin{tabular}{rcllllll}
			\hline\noalign{\smallskip}
\multicolumn{1}{c}{\textbf{planet}} & 
\multicolumn{1}{c}{\textbf{$m_{*} (M_{sol})$}} & 
\multicolumn{1}{c}{\textbf{$T (days)$}} & 
\multicolumn{1}{c}{\textbf{$K (ms^{-1})$}} & 
\multicolumn{1}{c}{\textbf{$e$}} & 
\multicolumn{1}{c}{\textbf{$s$}} & 
\multicolumn{1}{c}{\textbf{$m_p (M_{Jup})$}} & 
\multicolumn{1}{c}{\textbf{$a (AU)$}} \\
			\hline\hline

HD169830 b	&	1.41	&	$	225.61	^{+	0.29	}_{-	0.31	}$	&	$	83.01	^{+	3.29	}_{-	3.25	}$	&	$	0.37	^{+	0.03	}_{-	0.03	}$	&	$	1.46	^{+	2.36	}_{-	1.13	}$	&	$	2.91	^{+	0.11	}_{-	0.11	}$	&	$	0.81	^{+	0.00	}_{-	0.00	}$	\\
c	&		&	$	139.53	^{+	4188.83	}_{-	137.98	}$	&	$	3.39	^{+	9.48	}_{-	2.78	}$	&	$	0.55	^{+	0.33	}_{-	0.37	}$	&								&	$	0.05	^{+	0.43	}_{-	0.05	}$	&	$	0.59	^{+	5.24	}_{-	0.56	}$	\\
HD171028 b	&	0.99	&	$	498.38	^{+	29.74	}_{-	41.02	}$	&	$	72.24	^{+	80.06	}_{-	23.28	}$	&	$	0.61	^{+	0.15	}_{-	0.11	}$	&	$	1.77	^{+	0.66	}_{-	0.47	}$	&	$	2.27	^{+	1.75	}_{-	0.69	}$	&	$	1.23	^{+	0.05	}_{-	0.07	}$	\\
c	&		&	$	267.54	^{+	7.12	}_{-	4.33	}$	&	$	47.44	^{+	5.43	}_{-	2.78	}$	&	$	0.57	^{+	0.13	}_{-	0.08	}$	&								&	$	1.23	^{+	0.18	}_{-	0.19	}$	&	$	0.81	^{+	0.01	}_{-	0.01	}$	\\
HD183263 b	&	1.12	&	$	625.64	^{+	0.96	}_{-	0.59	}$	&	$	86.82	^{+	1.28	}_{-	1.28	}$	&	$	0.38	^{+	0.01	}_{-	0.01	}$	&	$	3.54	^{+	0.57	}_{-	0.48	}$	&	$	3.65	^{+	0.05	}_{-	0.05	}$	&	$	1.49	^{+	0.00	}_{-	0.00	}$	\\
c	&		&	$	4630.04	^{+	1456.78	}_{-	1072.15	}$	&	$	74.41	^{+	42.59	}_{-	19.98	}$	&	$	0.07	^{+	0.09	}_{-	0.05	}$	&								&	$	6.57	^{+	4.72	}_{-	2.18	}$	&	$	5.65	^{+	1.13	}_{-	0.91	}$	\\
HD185269 b	&	1.30	&	$	6.84	^{+	0.00	}_{-	0.00	}$	&	$	89.81	^{+	4.33	}_{-	4.08	}$	&	$	0.28	^{+	0.03	}_{-	0.04	}$	&	$	7.14	^{+	2.10	}_{-	2.25	}$	&	$	0.96	^{+	0.04	}_{-	0.04	}$	&	$	0.08	^{+	0.00	}_{-	0.00	}$	\\
c	&		&	$	32.98	^{+	2639.27	}_{-	31.94	}$	&	$	4.17	^{+	9.25	}_{-	3.46	}$	&	$	0.55	^{+	0.32	}_{-	0.36	}$	&								&	$	0.05	^{+	0.19	}_{-	0.04	}$	&	$	0.22	^{+	3.89	}_{-	0.20	}$	\\
HD187123 b	&	1.04	&	$	3.10	^{+	0.00	}_{-	0.00	}$	&	$	69.57	^{+	0.49	}_{-	0.49	}$	&	$	0.01	^{+	0.01	}_{-	0.01	}$	&	$	0.49	^{+	0.53	}_{-	0.35	}$	&	$	0.51	^{+	0.00	}_{-	0.00	}$	&	$	0.04	^{+	0.00	}_{-	0.00	}$	\\
c	&		&	$	5502.18	^{+	2809.80	}_{-	1322.20	}$	&	$	28.50	^{+	5.09	}_{-	3.06	}$	&	$	0.32	^{+	0.12	}_{-	0.08	}$	&								&	$	2.42	^{+	0.67	}_{-	0.40	}$	&	$	6.18	^{+	1.96	}_{-	1.03	}$	\\
HD189733 b	&	0.81	&	$	2.22	^{+	0.00	}_{-	0.00	}$	&	$	198.36	^{+	4.00	}_{-	4.07	}$	&	$	0.02	^{+	0.02	}_{-	0.02	}$	&	$	11.26	^{+	0.98	}_{-	0.87	}$	&	$	1.11	^{+	0.02	}_{-	0.02	}$	&	$	0.03	^{+	0.00	}_{-	0.00	}$	\\
c	&		&	$	3.92	^{+	0.00	}_{-	0.00	}$	&	$	25.58	^{+	4.20	}_{-	3.99	}$	&	$	0.96	^{+	0.01	}_{-	0.01	}$	&								&	$	0.05	^{+	0.01	}_{-	0.01	}$	&	$	0.05	^{+	0.00	}_{-	0.00	}$	\\
HD190228 b	&	1.82	&	$	149.82	^{+	3434.51	}_{-	147.94	}$	&	$	2.23	^{+	5.57	}_{-	1.79	}$	&	$	0.55	^{+	0.33	}_{-	0.37	}$	&	$	1.20	^{+	1.82	}_{-	0.91	}$	&	$	0.05	^{+	0.30	}_{-	0.04	}$	&	$	0.67	^{+	4.92	}_{-	0.64	}$	\\
c	&		&	$	1140.04	^{+	16.47	}_{-	15.62	}$	&	$	92.35	^{+	5.40	}_{-	3.72	}$	&	$	0.52	^{+	0.05	}_{-	0.04	}$	&								&	$	6.07	^{+	4.92	}_{-	0.64	}$	&	$	2.61	^{+	0.03	}_{-	0.02	}$	\\
HD190360 b	&	0.98	&	$	2925.88	^{+	36.43	}_{-	41.38	}$	&	$	18.84	^{+	2.95	}_{-	3.16	}$	&	$	0.34	^{+	0.13	}_{-	0.12	}$	&	$	5.31	^{+	1.64	}_{-	2.14	}$	&	$	1.23	^{+	0.16	}_{-	0.20	}$	&	$	3.98	^{+	0.03	}_{-	0.04	}$	\\
c	&		&	$	184.70	^{+	3848.83	}_{-	181.67	}$	&	$	5.10	^{+	10.61	}_{-	4.14	}$	&	$	0.55	^{+	0.32	}_{-	0.35	}$	&								&	$	0.09	^{+	0.49	}_{-	0.08	}$	&	$	0.63	^{+	4.30	}_{-	0.59	}$	\\
HD190647 b	&	1.10	&	$	1036.28	^{+	8.27	}_{-	14.95	}$	&	$	36.17	^{+	1.89	}_{-	1.75	}$	&	$	0.17	^{+	0.03	}_{-	0.04	}$	&	$	0.59	^{+	0.68	}_{-	0.43	}$	&	$	1.89	^{+	0.10	}_{-	0.10	}$	&	$	2.07	^{+	0.01	}_{-	0.02	}$	\\
c	&		&	$	422.54	^{+	3134.94	}_{-	411.93	}$	&	$	2.88	^{+	13.42	}_{-	2.23	}$	&	$	0.55	^{+	0.30	}_{-	0.36	}$	&								&	$	0.08	^{+	0.77	}_{-	0.07	}$	&	$	1.14	^{+	3.57	}_{-	1.04	}$	\\
HD195019 b	&	1.02	&	$	45.87	^{+	11.86	}_{-	0.76	}$	&	$	176.87	^{+	70.77	}_{-	43.26	}$	&	$	0.79	^{+	0.11	}_{-	0.17	}$	&	$	149.90	^{+	9.71	}_{-	8.67	}$	&	$	1.96	^{+	0.64	}_{-	0.42	}$	&	$	0.25	^{+	0.04	}_{-	0.00	}$	\\
c	&		&	$	51.93	^{+	0.08	}_{-	0.10	}$	&	$	159.88	^{+	33.08	}_{-	34.85	}$	&	$	0.74	^{+	0.07	}_{-	0.08	}$	&								&	$	1.95	^{+	0.52	}_{-	0.45	}$	&	$	0.27	^{+	0.00	}_{-	0.00	}$	\\
HD202206 b	&	1.07	&	$	255.87	^{+	0.08	}_{-	0.08	}$	&	$	706.93	^{+	119.27	}_{-	65.64	}$	&	$	0.37	^{+	0.02	}_{-	0.02	}$	&	$	23.09	^{+	2.18	}_{-	1.91	}$	&	$	21.46	^{+	3.75	}_{-	2.07	}$	&	$	0.81	^{+	0.00	}_{-	0.00	}$	\\
c	&		&	$	258.20	^{+	1.14	}_{-	1.02	}$	&	$	291.17	^{+	164.81	}_{-	66.27	}$	&	$	0.26	^{+	0.06	}_{-	0.06	}$	&								&	$	9.23	^{+	4.97	}_{-	2.03	}$	&	$	0.81	^{+	0.00	}_{-	0.00	}$	\\
HD20868 b	&	0.78	&	$	380.86	^{+	0.08	}_{-	0.08	}$	&	$	100.22	^{+	0.45	}_{-	0.46	}$	&	$	0.76	^{+	0.00	}_{-	0.00	}$	&	$	0.74	^{+	0.43	}_{-	0.48	}$	&	$	1.99	^{+	0.01	}_{-	0.01	}$	&	$	0.95	^{+	0.00	}_{-	0.00	}$	\\
c	&		&	$	111.05	^{+	1942.40	}_{-	110.42	}$	&	$	1.59	^{+	2.19	}_{-	1.07	}$	&	$	0.60	^{+	0.30	}_{-	0.38	}$	&								&	$	0.02	^{+	0.07	}_{-	0.01	}$	&	$	0.42	^{+	2.49	}_{-	0.40	}$	\\
HD209458 b	&	1.13	&	$	3.52	^{+	0.00	}_{-	0.00	}$	&	$	84.27	^{+	0.85	}_{-	0.84	}$	&	$	0.01	^{+	0.01	}_{-	0.01	}$	&	$	2.91	^{+	0.79	}_{-	0.88	}$	&	$	0.68	^{+	0.01	}_{-	0.01	}$	&	$	0.05	^{+	0.00	}_{-	0.00	}$	\\
c	&		&	$	264.63	^{+	1050.15	}_{-	259.91	}$	&	$	2.85	^{+	3.04	}_{-	2.03	}$	&	$	0.54	^{+	0.34	}_{-	0.37	}$	&								&	$	0.07	^{+	0.11	}_{-	0.06	}$	&	$	0.84	^{+	1.61	}_{-	0.78	}$	\\
HD212301 b	&	1.05	&	$	2356.61	^{+	6274.66	}_{-	2104.70	}$	&	$	26.81	^{+	76.37	}_{-	17.28	}$	&	$	0.52	^{+	0.35	}_{-	0.36	}$	&	$	5.42	^{+	1.58	}_{-	1.26	}$	&	$	1.22	^{+	4.93	}_{-	0.89	}$	&	$	3.52	^{+	4.85	}_{-	2.73	}$	\\
c	&		&	$	2.25	^{+	0.00	}_{-	0.00	}$	&	$	56.85	^{+	1.92	}_{-	1.93	}$	&	$	0.09	^{+	0.04	}_{-	0.04	}$	&								&	$	0.38	^{+	0.01	}_{-	0.01	}$	&	$	0.03	^{+	0.00	}_{-	0.00	}$	\\
HD217107 b	&	1.11	&	$	7.13	^{+	0.00	}_{-	0.00	}$	&	$	138.34	^{+	1.11	}_{-	1.11	}$	&	$	0.13	^{+	0.01	}_{-	0.01	}$	&	$	10.19	^{+	0.64	}_{-	0.59	}$	&	$	1.39	^{+	0.01	}_{-	0.01	}$	&	$	0.08	^{+	0.00	}_{-	0.00	}$	\\
c	&		&	$	4106.23	^{+	248.05	}_{-	113.88	}$	&	$	39.48	^{+	13.92	}_{-	3.49	}$	&	$	0.56	^{+	0.04	}_{-	0.03	}$	&								&	$	2.77	^{+	1.00	}_{-	0.27	}$	&	$	5.20	^{+	0.21	}_{-	0.10	}$	\\
HD219828 b	&	1.24	&	$	3.83	^{+	0.00	}_{-	0.00	}$	&	$	7.22	^{+	0.59	}_{-	0.62	}$	&	$	0.09	^{+	0.09	}_{-	0.06	}$	&	$	1.78	^{+	0.55	}_{-	0.39	}$	&	$	0.06	^{+	0.01	}_{-	0.01	}$	&	$	0.05	^{+	0.00	}_{-	0.00	}$	\\
c	&		&	$	956.74	^{+	1301.58	}_{-	366.96	}$	&	$	78.15	^{+	131.30	}_{-	45.22	}$	&	$	0.53	^{+	0.17	}_{-	0.20	}$	&								&	$	3.55	^{+	10.00	}_{-	2.29	}$	&	$	2.04	^{+	1.58	}_{-	0.56	}$	\\
HD221287 b	&	1.30	&	$	0.60	^{+	525.67	}_{-	0.00	}$	&	$	9.66	^{+	6.31	}_{-	8.16	}$	&	$	0.41	^{+	0.39	}_{-	0.28	}$	&	$	8.00	^{+	2.88	}_{-	2.87	}$	&	$	0.05	^{+	0.11	}_{-	0.04	}$	&	$	0.02	^{+	1.38	}_{-	0.00	}$	\\
c	&		&	$	455.12	^{+	6.45	}_{-	4.50	}$	&	$	71.62	^{+	16.56	}_{-	7.17	}$	&	$	0.13	^{+	0.12	}_{-	0.08	}$	&								&	$	3.21	^{+	0.65	}_{-	0.32	}$	&	$	1.26	^{+	0.01	}_{-	0.01	}$	\\
HD224693 b	&	1.30	&	$	26.75	^{+	0.03	}_{-	0.03	}$	&	$	38.90	^{+	1.77	}_{-	1.72	}$	&	$	0.04	^{+	0.04	}_{-	0.03	}$	&	$	1.81	^{+	1.07	}_{-	1.11	}$	&	$	0.68	^{+	0.03	}_{-	0.03	}$	&	$	0.19	^{+	0.00	}_{-	0.00	}$	\\
c	&		&	$	7882.48	^{+	4298.55	}_{-	3049.23	}$	&	$	1849.37	^{+	94.81	}_{-	501.16	}$	&	$	0.68	^{+	0.12	}_{-	0.11	}$	&								&	$	153.48	^{+	42.92	}_{-	58.00	}$	&	$	8.46	^{+	2.85	}_{-	2.35	}$	\\
HD23127 b	&	1.13	&	$	1124.66	^{+	841.10	}_{-	1121.44	}$	&	$	8.38	^{+	19.55	}_{-	7.33	}$	&	$	0.46	^{+	0.34	}_{-	0.24	}$	&	$	10.66	^{+	2.22	}_{-	1.81	}$	&	$	0.15	^{+	1.28	}_{-	0.14	}$	&	$	2.20	^{+	0.99	}_{-	2.16	}$	\\
c	&		&	$	1219.14	^{+	36.54	}_{-	1099.09	}$	&	$	25.84	^{+	4.15	}_{-	22.15	}$	&	$	0.44	^{+	0.17	}_{-	0.14	}$	&								&	$	1.31	^{+	0.23	}_{-	1.26	}$	&	$	2.33	^{+	0.05	}_{-	1.83	}$	\\
HD2638 b	&	0.93	&	$	3.44	^{+	0.00	}_{-	0.00	}$	&	$	67.54	^{+	0.87	}_{-	0.84	}$	&	$	0.01	^{+	0.01	}_{-	0.01	}$	&	$	2.52	^{+	0.98	}_{-	0.94	}$	&	$	0.48	^{+	0.01	}_{-	0.01	}$	&	$	0.04	^{+	0.00	}_{-	0.00	}$	\\
c	&		&	$	26.04	^{+	1978.12	}_{-	25.00	}$	&	$	4.45	^{+	9.20	}_{-	2.36	}$	&	$	0.41	^{+	0.39	}_{-	0.27	}$	&								&	$	0.05	^{+	0.34	}_{-	0.03	}$	&	$	0.17	^{+	2.87	}_{-	0.15	}$	\\
HD27442 b	&	1.20	&	$	410.58	^{+	866.97	}_{-	407.42	}$	&	$	6.33	^{+	26.90	}_{-	5.26	}$	&	$	0.30	^{+	0.52	}_{-	0.26	}$	&	$	2.57	^{+	1.53	}_{-	1.47	}$	&	$	0.14	^{+	1.23	}_{-	0.12	}$	&	$	1.15	^{+	1.30	}_{-	1.10	}$	\\
c	&		&	$	417.14	^{+	945.92	}_{-	9.45	}$	&	$	31.26	^{+	2.69	}_{-	28.78	}$	&	$	0.10	^{+	0.49	}_{-	0.06	}$	&								&	$	1.29	^{+	0.11	}_{-	1.22	}$	&	$	1.16	^{+	1.40	}_{-	0.02	}$	\\
HD27894 b	&	0.75	&	$	23.31	^{+	9.65	}_{-	0.23	}$	&	$	10.37	^{+	2.85	}_{-	1.71	}$	&	$	0.28	^{+	0.44	}_{-	0.22	}$	&	$	0.71	^{+	1.02	}_{-	0.53	}$	&	$	0.12	^{+	0.02	}_{-	0.02	}$	&	$	0.15	^{+	0.04	}_{-	0.00	}$	\\
c	&		&	$	17.98	^{+	0.03	}_{-	0.01	}$	&	$	56.34	^{+	2.56	}_{-	1.29	}$	&	$	0.05	^{+	0.03	}_{-	0.03	}$	&								&	$	0.60	^{+	0.03	}_{-	0.01	}$	&	$	0.12	^{+	0.00	}_{-	0.00	}$	\\
HD28185 b	&	0.99	&	$	381.37	^{+	1.22	}_{-	2.47	}$	&	$	161.15	^{+	8.85	}_{-	10.53	}$	&	$	0.04	^{+	0.03	}_{-	0.03	}$	&	$	4.00	^{+	2.23	}_{-	2.54	}$	&	$	5.71	^{+	0.32	}_{-	0.38	}$	&	$	1.03	^{+	0.00	}_{-	0.00	}$	\\
c	&		&	$	508.09	^{+	1160.41	}_{-	198.94	}$	&	$	22.84	^{+	26.97	}_{-	11.21	}$	&	$	0.37	^{+	0.36	}_{-	0.21	}$	&								&	$	0.86	^{+	1.01	}_{-	0.49	}$	&	$	1.24	^{+	1.50	}_{-	0.35	}$	\\
HD285968 b	&	0.49	&	$	10.25	^{+	1457.68	}_{-	8.28	}$	&	$	3.84	^{+	7.55	}_{-	3.20	}$	&	$	0.43	^{+	0.41	}_{-	0.32	}$	&	$	1.91	^{+	1.90	}_{-	1.45	}$	&	$	0.03	^{+	0.05	}_{-	0.02	}$	&	$	0.07	^{+	1.92	}_{-	0.05	}$	\\
c	&		&	$	10.23	^{+	0.01	}_{-	1.62	}$	&	$	10.62	^{+	2.09	}_{-	5.78	}$	&	$	0.19	^{+	0.29	}_{-	0.14	}$	&								&	$	0.07	^{+	0.01	}_{-	0.04	}$	&	$	0.07	^{+	0.00	}_{-	0.01	}$	\\
HD330075 b	&	0.70	&	$	3.39	^{+	0.00	}_{-	0.00	}$	&	$	106.96	^{+	1.00	}_{-	1.00	}$	&	$	0.01	^{+	0.01	}_{-	0.00	}$	&	$	1.65	^{+	0.78	}_{-	0.68	}$	&	$	0.62	^{+	0.01	}_{-	0.01	}$	&	$	0.04	^{+	0.00	}_{-	0.00	}$	\\
c	&		&	$	655.66	^{+	6970.11	}_{-	589.87	}$	&	$	10.51	^{+	98.32	}_{-	8.53	}$	&	$	0.42	^{+	0.40	}_{-	0.33	}$	&								&	$	0.27	^{+	5.27	}_{-	0.24	}$	&	$	1.31	^{+	5.42	}_{-	1.03	}$	\\
HD33636 b	&	1.02	&	$	2127.71	^{+	11.43	}_{-	10.98	}$	&	$	773.84	^{+	123.39	}_{-	762.03	}$	&	$	0.89	^{+	0.03	}_{-	0.22	}$	&	$	0.64	^{+	0.71	}_{-	0.47	}$	&	$	22.86	^{+	3.33	}_{-	22.32	}$	&	$	3.26	^{+	0.01	}_{-	0.01	}$	\\
c	&		&	$	7841.32	^{+	3695.12	}_{-	3270.74	}$	&	$	1923.73	^{+	58.92	}_{-	174.57	}$	&	$	0.61	^{+	0.03	}_{-	0.01	}$	&								&	$	149.03	^{+	17.99	}_{-	27.71	}$	&	$	7.78	^{+	2.28	}_{-	2.35	}$	\\
HD3651 b	&	0.88	&	$	294.67	^{+	61.62	}_{-	195.79	}$	&	$	3.49	^{+	1.88	}_{-	1.60	}$	&	$	0.32	^{+	0.41	}_{-	0.23	}$	&	$	4.41	^{+	0.52	}_{-	0.46	}$	&	$	0.10	^{+	0.05	}_{-	0.06	}$	&	$	0.83	^{+	0.11	}_{-	0.43	}$	\\
c	&		&	$	62.25	^{+	0.03	}_{-	0.03	}$	&	$	16.14	^{+	1.46	}_{-	1.48	}$	&	$	0.60	^{+	0.05	}_{-	0.06	}$	&								&	$	0.23	^{+	0.01	}_{-	0.02	}$	&	$	0.29	^{+	0.00	}_{-	0.00	}$	\\
HD38529 b	&	1.48	&	$	14.31	^{+	0.00	}_{-	0.00	}$	&	$	54.97	^{+	1.70	}_{-	1.66	}$	&	$	0.17	^{+	0.03	}_{-	0.03	}$	&	$	13.11	^{+	0.93	}_{-	0.86	}$	&	$	0.84	^{+	0.03	}_{-	0.02	}$	&	$	0.13	^{+	0.00	}_{-	0.00	}$	\\
c	&		&	$	2148.41	^{+	5.95	}_{-	7.87	}$	&	$	170.83	^{+	1.92	}_{-	1.86	}$	&	$	0.34	^{+	0.01	}_{-	0.01	}$	&								&	$	13.23	^{+	0.14	}_{-	0.14	}$	&	$	3.71	^{+	0.01	}_{-	0.01	}$	\\
HD4203 b	&	1.13	&	$	438.04	^{+	7.20	}_{-	4.69	}$	&	$	56.24	^{+	29.91	}_{-	9.58	}$	&	$	0.69	^{+	0.13	}_{-	0.08	}$	&	$	1.39	^{+	1.65	}_{-	1.04	}$	&	$	1.72	^{+	0.53	}_{-	0.38	}$	&	$	1.18	^{+	0.01	}_{-	0.01	}$	\\
c	&		&	$	391.98	^{+	202.96	}_{-	200.21	}$	&	$	10.70	^{+	11.08	}_{-	3.76	}$	&	$	0.32	^{+	0.33	}_{-	0.23	}$	&								&	$	0.37	^{+	0.42	}_{-	0.20	}$	&	$	1.09	^{+	0.35	}_{-	0.41	}$	\\
HD4208 b	&	0.88	&	$	129.04	^{+	708.37	}_{-	121.12	}$	&	$	2.62	^{+	3.03	}_{-	1.80	}$	&	$	0.53	^{+	0.34	}_{-	0.39	}$	&	$	0.80	^{+	0.92	}_{-	0.58	}$	&	$	0.05	^{+	0.07	}_{-	0.04	}$	&	$	0.48	^{+	1.19	}_{-	0.40	}$	\\
c	&		&	$	829.27	^{+	9.57	}_{-	9.59	}$	&	$	18.79	^{+	0.82	}_{-	0.95	}$	&	$	0.06	^{+	0.05	}_{-	0.04	}$	&								&	$	0.80	^{+	0.04	}_{-	0.04	}$	&	$	1.66	^{+	0.01	}_{-	0.01	}$	\\
HD43691 b	&	1.38	&	$	516.72	^{+	5417.79	}_{-	471.62	}$	&	$	19.26	^{+	70.84	}_{-	11.52	}$	&	$	0.45	^{+	0.36	}_{-	0.31	}$	&	$	11.09	^{+	3.45	}_{-	2.98	}$	&	$	0.57	^{+	5.63	}_{-	0.39	}$	&	$	1.40	^{+	5.74	}_{-	1.13	}$	\\
c	&		&	$	36.96	^{+	0.04	}_{-	0.05	}$	&	$	125.00	^{+	4.49	}_{-	4.59	}$	&	$	0.12	^{+	0.04	}_{-	0.04	}$	&								&	$	2.52	^{+	0.10	}_{-	0.10	}$	&	$	0.24	^{+	0.00	}_{-	0.00	}$	\\
HD43848 b	&	0.93	&	$	2261.82	^{+	513.95	}_{-	2252.19	}$	&	$	44.31	^{+	612.06	}_{-	41.42	}$	&	$	0.67	^{+	0.16	}_{-	0.37	}$	&	$	4.49	^{+	4.32	}_{-	3.44	}$	&	$	1.41	^{+	26.25	}_{-	1.38	}$	&	$	3.29	^{+	0.48	}_{-	3.21	}$	\\
c	&		&	$	2305.82	^{+	158.97	}_{-	2261.64	}$	&	$	476.62	^{+	555.76	}_{-	471.61	}$	&	$	0.73	^{+	0.12	}_{-	0.35	}$	&								&	$	22.16	^{+	14.50	}_{-	22.11	}$	&	$	3.33	^{+	0.15	}_{-	3.10	}$	\\
HD46375 b	&	0.93	&	$	3.02	^{+	0.00	}_{-	0.00	}$	&	$	33.79	^{+	0.79	}_{-	0.81	}$	&	$	0.05	^{+	0.03	}_{-	0.03	}$	&	$	2.92	^{+	0.70	}_{-	0.71	}$	&	$	0.23	^{+	0.01	}_{-	0.01	}$	&	$	0.04	^{+	0.00	}_{-	0.00	}$	\\
c	&		&	$	30.41	^{+	931.01	}_{-	29.23	}$	&	$	2.93	^{+	4.85	}_{-	2.28	}$	&	$	0.65	^{+	0.25	}_{-	0.42	}$	&								&	$	0.03	^{+	0.08	}_{-	0.02	}$	&	$	0.19	^{+	1.67	}_{-	0.16	}$	\\

\multicolumn{8}{r}{\tiny{\textit{...continued on next page...}}}
		\end{tabular}
	\end{center}
\end{table*}

\begin{table*}
\scriptsize
\smallskip
	\begin{center}
		\begin{tabular}{rcllllll}
			\hline\noalign{\smallskip}
\multicolumn{1}{c}{\textbf{planet}} & 
\multicolumn{1}{c}{\textbf{$m_{*} (M_{sol})$}} & 
\multicolumn{1}{c}{\textbf{$T (days)$}} & 
\multicolumn{1}{c}{\textbf{$K (ms^{-1})$}} & 
\multicolumn{1}{c}{\textbf{$e$}} & 
\multicolumn{1}{c}{\textbf{$s$}} & 
\multicolumn{1}{c}{\textbf{$m_p (M_{Jup})$}} & 
\multicolumn{1}{c}{\textbf{$a (AU)$}} \\
			\hline\hline

HD47536 b	&	0.94	&	$	698.09	^{+	24.29	}_{-	686.96	}$	&	$	105.12	^{+	22.10	}_{-	98.90	}$	&	$	0.18	^{+	0.40	}_{-	0.14	}$	&	$	4.58	^{+	9.07	}_{-	3.83	}$	&	$	4.33	^{+	0.95	}_{-	4.28	}$	&	$	1.51	^{+	0.03	}_{-	1.41	}$	\\
c	&		&	$	713.17	^{+	4061.39	}_{-	662.35	}$	&	$	87.58	^{+	41.26	}_{-	81.24	}$	&	$	0.30	^{+	0.47	}_{-	0.20	}$	&								&	$	3.76	^{+	1.67	}_{-	3.69	}$	&	$	1.53	^{+	3.91	}_{-	1.27	}$	\\
HD49674 b	&	1.01	&	$	275.09	^{+	2808.10	}_{-	248.28	}$	&	$	4.23	^{+	4.57	}_{-	2.19	}$	&	$	0.64	^{+	0.30	}_{-	0.42	}$	&	$	2.56	^{+	1.04	}_{-	1.11	}$	&	$	0.10	^{+	0.10	}_{-	0.07	}$	&	$	0.83	^{+	3.33	}_{-	0.65	}$	\\
c	&		&	$	4.95	^{+	0.00	}_{-	0.00	}$	&	$	11.88	^{+	1.04	}_{-	1.00	}$	&	$	0.09	^{+	0.08	}_{-	0.06	}$	&								&	$	0.10	^{+	0.01	}_{-	0.01	}$	&	$	0.06	^{+	0.00	}_{-	0.00	}$	\\
HD50499 b	&	1.28	&	$	2450.67	^{+	54.08	}_{-	45.96	}$	&	$	21.82	^{+	1.91	}_{-	1.90	}$	&	$	0.25	^{+	0.07	}_{-	0.07	}$	&	$	2.95	^{+	1.02	}_{-	1.08	}$	&	$	1.65	^{+	0.15	}_{-	0.15	}$	&	$	3.86	^{+	0.06	}_{-	0.05	}$	\\
c	&		&	$	9734.20	^{+	3449.64	}_{-	3466.65	}$	&	$	57.00	^{+	109.73	}_{-	32.17	}$	&	$	0.54	^{+	0.30	}_{-	0.34	}$	&								&	$	5.58	^{+	6.81	}_{-	2.96	}$	&	$	9.69	^{+	2.17	}_{-	2.46	}$	\\
HD5319 b	&	1.60	&	$	684.85	^{+	13.55	}_{-	20.85	}$	&	$	37.35	^{+	9.92	}_{-	4.77	}$	&	$	0.11	^{+	0.12	}_{-	0.07	}$	&	$	5.43	^{+	1.03	}_{-	0.82	}$	&	$	2.20	^{+	0.54	}_{-	0.28	}$	&	$	1.78	^{+	0.02	}_{-	0.04	}$	\\
c	&		&	$	1541.64	^{+	1673.87	}_{-	412.57	}$	&	$	15.24	^{+	7.88	}_{-	3.29	}$	&	$	0.41	^{+	0.29	}_{-	0.28	}$	&								&	$	1.08	^{+	0.56	}_{-	0.23	}$	&	$	3.05	^{+	1.93	}_{-	0.57	}$	\\
HD63454 b	&	0.80	&	$	33.11	^{+	3825.01	}_{-	32.19	}$	&	$	3.07	^{+	7.74	}_{-	2.50	}$	&	$	0.57	^{+	0.33	}_{-	0.39	}$	&	$	5.54	^{+	1.33	}_{-	1.18	}$	&	$	0.03	^{+	0.15	}_{-	0.02	}$	&	$	0.19	^{+	4.28	}_{-	0.17	}$	\\
c	&		&	$	2.82	^{+	0.00	}_{-	0.00	}$	&	$	63.26	^{+	1.78	}_{-	1.79	}$	&	$	0.02	^{+	0.03	}_{-	0.02	}$	&								&	$	0.38	^{+	0.01	}_{-	0.01	}$	&	$	0.04	^{+	0.00	}_{-	0.00	}$	\\
HD68988 b	&	1.12	&	$	6.28	^{+	0.00	}_{-	0.00	}$	&	$	189.57	^{+	1.57	}_{-	1.59	}$	&	$	0.16	^{+	0.01	}_{-	0.01	}$	&	$	3.36	^{+	1.19	}_{-	1.11	}$	&	$	1.83	^{+	0.01	}_{-	0.01	}$	&	$	0.07	^{+	0.00	}_{-	0.00	}$	\\
c	&		&	$	4053.58	^{+	1411.19	}_{-	588.92	}$	&	$	68.13	^{+	13.26	}_{-	6.01	}$	&	$	0.16	^{+	0.09	}_{-	0.07	}$	&								&	$	5.69	^{+	1.68	}_{-	0.72	}$	&	$	5.17	^{+	1.14	}_{-	0.51	}$	\\
HD73267 b	&	0.89	&	$	1265.34	^{+	6.17	}_{-	9.40	}$	&	$	18.63	^{+	19.36	}_{-	8.74	}$	&	$	0.56	^{+	0.23	}_{-	0.53	}$	&	$	0.99	^{+	0.49	}_{-	0.55	}$	&	$	0.72	^{+	1.12	}_{-	0.43	}$	&	$	2.20	^{+	0.01	}_{-	0.01	}$	\\
c	&		&	$	1278.04	^{+	16.28	}_{-	23.90	}$	&	$	73.21	^{+	19.97	}_{-	5.33	}$	&	$	0.28	^{+	0.03	}_{-	0.09	}$	&								&	$	3.46	^{+	1.00	}_{-	0.23	}$	&	$	2.22	^{+	0.02	}_{-	0.03	}$	\\
HD73526 b	&	1.01	&	$	178.17	^{+	0.53	}_{-	1.50	}$	&	$	45.30	^{+	9.09	}_{-	6.41	}$	&	$	0.70	^{+	0.08	}_{-	0.09	}$	&	$	9.51	^{+	2.84	}_{-	2.23	}$	&	$	0.91	^{+	0.13	}_{-	0.11	}$	&	$	0.62	^{+	0.00	}_{-	0.00	}$	\\
c	&		&	$	193.42	^{+	0.47	}_{-	0.56	}$	&	$	114.30	^{+	8.71	}_{-	8.35	}$	&	$	0.52	^{+	0.04	}_{-	0.05	}$	&								&	$	2.79	^{+	0.16	}_{-	0.16	}$	&	$	0.66	^{+	0.00	}_{-	0.00	}$	\\
HD74156 b	&	1.24	&	$	51.65	^{+	0.01	}_{-	0.01	}$	&	$	116.23	^{+	3.58	}_{-	3.46	}$	&	$	0.65	^{+	0.01	}_{-	0.01	}$	&	$	8.55	^{+	0.85	}_{-	0.74	}$	&	$	1.88	^{+	0.04	}_{-	0.04	}$	&	$	0.29	^{+	0.00	}_{-	0.00	}$	\\
c	&		&	$	2519.02	^{+	20.76	}_{-	20.00	}$	&	$	109.95	^{+	13.05	}_{-	9.31	}$	&	$	0.41	^{+	0.05	}_{-	0.05	}$	&								&	$	7.74	^{+	1.12	}_{-	0.85	}$	&	$	3.89	^{+	0.02	}_{-	0.02	}$	\\
HD75289 b	&	1.19	&	$	3.51	^{+	0.00	}_{-	0.00	}$	&	$	53.94	^{+	1.37	}_{-	1.33	}$	&	$	0.02	^{+	0.02	}_{-	0.01	}$	&	$	0.73	^{+	1.03	}_{-	0.55	}$	&	$	0.45	^{+	0.01	}_{-	0.01	}$	&	$	0.05	^{+	0.00	}_{-	0.00	}$	\\
c	&		&	$	112.58	^{+	5535.62	}_{-	110.72	}$	&	$	2.51	^{+	10.42	}_{-	2.03	}$	&	$	0.54	^{+	0.34	}_{-	0.37	}$	&								&	$	0.04	^{+	0.61	}_{-	0.04	}$	&	$	0.48	^{+	6.09	}_{-	0.45	}$	\\
HD75898 b	&	1.30	&	$	419.62	^{+	10.28	}_{-	8.47	}$	&	$	69.03	^{+	9.74	}_{-	8.65	}$	&	$	0.10	^{+	0.08	}_{-	0.07	}$	&	$	5.21	^{+	1.49	}_{-	1.04	}$	&	$	3.00	^{+	0.43	}_{-	0.34	}$	&	$	1.20	^{+	0.02	}_{-	0.02	}$	\\
c	&		&	$	368.23	^{+	6020.36	}_{-	59.73	}$	&	$	32.11	^{+	86.77	}_{-	16.18	}$	&	$	0.51	^{+	0.32	}_{-	0.34	}$	&								&	$	1.34	^{+	5.41	}_{-	0.77	}$	&	$	1.10	^{+	6.26	}_{-	0.12	}$	\\
HD76700 b	&	1.13	&	$	4.44	^{+	368.07	}_{-	0.47	}$	&	$	12.00	^{+	15.68	}_{-	10.10	}$	&	$	0.23	^{+	0.54	}_{-	0.19	}$	&	$	7.53	^{+	3.00	}_{-	4.24	}$	&	$	0.22	^{+	0.07	}_{-	0.19	}$	&	$	0.06	^{+	1.00	}_{-	0.00	}$	\\
c	&		&	$	1.33	^{+	0.00	}_{-	0.01	}$	&	$	21.74	^{+	4.32	}_{-	18.71	}$	&	$	0.14	^{+	0.43	}_{-	0.10	}$	&								&	$	0.13	^{+	0.03	}_{-	0.11	}$	&	$	0.02	^{+	0.00	}_{-	0.00	}$	\\
HD80606 b	&	0.96	&	$	111.44	^{+	0.00	}_{-	0.00	}$	&	$	857.60	^{+	577.43	}_{-	312.73	}$	&	$	0.97	^{+	0.02	}_{-	0.02	}$	&	$	2.84	^{+	4.11	}_{-	2.27	}$	&	$	4.30	^{+	0.86	}_{-	0.56	}$	&	$	0.45	^{+	0.00	}_{-	0.00	}$	\\
c	&		&	$	0.49	^{+	0.00	}_{-	0.00	}$	&	$	18.43	^{+	4.06	}_{-	3.77	}$	&	$	0.41	^{+	0.20	}_{-	0.29	}$	&								&	$	0.06	^{+	0.01	}_{-	0.01	}$	&	$	0.01	^{+	0.00	}_{-	0.00	}$	\\
HD81040 b	&	0.96	&	$	1100.01	^{+	7.14	}_{-	8.03	}$	&	$	178.02	^{+	11.03	}_{-	12.52	}$	&	$	0.48	^{+	0.08	}_{-	0.06	}$	&	$	8.20	^{+	12.12	}_{-	6.91	}$	&	$	7.75	^{+	0.37	}_{-	0.58	}$	&	$	2.06	^{+	0.01	}_{-	0.01	}$	\\
c	&		&	$	207.07	^{+	265.81	}_{-	177.01	}$	&	$	33.11	^{+	13.01	}_{-	12.96	}$	&	$	0.25	^{+	0.44	}_{-	0.18	}$	&								&	$	0.87	^{+	0.34	}_{-	0.63	}$	&	$	0.68	^{+	0.50	}_{-	0.49	}$	\\
HD82943  b	&	1.13	&	$	445.99	^{+	4.32	}_{-	4.96	}$	&	$	37.75	^{+	3.92	}_{-	3.57	}$	&	$	0.25	^{+	0.14	}_{-	0.17	}$	&	$	3.83	^{+	1.51	}_{-	1.36	}$	&	$	1.48	^{+	0.10	}_{-	0.11	}$	&	$	1.19	^{+	0.01	}_{-	0.01	}$	\\
c	&		&	$	221.52	^{+	1.41	}_{-	1.59	}$	&	$	62.85	^{+	9.22	}_{-	7.35	}$	&	$	0.35	^{+	0.05	}_{-	0.05	}$	&								&	$	1.90	^{+	0.32	}_{-	0.25	}$	&	$	0.75	^{+	0.00	}_{-	0.00	}$	\\
HD8574 b	&	1.12	&	$	7931.89	^{+	4630.89	}_{-	4011.00	}$	&	$	59.61	^{+	114.58	}_{-	41.65	}$	&	$	0.52	^{+	0.30	}_{-	0.33	}$	&	$	3.86	^{+	3.52	}_{-	3.02	}$	&	$	4.80	^{+	8.94	}_{-	3.38	}$	&	$	8.08	^{+	2.90	}_{-	3.03	}$	\\
c	&		&	$	227.24	^{+	0.70	}_{-	0.69	}$	&	$	65.11	^{+	3.48	}_{-	3.41	}$	&	$	0.30	^{+	0.04	}_{-	0.04	}$	&								&	$	2.01	^{+	0.10	}_{-	0.10	}$	&	$	0.76	^{+	0.00	}_{-	0.00	}$	\\
HD86081 b	&	1.21	&	$	2.14	^{+	0.00	}_{-	0.00	}$	&	$	207.62	^{+	0.84	}_{-	0.84	}$	&	$	0.01	^{+	0.01	}_{-	0.00	}$	&	$	0.68	^{+	0.84	}_{-	0.50	}$	&	$	1.49	^{+	0.01	}_{-	0.01	}$	&	$	0.03	^{+	0.00	}_{-	0.00	}$	\\
c	&		&	$	554.39	^{+	5438.59	}_{-	549.17	}$	&	$	2.78	^{+	22.00	}_{-	2.24	}$	&	$	0.54	^{+	0.34	}_{-	0.37	}$	&								&	$	0.06	^{+	1.29	}_{-	0.06	}$	&	$	1.41	^{+	5.47	}_{-	1.34	}$	\\
HD89307 b	&	0.99	&	$	2173.22	^{+	50.00	}_{-	53.20	}$	&	$	34.25	^{+	21.54	}_{-	26.30	}$	&	$	0.33	^{+	0.41	}_{-	0.23	}$	&	$	3.05	^{+	5.33	}_{-	2.46	}$	&	$	2.05	^{+	0.83	}_{-	1.67	}$	&	$	3.27	^{+	0.05	}_{-	0.05	}$	\\
c	&		&	$	318.89	^{+	7411.65	}_{-	317.49	}$	&	$	9.35	^{+	85.84	}_{-	8.33	}$	&	$	0.49	^{+	0.36	}_{-	0.33	}$	&								&	$	0.12	^{+	6.16	}_{-	0.11	}$	&	$	0.91	^{+	6.72	}_{-	0.89	}$	\\
HD93083 b	&	0.70	&	$	241.82	^{+	3377.10	}_{-	225.82	}$	&	$	4.26	^{+	9.78	}_{-	2.78	}$	&	$	0.51	^{+	0.34	}_{-	0.34	}$	&	$	1.56	^{+	0.89	}_{-	0.83	}$	&	$	0.09	^{+	0.40	}_{-	0.08	}$	&	$	0.67	^{+	3.42	}_{-	0.56	}$	\\
c	&		&	$	143.85	^{+	1.68	}_{-	1.71	}$	&	$	17.87	^{+	1.46	}_{-	1.47	}$	&	$	0.12	^{+	0.07	}_{-	0.07	}$	&								&	$	0.36	^{+	0.03	}_{-	0.03	}$	&	$	0.48	^{+	0.00	}_{-	0.00	}$	\\
HR 810 b	&	1.11	&	$	302.20	^{+	1507.67	}_{-	299.19	}$	&	$	18.06	^{+	39.95	}_{-	16.02	}$	&	$	0.38	^{+	0.46	}_{-	0.29	}$	&	$	16.54	^{+	4.06	}_{-	3.58	}$	&	$	0.48	^{+	1.53	}_{-	0.45	}$	&	$	0.91	^{+	2.10	}_{-	0.87	}$	\\
c	&		&	$	302.15	^{+	2.91	}_{-	281.12	}$	&	$	55.27	^{+	7.21	}_{-	47.77	}$	&	$	0.14	^{+	0.34	}_{-	0.10	}$	&								&	$	1.93	^{+	0.26	}_{-	1.86	}$	&	$	0.91	^{+	0.01	}_{-	0.76	}$	\\
kappa CrB b	&	1.80	&	$	1185.12	^{+	158.87	}_{-	1182.23	}$	&	$	15.83	^{+	9.29	}_{-	14.64	}$	&	$	0.18	^{+	0.60	}_{-	0.15	}$	&	$	4.75	^{+	0.90	}_{-	0.84	}$	&	$	0.95	^{+	1.00	}_{-	0.93	}$	&	$	2.67	^{+	0.23	}_{-	2.62	}$	\\
c	&		&	$	1198.66	^{+	128.88	}_{-	1181.54	}$	&	$	22.27	^{+	2.93	}_{-	21.04	}$	&	$	0.14	^{+	0.57	}_{-	0.11	}$	&								&	$	1.71	^{+	0.25	}_{-	1.69	}$	&	$	2.69	^{+	0.19	}_{-	2.53	}$	\\
NGC 2423 3 b	&	2.40	&	$	234.98	^{+	3.77	}_{-	3.09	}$	&	$	55.95	^{+	7.84	}_{-	7.69	}$	&	$	0.09	^{+	0.11	}_{-	0.06	}$	&	$	31.88	^{+	4.64	}_{-	3.78	}$	&	$	3.02	^{+	0.42	}_{-	0.42	}$	&	$	1.00	^{+	0.01	}_{-	0.01	}$	\\
c	&		&	$	357.54	^{+	4.89	}_{-	5.52	}$	&	$	107.45	^{+	56.48	}_{-	16.98	}$	&	$	0.47	^{+	0.13	}_{-	0.13	}$	&								&	$	5.91	^{+	3.13	}_{-	0.96	}$	&	$	1.32	^{+	0.01	}_{-	0.01	}$	\\
NGC 4349 127 b	&	3.90	&	$	678.17	^{+	7.23	}_{-	7.34	}$	&	$	188.39	^{+	15.39	}_{-	12.26	}$	&	$	0.14	^{+	0.05	}_{-	0.05	}$	&	$	14.64	^{+	4.03	}_{-	3.24	}$	&	$	19.97	^{+	1.73	}_{-	1.42	}$	&	$	2.38	^{+	0.02	}_{-	0.02	}$	\\
c	&		&	$	213.34	^{+	3126.70	}_{-	210.52	}$	&	$	10.89	^{+	49.78	}_{-	9.60	}$	&	$	0.53	^{+	0.32	}_{-	0.34	}$	&								&	$	0.37	^{+	6.09	}_{-	0.34	}$	&	$	1.10	^{+	5.78	}_{-	1.04	}$	\\
tau Boo b	&	1.34	&	$	3.31	^{+	0.00	}_{-	0.00	}$	&	$	467.29	^{+	5.35	}_{-	5.14	}$	&	$	0.02	^{+	0.01	}_{-	0.01	}$	&	$	27.05	^{+	5.09	}_{-	4.69	}$	&	$	4.17	^{+	0.05	}_{-	0.05	}$	&	$	0.05	^{+	0.00	}_{-	0.00	}$	\\
c	&		&	$	11695.40	^{+	2209.03	}_{-	2416.40	}$	&	$	142.19	^{+	14.64	}_{-	13.08	}$	&	$	0.39	^{+	0.08	}_{-	0.11	}$	&								&	$	17.61	^{+	1.84	}_{-	1.68	}$	&	$	11.12	^{+	1.36	}_{-	1.59	}$	\\
TrES-3 b	&	0.92	&	$	1.34	^{+	0.09	}_{-	0.05	}$	&	$	348.72	^{+	23.24	}_{-	75.06	}$	&	$	0.08	^{+	0.20	}_{-	0.06	}$	&	$	9.34	^{+	23.48	}_{-	8.01	}$	&	$	1.77	^{+	0.12	}_{-	0.45	}$	&	$	0.02	^{+	0.00	}_{-	0.00	}$	\\
c	&		&	$	50.79	^{+	2572.53	}_{-	49.53	}$	&	$	87.40	^{+	641.98	}_{-	83.99	}$	&	$	0.46	^{+	0.35	}_{-	0.34	}$	&								&	$	1.26	^{+	11.74	}_{-	1.22	}$	&	$	0.26	^{+	3.35	}_{-	0.24	}$	\\
WASP-2 b	&	0.88	&	$	2.15	^{+	0.00	}_{-	0.00	}$	&	$	159.33	^{+	8.66	}_{-	7.16	}$	&	$	0.22	^{+	0.12	}_{-	0.13	}$	&	$	2.63	^{+	6.47	}_{-	2.12	}$	&	$	0.90	^{+	0.04	}_{-	0.04	}$	&	$	0.03	^{+	0.00	}_{-	0.00	}$	\\
c	&		&	$	265.16	^{+	4448.69	}_{-	263.05	}$	&	$	10.34	^{+	144.62	}_{-	9.20	}$	&	$	0.53	^{+	0.34	}_{-	0.35	}$	&								&	$	0.15	^{+	5.19	}_{-	0.14	}$	&	$	0.77	^{+	4.50	}_{-	0.74	}$	\\
WASP-3 b	&	1.22	&	$	1.85	^{+	0.00	}_{-	0.00	}$	&	$	249.49	^{+	11.68	}_{-	11.41	}$	&	$	0.08	^{+	0.05	}_{-	0.05	}$	&	$	3.58	^{+	10.43	}_{-	2.98	}$	&	$	1.71	^{+	0.08	}_{-	0.08	}$	&	$	0.03	^{+	0.00	}_{-	0.00	}$	\\
c	&		&	$	230.52	^{+	4532.94	}_{-	227.26	}$	&	$	10.38	^{+	92.25	}_{-	9.17	}$	&	$	0.54	^{+	0.33	}_{-	0.36	}$	&								&	$	0.19	^{+	3.82	}_{-	0.18	}$	&	$	0.79	^{+	5.13	}_{-	0.74	}$	\\
WASP-4 b	&	0.91	&	$	1.34	^{+	0.00	}_{-	0.00	}$	&	$	243.72	^{+	9.77	}_{-	10.23	}$	&	$	0.04	^{+	0.03	}_{-	0.03	}$	&	$	3.70	^{+	8.51	}_{-	3.07	}$	&	$	1.24	^{+	0.05	}_{-	0.05	}$	&	$	0.02	^{+	0.00	}_{-	0.00	}$	\\
c	&		&	$	710.90	^{+	2716.90	}_{-	618.39	}$	&	$	174.77	^{+	810.23	}_{-	163.03	}$	&	$	0.54	^{+	0.28	}_{-	0.36	}$	&								&	$	5.46	^{+	36.64	}_{-	5.32	}$	&	$	1.51	^{+	2.80	}_{-	1.12	}$	\\
WASP-5 b	&	1.01	&	$	1.63	^{+	0.00	}_{-	0.00	}$	&	$	282.05	^{+	9.68	}_{-	9.28	}$	&	$	0.07	^{+	0.04	}_{-	0.04	}$	&	$	2.89	^{+	7.23	}_{-	2.35	}$	&	$	1.64	^{+	0.05	}_{-	0.05	}$	&	$	0.03	^{+	0.00	}_{-	0.00	}$	\\
c	&		&	$	338.95	^{+	4577.80	}_{-	334.17	}$	&	$	10.05	^{+	107.77	}_{-	8.95	}$	&	$	0.53	^{+	0.33	}_{-	0.37	}$	&								&	$	0.17	^{+	4.54	}_{-	0.16	}$	&	$	0.95	^{+	4.72	}_{-	0.90	}$	\\
XO-1 b	&	1.03	&	$	3.94	^{+	0.00	}_{-	0.00	}$	&	$	118.26	^{+	13.15	}_{-	11.76	}$	&	$	0.10	^{+	0.12	}_{-	0.07	}$	&	$	3.26	^{+	8.68	}_{-	2.68	}$	&	$	0.93	^{+	0.10	}_{-	0.09	}$	&	$	0.05	^{+	0.00	}_{-	0.00	}$	\\
c	&		&	$	253.27	^{+	4507.93	}_{-	251.52	}$	&	$	9.60	^{+	86.27	}_{-	8.36	}$	&	$	0.56	^{+	0.32	}_{-	0.37	}$	&								&	$	0.14	^{+	3.70	}_{-	0.13	}$	&	$	0.79	^{+	4.80	}_{-	0.76	}$	\\
XO-2 b	&	0.97	&	$	2.62	^{+	0.00	}_{-	0.00	}$	&	$	85.83	^{+	10.59	}_{-	10.43	}$	&	$	0.18	^{+	0.15	}_{-	0.13	}$	&	$	3.52	^{+	9.57	}_{-	2.90	}$	&	$	0.55	^{+	0.07	}_{-	0.07	}$	&	$	0.04	^{+	0.00	}_{-	0.00	}$	\\
c	&		&	$	188.83	^{+	4420.68	}_{-	186.36	}$	&	$	9.80	^{+	95.98	}_{-	8.70	}$	&	$	0.52	^{+	0.35	}_{-	0.35	}$	&								&	$	0.14	^{+	3.68	}_{-	0.13	}$	&	$	0.64	^{+	4.73	}_{-	0.60	}$	\\
XO-3 b	&	1.41	&	$	3.60	^{+	0.01	}_{-	0.01	}$	&	$	1133.58	^{+	83.26	}_{-	78.68	}$	&	$	0.20	^{+	0.06	}_{-	0.06	}$	&	$	203.12	^{+	24.39	}_{-	20.46	}$	&	$	10.51	^{+	0.82	}_{-	0.77	}$	&	$	0.05	^{+	0.00	}_{-	0.00	}$	\\
c	&		&	$	3.38	^{+	0.01	}_{-	0.01	}$	&	$	2272.98	^{+	98.15	}_{-	97.47	}$	&	$	0.03	^{+	0.03	}_{-	0.02	}$	&								&	$	21.09	^{+	0.91	}_{-	0.91	}$	&	$	0.05	^{+	0.00	}_{-	0.00	}$	\\
XO-4 b	&	1.32	&	$	4.13	^{+	0.00	}_{-	0.00	}$	&	$	191.72	^{+	89.73	}_{-	32.29	}$	&	$	0.32	^{+	0.25	}_{-	0.21	}$	&	$	4.44	^{+	13.91	}_{-	3.74	}$	&	$	1.74	^{+	0.51	}_{-	0.26	}$	&	$	0.06	^{+	0.00	}_{-	0.00	}$	\\
c	&		&	$	235.76	^{+	4051.59	}_{-	233.46	}$	&	$	14.93	^{+	186.58	}_{-	13.58	}$	&	$	0.52	^{+	0.34	}_{-	0.35	}$	&								&	$	0.27	^{+	8.47	}_{-	0.25	}$	&	$	0.82	^{+	4.85	}_{-	0.78	}$	\\

\hline
		\end{tabular}
	\end{center}
\end{table*}

\begin{table*}
\caption{The number of published planets compared with the best-fit model from this analysis (i.e. the flags and reduced chi-square ratios for the results for each system). The `candidates' column shows the current number of confirmed planets (from \textit{http://www.exoplanet.eu} and \textit{http://exoplanets.org}, as of 2011 August 01), and the `visual quality flag' (assigned by eye) is the best \textsc{exofit} model, where `1' signifies that the 1-planet fit is best, and `2' means that the 2-planet fit is best. `3' means that both 1- and 2-planet solutions provide equally good or bad fits, and this class is again subdivided into `3a' and `3b', as explained in Section 6.2. Also shown is the log-likelihood ratio of the chi-square values, $R$, as defined in Section 5. Our visual flag assignments are validated somewhat by noting that in $99\%$ of systems the visual flag and chi-square results agree (or at least are not contradictory, for the class 3 cases). Those systems denoted by `-' are those where there were not sufficient degrees of freedom to calculate a value for the log likelihood ratio. The prior flag is also shown, where flag \textit{N} indicates that the analysis was performed using the normal priors shown in Tables 1 and 2, and flag \textit{D} indicates an analysis with different priors as shown in Table 3.}
\label{tab:flags}
\smallskip
\scriptsize

	\begin{center}
		\begin{tabular}{rcccc}
			\hline\noalign{\smallskip}
\multicolumn{1}{c}{\textbf{system}} & 
\multicolumn{1}{c}{\textbf{number of candidates}} & 
\multicolumn{1}{c}{\textbf{visual quality flag}} & 
\multicolumn{1}{c}{\textbf{$R$}} &
\multicolumn{1}{c}{\textbf{flag for period}} \\
 & 
\multicolumn{1}{c}{\textbf{from literature}} & 
 & &
\multicolumn{1}{c}{\textbf{prior used}} \\
			\hline\hline

BD-17 63	&	1	&	1	&	25.06	&	N	\\
ChaHa8	&	1	&	1	&	-	&	N	\\
epsilon Eri	&	1	&	3a	&	-0.84	&	D	\\
epsilon Tau	&	1	&	1	&	4.62	&	N	\\
gamma Cep	&	1	&	2	&	-398.83	&	D	\\
GJ3021	&	1	&	3b	&	0.28	&	N	\\
GJ317	&	1	&	1	&	47.75	&	N	\\
GJ674	&	1	&	3a	&	-8.44	&	N	\\
GJ849	&	1	&	3a	&	8.20	&	D	\\
GJ86	&	1	&	2	&	-1554.76	&	D	\\
HAT-P-6	&	1	&	1	&	1450.29	&	N	\\
HAT-P-8	&	1	&	3b	&	-	&	N	\\
HAT-P-9	&	1	&	3b	&	0.95	&	D	\\
HD101930	&	1	&	3b	&	6.12	&	N	\\
HD108874	&	2	&	2	&	-8.23	&	N	\\
HD11506	&	2	&	2	&	-3.36	&	N	\\
HD118203	&	1	&	1	&	80.12	&	D	\\
HD12661	&	2	&	2	&	-15.22	&	D	\\
HD128311	&	2	&	2	&	-32.30	&	D	\\
HD131664	&	1	&	1	&	3.34	&	D	\\
HD132406	&	1	&	1	&	234.41	&	N	\\
HD142	&	1	&	3a	&	-5.87	&	D	\\
HD142022	&	1	&	3b	&	0.43	&	N	\\
HD149143	&	1	&	3b	&	1.49	&	N	\\
HD154345	&	1	&	1	&	44.33	&	N	\\
HD155358	&	2	&	2	&	-2.90	&	N	\\
HD162020	&	1	&	3b	&	102.59	&	D	\\
HD168443	&	2	&	2	&	-994.46	&	D	\\
HD169830	&	2	&	3b	&	0.25	&	D	\\
HD171028	&	1	&	3a	&	18.72	&	N	\\
HD183263	&	2	&	2	&	-187.35	&	D	\\
HD185269	&	1	&	3b	&	0.83	&	N	\\
HD187123	&	2	&	3b	&	-29.06	&	D	\\
HD189733	&	1	&	3a	&	-140.98	&	D	\\
HD190228	&	1	&	3b	&	0.08	&	N	\\
HD190360	&	2	&	3b	&	0.04	&	D	\\
HD190647	&	1	&	3b	&	5.01	&	N	\\
HD195019	&	1	&	1	&	771.62	&	N	\\
HD202206	&	2	&	3b	&	-4.59	&	D	\\
HD20868	&	1	&	2	&	-252.94	&	D	\\
HD209458	&	1	&	3b	&	-0.03	&	D	\\
HD212301	&	1	&	1	&	5.15	&	N	\\
HD217107	&	2	&	3a	&	-59.88	&	D	\\
HD219828	&	1	&	3a	&	-352.73	&	D	\\
HD221287	&	1	&	1	&	141.08	&	N	\\
HD224693	&	1	&	3b	&	20.92	&	N	\\
HD23127	&	1	&	1	&	69.47	&	N	\\
HD2638	&	1	&	3b	&	-0.97	&	N	\\
HD27442	&	1	&	3b	&	10.37	&	N	\\
HD27894	&	1	&	1	&	53.60	&	N	\\
HD28185	&	1	&	1	&	10.27	&	D	\\
HD285968	&	1	&	1	&	0.42	&	N	\\
HD330075	&	1	&	3b	&	745.43	&	D	\\
HD33636	&	1	&	1	&	104.05	&	D	\\
HD3651	&	1	&	3b	&	-0.06	&	N	\\
HD38529	&	2	&	2	&	-41.83	&	D	\\
HD4203	&	1	&	3a	&	4.92	&	N	\\
HD4208	&	1	&	3b	&	0.65	&	N	\\
HD43691	&	1	&	3b	&	5.07	&	N	\\
HD43848	&	1	&	1	&	-	&	N	\\
HD46375	&	1	&	3b	&	-0.04	&	D	\\
HD47536	&	2	&	1	&	17.15	&	D	\\
HD49674	&	1	&	3b	&	0.24	&	N	\\
HD50499	&	1	&	2	&	-6.74	&	D	\\

\multicolumn{5}{r}{\tiny{\textit{continued on next page...}}}
		\end{tabular}
	\end{center}
\end{table*}

\begin{table*}
\scriptsize
\smallskip
	\begin{center}
		\begin{tabular}{rcccc}
			\hline
\multicolumn{1}{c}{\textbf{system}} & 
\multicolumn{1}{c}{\textbf{number of candidates}} & 
\multicolumn{1}{c}{\textbf{visual quality flag}} & 
\multicolumn{1}{c}{$R$} &
\multicolumn{1}{c}{\textbf{flag for period}} \\
 & 
\multicolumn{1}{c}{\textbf{from literature}} & 
 & &
\multicolumn{1}{c}{\textbf{prior used}} \\
			\hline\hline

HD5319	&	1	&	2	&	-12.72	&	D	\\
HD63454	&	1	&	3b	&	2.57	&	N	\\
HD68988	&	1	&	2	&	-100.51	&	D	\\
HD73267	&	1	&	3b	&	7.60	&	D	\\
HD73526	&	2	&	2	&	-7.43	&	N	\\
HD74156	&	2	&	2	&	-241.14	&	D	\\
HD75289	&	1	&	1	&	0.14	&	N	\\
HD75898	&	1	&	3a	&	28.79	&	N	\\
HD76700	&	1	&	1	&	14.28	&	N	\\
HD80606	&	1	&	1	&	4.22	&	D	\\
HD81040	&	1	&	3b	&	-1.13	&	N	\\
HD82943 	&	2	&	2	&	-319.74	&	N	\\
HD8574	&	1	&	1	&	-0.03	&	N	\\
HD86081	&	1	&	1	&	0.19	&	D	\\
HD89307	&	1	&	3b	&	-	&	D	\\
HD93083	&	1	&	3b	&	32.00	&	N	\\
HR 810	&	1	&	3b	&	8.71	&	N	\\
kappa CrB	&	1	&	1	&	8.87	&	N	\\
NGC 2423 3	&	1	&	3b	&	52.75	&	N	\\
NGC4 349 127	&	1	&	3b	&	24.85	&	N	\\
tau Boo	&	1	&	3b	&	-8.90	&	D	\\
TrES-3	&	1	&	2	&	-	&	D	\\
WASP-2	&	1	&	3b	&	-	&	D	\\
WASP-3	&	1	&	3b	&	-	&	D	\\
WASP-4	&	1	&	3b	&	-33.89	&	N	\\
WASP-5	&	1	&	3b	&	-	&	N	\\
XO-1	&	1	&	1	&	-	&	D	\\
XO-2	&	1	&	3b	&	-	&	D	\\
XO-3	&	1	&	1	&	20.24	&	N	\\
XO-4	&	1	&	3b	&	-	&	D	\\

\hline
		\end{tabular}
	\end{center}
\end{table*}

\label{lastpage}

\end{document}